\documentclass{ijuc}
\usepackage{graphicx} 
\usepackage[square,numbers,sort]{natbib}  
\usepackage{svg}
\usepackage{url}
\usepackage{array}
\usepackage{makecell}
\usepackage{amssymb}
\usepackage{amsmath}
\usepackage{multirow}
\usepackage{multicol}
\usepackage{subcaption}
\usepackage{float}
\usepackage{longtable}

\title{A Sensitivity Analysis of Cellular Automata and Heterogeneous Topology Networks: Partially-Local Cellular Automata and Homogeneous Homogeneous Random Boolean Networks}

\author{Tom Eivind Glover\inst{1}\email{tomglove@oslomet.no}
\and Ruben Jahren\inst{1} 
\and Francesco Martinuzzi\inst{4,5,6}
\and Pedro Gon\c{c}alves Lind \inst{1,3}
\and Stefano Nichele \inst{1,2}
}

\institute{Department of Computer Science,  OsloMet - Oslo Metropolitan University, Oslo, Norway
\and
Department of Computer Science and Communication,  Østfold University College, Halden, Norway
\and
Simula Research Laboratory, Numerical Analysis and Scientific Computing, 0164 Oslo, Norway
\and
Center for Scalable Data Analytics and Artificial Intelligence,  Leipzig University, Leipzig, Germany
\and
Institute for Earth System Science \& Remote Sensing, Leipzig University, Leipzig, Germany
\and
Remote Sensing Centre for Earth System Research,  Leipzig University, Leipzig, Germany
}

\begin{document}

\maketitle

\begin{abstract}
    Elementary Cellular Automata (ECA) are a well-studied computational universe that is, despite its simple configurations, capable of impressive computational variety. Harvesting this computation in a useful way has historically shown itself to be difficult, but if combined with reservoir computing (RC), this becomes much more feasible. Furthermore, RC and ECA enable energy-efficient AI, making the combination a promising concept for Edge AI. In this work, we contrast ECA to substrates of Partially-Local CA (PLCA) and Homogeneous Homogeneous Random Boolean Networks (HHRBN). They are, in comparison, the topological heterogeneous counterparts of ECA. This represents a step from ECA towards more biological-plausible substrates. We analyse these substrates by testing on an RC benchmark (5-bit memory), using Temporal Derrida plots to estimate the sensitivity and assess the defect collapse rate. We find that, counterintuitively, disordered topology does not necessarily mean disordered computation. There are countering computational "forces" of topology imperfections leading to a higher collapse rate (order) and yet, if accounted for, an increased sensitivity to the initial condition. These observations together suggest a shrinking critical range. 
\end{abstract}


\section*{List of Abbreviations}
\begin{longtable}{p{0.2\textwidth} p{0.8\textwidth}}
    CA & Cellular Automata \\
    ECA & Elementary Cellular Automata \\
    RBN & Random Boolean Networks \\
    HHRBN & Homogeneous Homogeneous Random Boolean Networks  \\
    ME & Minimum Equivalence \\
    PLCA & Partially-local CA  \\
    RC & Reservoir Computing \\
    ReCA & Reservoir Computing with Cellular Automata \\
    TDP & Temporal Derrida Plot \\
    ESN & Echo State Networks \\
    LSM & Liquid State Machines \\
    pRNG & pseudo-Random Number Generator \\
    RNG &  Random Number Generator\\
\end{longtable}

\section{Introduction}

Standard Artificial Intelligence (AI) approaches rely on high-performance computing such as with cloud or cluster computing. However, these are very energy-intensive resources, and many popular models are energy-intensive in training \citep{strubell2019energy} and inference \citep{luccioni2023power}. Conversely, biological intelligence has made highly energy-effective solutions, e.g. the brain. Despite operating under conditions such as increased decentralisation, asynchronisation, and slower signal propagation, biological intelligence has achieved highly energy-efficient solutions, with the human brain operating at approximately 25-30 watts \citep{neumann1966theory, howard2012energy}. These observations indicate that there is still much to learn and gain from studying biological intelligence. 

\begin{figure}[h]
\begin{center}
\includegraphics[width=1\columnwidth]{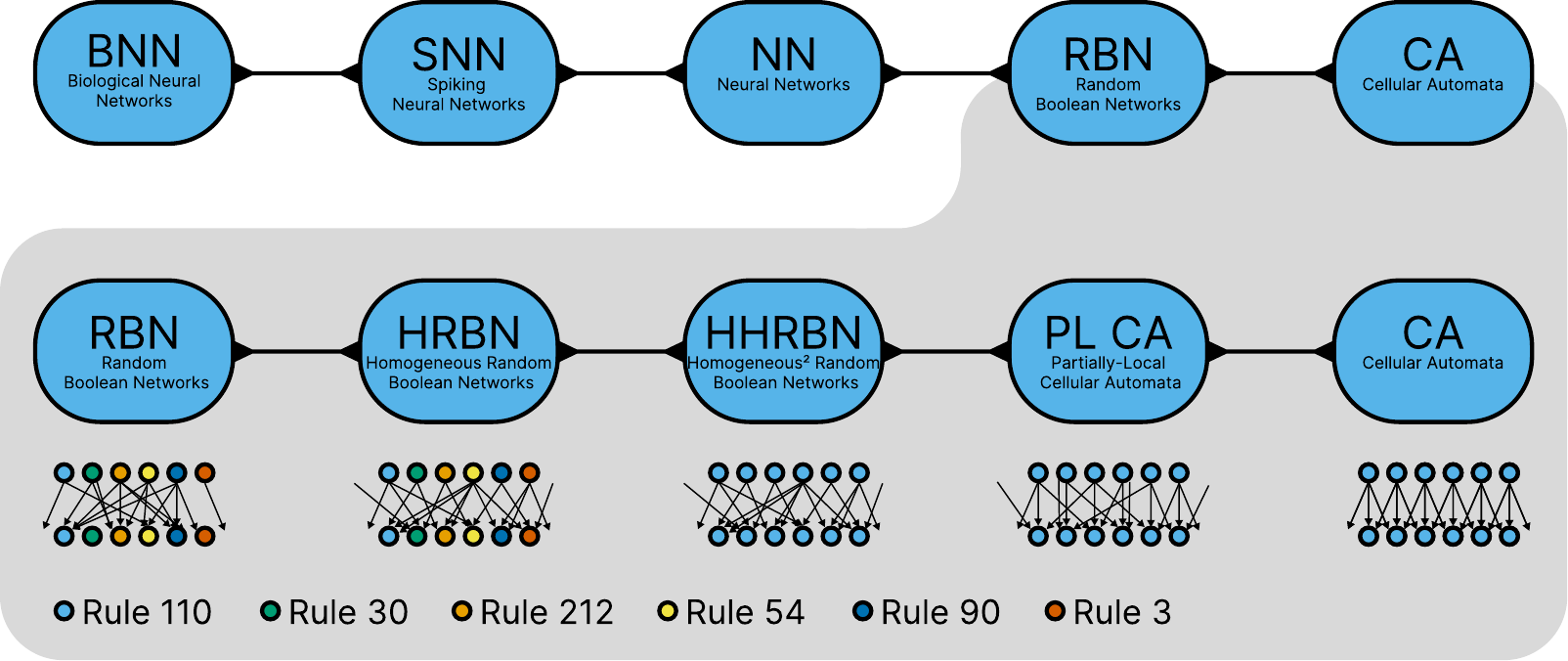} 
\caption{There is a big difference between CA and BNN. The difference can be viewed as a series of discrete steps between substrates, but even between RBN and CA, there are many discrete steps. This figure illustrates the different substrates as a direct path, but note that the steps from CA to RBN could have been done differently than illustrated. Also note, that this is a simplified imperfect model of the substrate space between ECA and BNN.}
\label{fig:SubstrateOverview}
\end{center}
\end{figure}

In this work, we focus on unconventional computational models, such as Cellular Automata (CA) or Random Boolean Networks (RBN), which utilise Boolean logic between local cells (nodes). This reliance on Boolean logic enables easy hardware implementation, as the operations can be implemented in circuitry or an FPGA, allowing energy-efficient inference of the model. It is possible to create an abstract pathway from CA to Biological Neural Networks (BNN); one example can be seen in Figure \ref{fig:SubstrateOverview}, and this pathway would require many steps. CA is a special case of RBN where the neighbour connections are entirely regular, and every cell has the same activation function. Viewed from the other direction, RBN is a CA with random neighbourhood and random rules per cell (node). RBN is a well-known simple biological model of the Gene regulatory network \citep{mitchell2009complexity, kauffman_metabolic_1969}, which is an intelligence- and computational space used for solving, among other things, morphological problems. Though there are multiple discrete steps between CA and RBN, as can be seen in Figure \ref{fig:SubstrateOverview}, we limit ourselves to exploring substrates between CA and Homogeneous Homogeneous RBN (HHRBN); these are the substrates that have the same rule (Homogeneous) in every cell. Essentially, we compare homogeneous topology networks to heterogeneous topology networks. 

Note that many of the modern directions of these CA and RBN computational models are moving into the continuous domain, such as for CA, the continuous models of Lenia \citep{Chan2019Lenia} and Neural CA \citep{mordvintsev2020growing} are having much success modelling biological processes and biological like behaviours. Similarly, Random Boolean Networks (RBN) are moving into the continuous domain, such as continuous RBN \citep{vohradsky2001neural} or stochastic RBN \citep{ribeiro2006general, elowitz2002stochastic}, and they are argued to be more biologically plausible models. Although we want to encourage these explorations and essential directions, relying on the continuous domain currently means either running on specialised hardware or taking an energy efficiency loss, as floating point calculations are much more costly than integer or, even further, binary calculations. The future is still being determined, but when it comes to specialised hardware like neuromorphic chips \citep{orchard2021efficient}, they are not expected to replace traditional hardware. Unless there is a significant breakthrough, specialised hardware like neuromorphic chips or quantum computing will likely exist alongside and in cooperation with conventional computational systems. Therefore, the energy-efficient binary models are still worth developing.   

In this work, we stay within the binary domain for energy efficiency. We explore and compare three binary models for computation, namely CA and models intermediate to RBN (PLCA, HHRBN). We do this mainly to understand them as computational models better. Firstly, We combine them with Reservoir Computing (RC), an energy-efficient "substrate-independent" training method. The combination can potentially be an energy-efficient model in training and inference. We run these models on a simple 5-bit memory benchmark and find that ECA generally performs better. We follow up by measuring the sensitivity (a necessary but not sufficient condition of chaos) of the networks using a Temporal Derrida plot. We find that, in general, HHRBN and PLCA slightly increase the sensitivity. Yet, by analysing the defect collapse rate, we find that the substrates as a whole also skew towards ordered behaviour as there are mitigating circumstances like how the random topology leads to imperfect connectivity that leads to a stronger attractor. This signifies that we observe a shrinking critical range in PLCA and HHRBN. Though HHRBN are disordered in the topology, the behaviour of the network from these effects is not as disordered as it might be natural to assume. This means counter-intuitively that regular connections can more reliably reach a higher level of disorder.

\section{Background and Related Work}
In this section, we will detail the background, related work, and theory relevant to this paper. This paper connects many fields, uses several substrates, and relies on several empirical and theoretical methods. This has made this section necessarily extensive to provide a comprehensive overview. In general this sections begins by explaining the different substrates, followed by more theoretical overview of said substrates as well as relevant concepts. Finally, the relevant related work is presented. 
\subsection{Cellular Automata}
Cellular Automata (CA) are a simple model consisting of a grid of cells possessing a limited set of $k$ discrete states placed on a uniformly connected grid, typically in 1 or 2 dimensions. The cell state changes iteratively, depending on the state of the neighbours. Which neighbour state combination results in which next state is determined by a lookup table, typically called the Transition Table (TT). CA was first used to study self-replication by John von Neumann in 1940 but published in 1966 \citep{neumann1966theory}. It can be considered an idealised system for parallel and decentralised computation \citep{mitchell2001life}.
\subsubsection{Elementary Cellular Automata (ECA)}

\begin{figure}[h]
\begin{center}
\includegraphics[width=1\columnwidth]{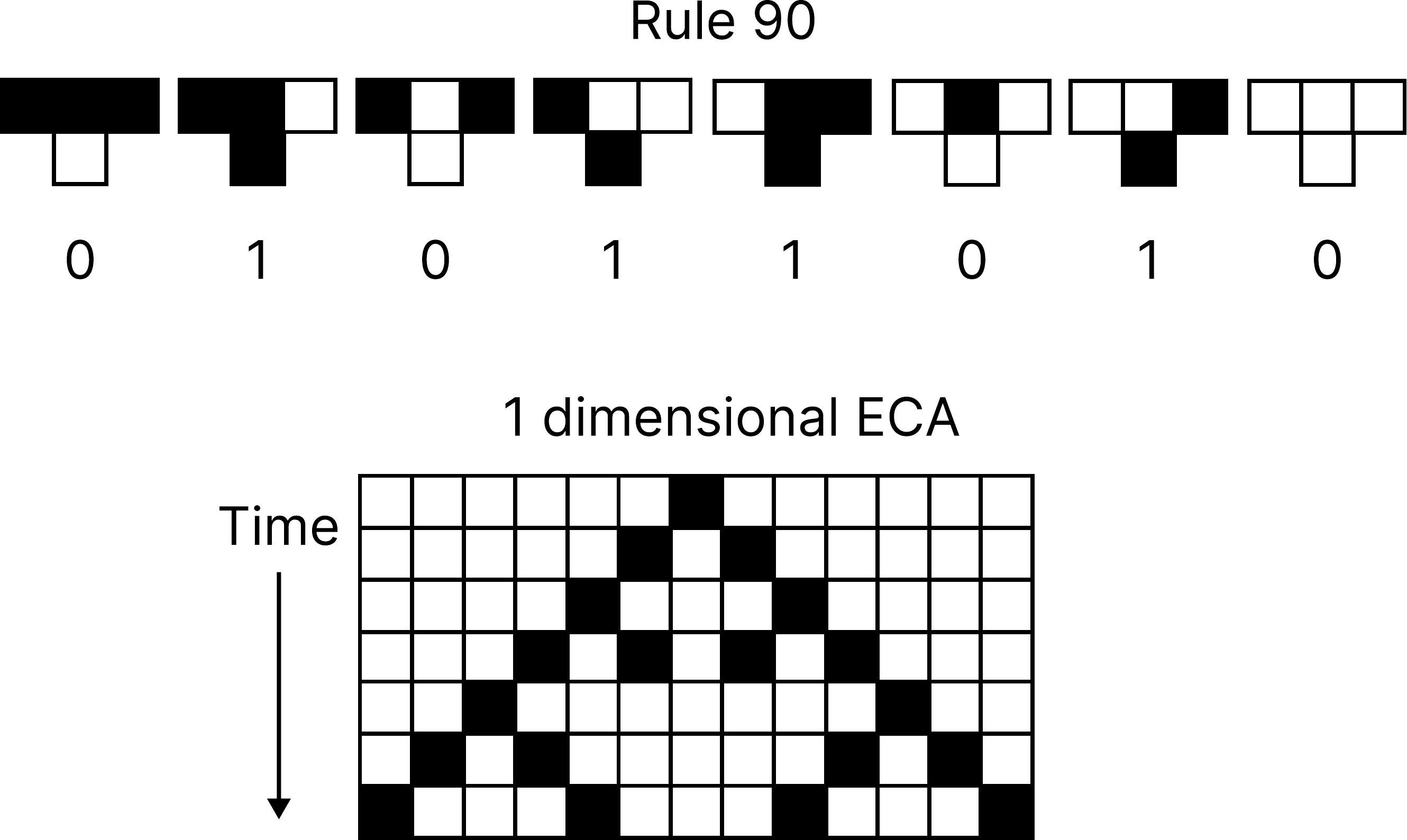} 
\caption{Example of 1 dimensional CA with rule 90 with TT, starting from a central cell on, executing 7 time-steps.}
\label{fig:ECAex}
\end{center}
\end{figure}

Elementary Cellular automata (ECA) is a subset of CA in 1-dimension, binary states ($S=2$) and 3 neighbours ($K=3$) (left, right and centre). Therefore, ECA only has $S^{S^K} = 2^{2^3} = 256$ possible rules, and the whole set of these is often named the rule-space. It is a convention to name individual rules in a rule-space after the output states of the TT $\mbox{Binary}(01011010) = \mbox{Decimal}(90)$. CA is deterministic, and the rule, together with the initial condition, leads the CA into a set of subsequent states called the trajectory. An example of rule 90 can be seen in Figure \ref{fig:ECAex}. Rule 110 has even been shown to be computationally universal \citep{cook2004universality}, but one can question whether that is a useful definition of computation for a parallel and distributed computational substrate \citep{barbora2021hirarchy}. 
\subsubsection{Two Dimensional CA (2D CA)}
\begin{figure}[h]
\begin{center}
\includegraphics[width=0.6\columnwidth]{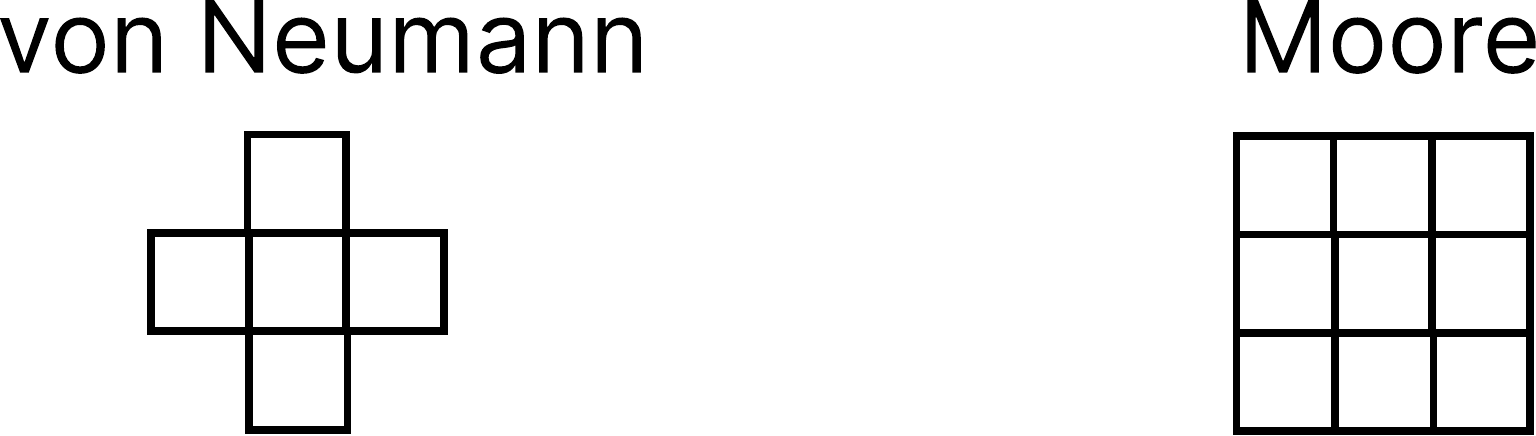} 
\caption{Common 2-dimensional neighbourhood schemes.}
\label{fig:2DCANeightbourhoods}
\end{center}
\end{figure}

Beyond ECA are many other types of CA, such as 2-dimensional CA, where instead of configuring the cells in a 1-dimensional line, they are now configured as a 2D surface. In 2D CA, the most typical neighbourhood scheme is one of two configurations in Figure \ref{fig:2DCANeightbourhoods}.

The Rule space of 2D CA is quite large, especially with a Moore neighbourhood. This space have $2^{2^9} = 1.32 * 10^{154}$ different Rules. It is too large to search exhaustively, but explorations into 2D CA are often limited to totalistic or outer-totalistic rules. Totalistic rules mean the rule does not distinguish which neighbours are in which state, but rather "counts" the number of neighbours with a specific state. 2D CA, with a Moore neighbourhoods, has only ten states to differentiate 0,1,...9 alive neighbours. Only $2^{10} = 1024$ totalistic rules exist in this rule-space. Outer-totalistic does the same but differs on the central cell. This means if the central cell is "dead" there are 9 (0,1,...8) different totalistic states the outer neighbours can be in and likewise, if the central cell is "alive". This means there are $2*9=18$ states for outer-totalistic rules to differentiate. Therefore, there are $2^{18} = 262 144$ different rules in this rule-space, though this can be somewhat reduced with symmetry equivalence classes \citep{glover2023minimum}. 

\begin{figure}[h]
\begin{center}
\includegraphics[width=0.65\columnwidth]{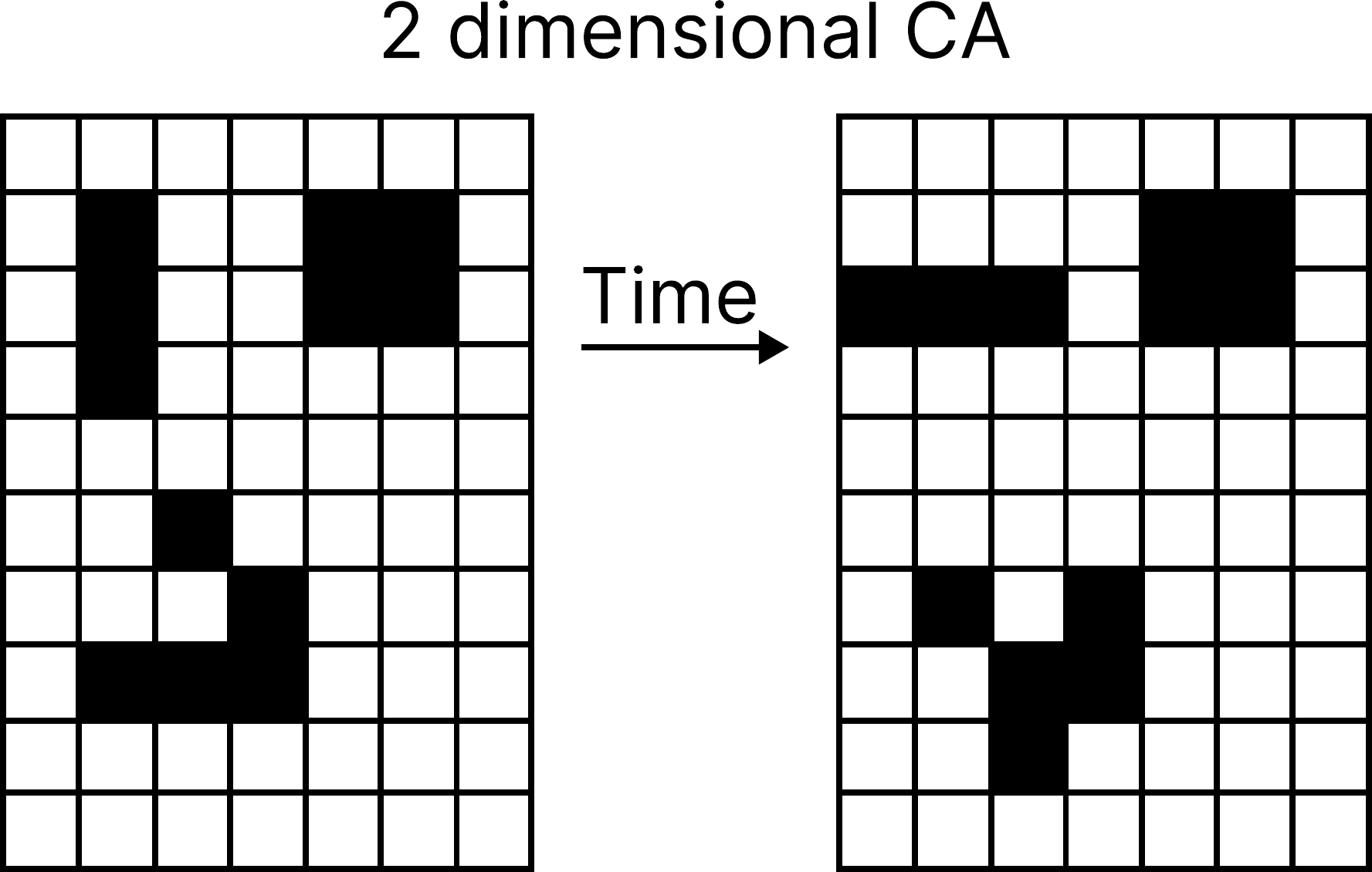} 
\caption{Single time-step of a 2-dimensional CA with Conway's Game of Life rules. It features an oscillating blinker, a stable block, and a spaceship glider.}
\label{fig:2DCAex}
\end{center}
\end{figure}

The most famous version of 2D CA is Game of Life (GoL) \citep{games1970fantastic}, Figure \ref{fig:2DCAex}. GoL is an outer-totalistic rule, and it works in the following manner: at each CA step, the next state changes depending on the following rules 
\begin{itemize}
    \item Any living cell dies if it has two or fewer living neighbours.
    \item Any Living cell persists if it has two or three living neighbours.
    \item Any living cell dies if it has more than three live neighbours.
    \item Any dead cell becomes alive if it has exactly three living neighbours.
\end{itemize}
In \citep{rendell2011universal}, it was also demonstrated that Game of Life is Turing complete. 
GoL can be expressed in a more general form, which is the convention for the outer-totalistic 2D CA. GoL would be in the form of B3/S23, where the Birth "B" component expresses the sum of neighbours needed to come alive, and the Survives "S" component expresses the sum of neighbours needed to stay alive. If a sum falls outside this value, the cell will die or remain dead.

\subsubsection{"Life-Like" Rules} 
\label{sec:lifeLikeBackground}
The 2D outer-totalistic binary CA rule space is often called the "life-like" \citep{conwaylife} rules-space, and beyond GoL, there are several other rules that are said to have "life-like" properties.
In \citep{eppstein2010growth}, the goal was to identify rules that could support similar "life-like" structures that can be constructed in GoL. In addition to identifying new ones, this article also provides an overview of many previously studied life-like rules that support structures such as replicators, oscillators, and spaceships. 
\subsection{Random Boolean Networks}
\begin{figure}[h]
\begin{center}
\includegraphics[width=1\columnwidth]{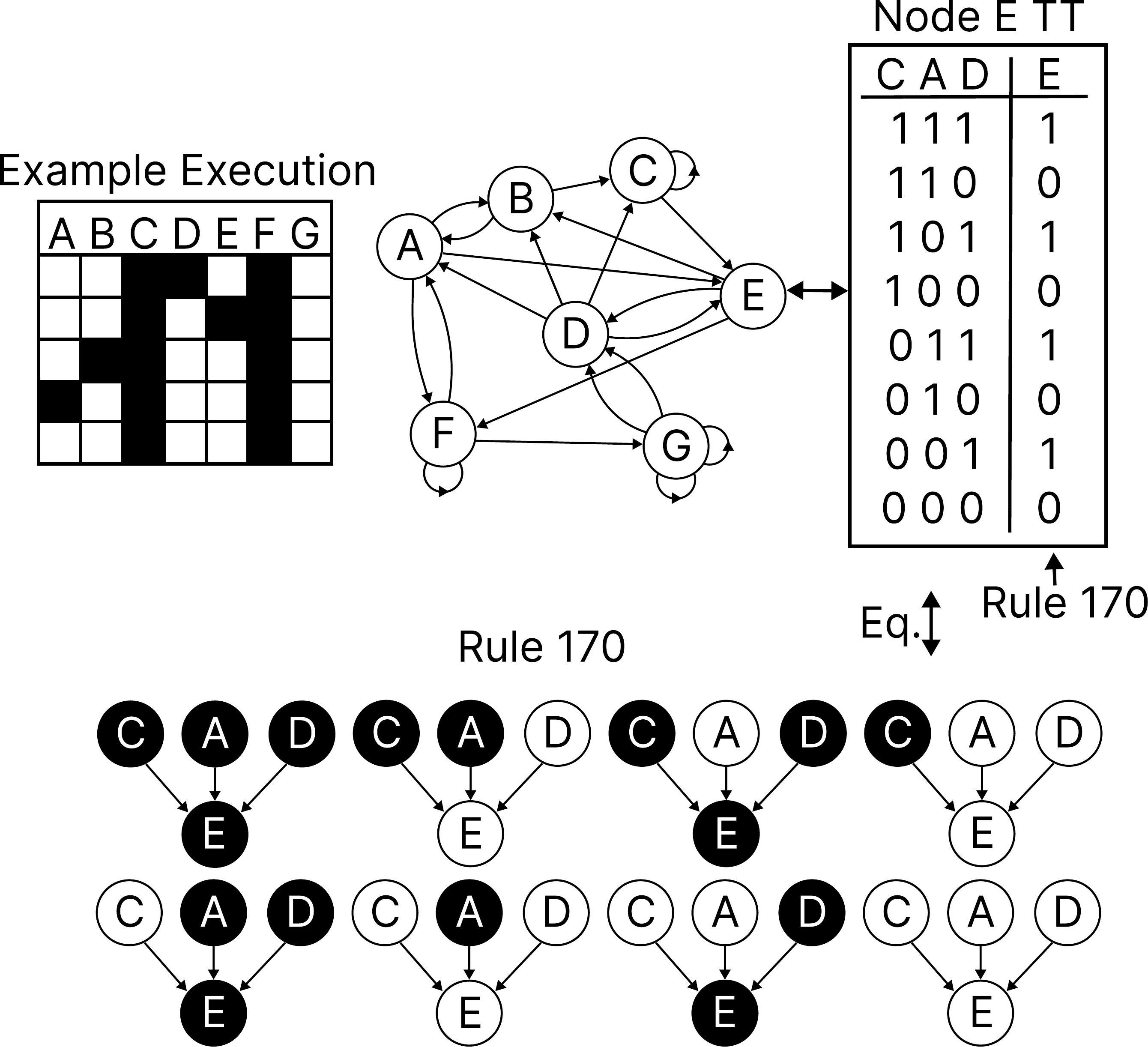} 
\caption{Example of an RBN with 7 Nodes and 3 neighbours, with a transition Table in two forms and a short execution example.}
\label{fig:RBNex}
\end{center}
\end{figure}

The RBN is similar to a CA yet has two key differences. Firstly, in the RBN, the grid neighbour connections are not regular but randomly set up. Secondly, every node (cell) typically has a random TT, often called an Activation function or Boolean function. This type of RBN is also sometimes called Classical RBN (CRBN) \citep{gershenson2002classification}. 
The number of direct neighbours can be random, semi-random or constant. The latter is called homogeneous RBN \citep{gershenson2002classification}, an example is given in Figure \ref{fig:RBNex}. 

As with CAs, several extensions exist beyond the original RBN, such as Continuous RBN \citep{vohradsky2001neural} or stochastic RBN \citep{ribeiro2006general, elowitz2002stochastic}. While Kauffman first developed the CRBN to model gene regulatory networks, these more modern extensions to the RBN model better emulate the biological activity of development \citep{vohradsky2001neural, elowitz2002stochastic}. However, the CRBN were discovered early on \citep{kauffman1971homeostasis} to contain a limited number of stable states, or attractors, from which the system would settle down to following a random initialisation. The basin of attraction reduces numerous initial states to a few stable cycles or fixed points. 

RBN can be defined as the following. A set of $N$ nodes connected randomly to $K$ number of other nodes, the specific connections for a given node can be denoted by $K_N$. The nodes can be in one of the two binary states, and every $N$ has the a random activation function $f_a$ (TT), out of $2^{2^K}$ possible rule setups. 

\subsection{RBN Classification}
ECA is often partitioned and classified into several different categories or traits. In \citep{martinez_note_2013}, a good overview of many common or well-known ones can be found. 

Similarly, RBN can be classified by their behaviour, i.e. ordered, complex or chaotic \citep{gershenson2002classification, kauffman1993origins}. Depending on the value of $N$ and $K$, the behaviour might differ, and one alternative name for RBN is the NK model. In \citep{kauffman1990requirements}, Kauffmann added another parameter $P$, which can organise the rule-space. The rule has a given P parameter value based on the number of neighbourhood combinations resulting in a 1 or a 0. In later work \citep{kauffman1993origins}, the larger distribution dominates, meaning $P\geq0.5$. Figure \ref{fig:RBNex} has $P=0.5$. One can use this parameter to control the behaviour. $P$ close to 1 would likely result in ordered behaviour, and $P$ close to 0.5 would likely result in chaotic behaviour. In between these, a critical (complex) $P_c$ behaviour might be found in the phase transition between order and chaos. This point or border is often also called the edge of chaos. The work is reminiscent of CA work in \citep{langton1990computation}, which we will introduce later. The $P$ is the same as $\lambda$ is in Langton's work, and in this work, we will use the $\lambda$ notation for both. 

Another way to categorise RBN and CA is to look at the basin of attraction. \citep{wuensche_global_1992, wuensche_basin_1992, wuensche_ghost_1994, wuensche1997attractor} did extensive work in both RBN and CA and their basin of attractions. What opened up this possibility was a method that could calculate backwards from a state. Take a cell in a state and consider what possible local neighbourhood configurations would result in this state. These are the possible previous states (preimage) for the neighbourhood. Finally, this can be applied to all the cells and limit the possibilities between cells by constraining satisfaction. The possible preimages often collapse to very few, making it possible to calculate the basin of attraction quickly.  

\subsection{Intermediate Substrates}

A system can be in a range of possible states that would be somewhere between CA and RBN. This paper will discuss a substrate with homogeneous rules but random neighbour wiring. This is what we define as HHRBN. HHRBN to distinguish it from what is in \citep{gershenson2002classification} called HRBN. It is called HHRBN rather than non-local CA because the substrate seems to behave more like RBN than CA, and the equivalence in this substrate is more applicable in RBN than CA. 

In \citep{li_phenomenology_1992}, Li worked with systems where all cells had the same activation function (TT), but the neighbour connections were in various configurations. In this work Li classified the different connection schemes as between non-local (random) and partially-local (central self-reference) as well as non-distinct and distinct input/output (uniform number of outputs). Li then classifies the rule-space for these substrates using mean field approximation and shows they are very neatly classified, particularly non-local CA (HHRBN). 

Much earlier \citep{walker_study_1965,walker_behavior_1971,walker_temporal_1966} studied a system that Li would classify as partially-local CA. 

HHRBN has additional commonly used names beyond non-local CA \citep{li_phenomenology_1992}, such as Graph CA \citep{marr2009outer, grattarola2021learning} or (Cellular) Automata Networks \citep{bhattacharjee2020survey}

In \citep{wuensche_basin_1992}, Wuensche examined substrates between CA and RBN, including non-local CA and other disordered CA. He defines disordered CA as a super-set of CA, which includes non-local CA and mixed rule CA. Furthermore, Wuensche calculates these networks' basin of attraction fields and demonstrates how rewiring the network can train or modify the basin of attraction.

Mixed rule CA is also known as Non-uniform CA \citep{cattaneo2009non, bhattacharjee2020survey} or hybrid CA \citep{bhattacharjee2020survey}.

\subsection{Minimum Equivalent (ME)}
In ECA, RBN and everything in between, we find equivalence classes that effectively reduce the number of unique rules for a given substrate. The List of rules that make up the minimum set of unique rules is called the Minimum equivalent (ME). These rules can be used as a smaller replacement for the entire computational space of a specific substrate.  
\subsubsection{Minimum Equivalence (ME) in Elementary Cellular Automata (ECA)}
\begin{table}[h]
\scriptsize
\centerline{
\begin{tabular}{|c|c||c|c||c|c|}
\hline
 Rule & Equivalent & Rule & Equivalent & Rule & Equivalent\\
\hline\hline
0 & 255 & 35 & 49,59,115 & 108 & 201 \\
\hline
1 & 127 & 36 & 219 & 110 & 124,137,193 \\
\hline
2 & 16,191,247 & 37 & 91 & 122 & 161 \\
\hline
3 & 17,63,119 & 38 & 52,155,211 & 126 & 129 \\
\hline
4 & 223 & 40 & 96,235,249 & 128 & 254 \\
\hline
5 & 95 & 41 & 97,107,121 & 130 & 144,190,246 \\
\hline
6 & 20,159,215 & 42 & 112,171,241 & 132 & 222 \\
\hline
7 & 21,31,87 & 43 & 113 & 134 & 148,158,214 \\
\hline
8 & 64,239,253 & 44 & 100,203,217 & 136 & 192,238,252 \\
\hline
9 & 65,111,125 & 45 & 75,89,101 & 138 & 174,208,244 \\
\hline
10 & 80,175,245 & 46 & 116,139,209 & 140 & 196,206,220 \\
\hline
11 & 47,81,117 & 50 & 179 & 142 & 212 \\
\hline
12 & 68,207,221 & 51 & & 146 & 182 \\
\hline
13 & 69,79,93 & 54 & 147 & 150 & \\
\hline
14 & 84,143,213 & 56 & 98,185,227 & 152 & 188,194,230 \\
\hline
15 & 85 & 57 & 99 & 154 & 166,180,210 \\
\hline
18 & 183 & 58 & 114,163,177 & 156 & 198 \\
\hline
19 & 55 & 60 & 102,153,195 & 160 & 250 \\
\hline
22 & 151 & 62 & 118,131,145 & 162 & 176,186,242 \\
\hline
23 & & 72 & 237 & 164 & 218 \\
\hline
24 & 66,189,231 & 73 & 109 & 168 & 224,234,248 \\
\hline
25 & 61,67,103 & 74 & 88,173,229 & 170 & 240 \\
\hline
26 & 82,167,181 & 76 & 205 & 172 & 202,216,228 \\
\hline
27 & 39,53,83 & 77 & & 178 & \\
\hline
28 & 70,157,199 & 78 & 92,141,197 & 184 & 226 \\
\hline
29 & 71 & 90 & 165 & 200 & 236 \\
\hline
30 & 86,135,149 & 94 & 133 & 204 & \\
\hline
32 & 251 & 104 & 233 & 232 & \\
\hline
33 & 123 & 105 & & & \\
\hline
34 & 48,187,243 & 106 & 120,169,225 & & \\
\hline

\end{tabular}}
\caption{The group of equivalent rules for ECA.  }
\label{tab:CAEQ}
\end{table}

ECA consists of $2^{2^3}=256$ rules, but due to symmetries and other properties, there are only 88 rules that are considered unique. The reason is that all excluded rules can be transformed into one of the 88 unique rules by one of the following trivial methods. 

\begin{itemize}
    \item reflection: switching left and right
    \item complement: switching 0 and 1
    \item reflection and complement: the combination of both transformations
\end{itemize}

\begin{figure}[h]
    \begin{center}
        \includegraphics[width=1\columnwidth]{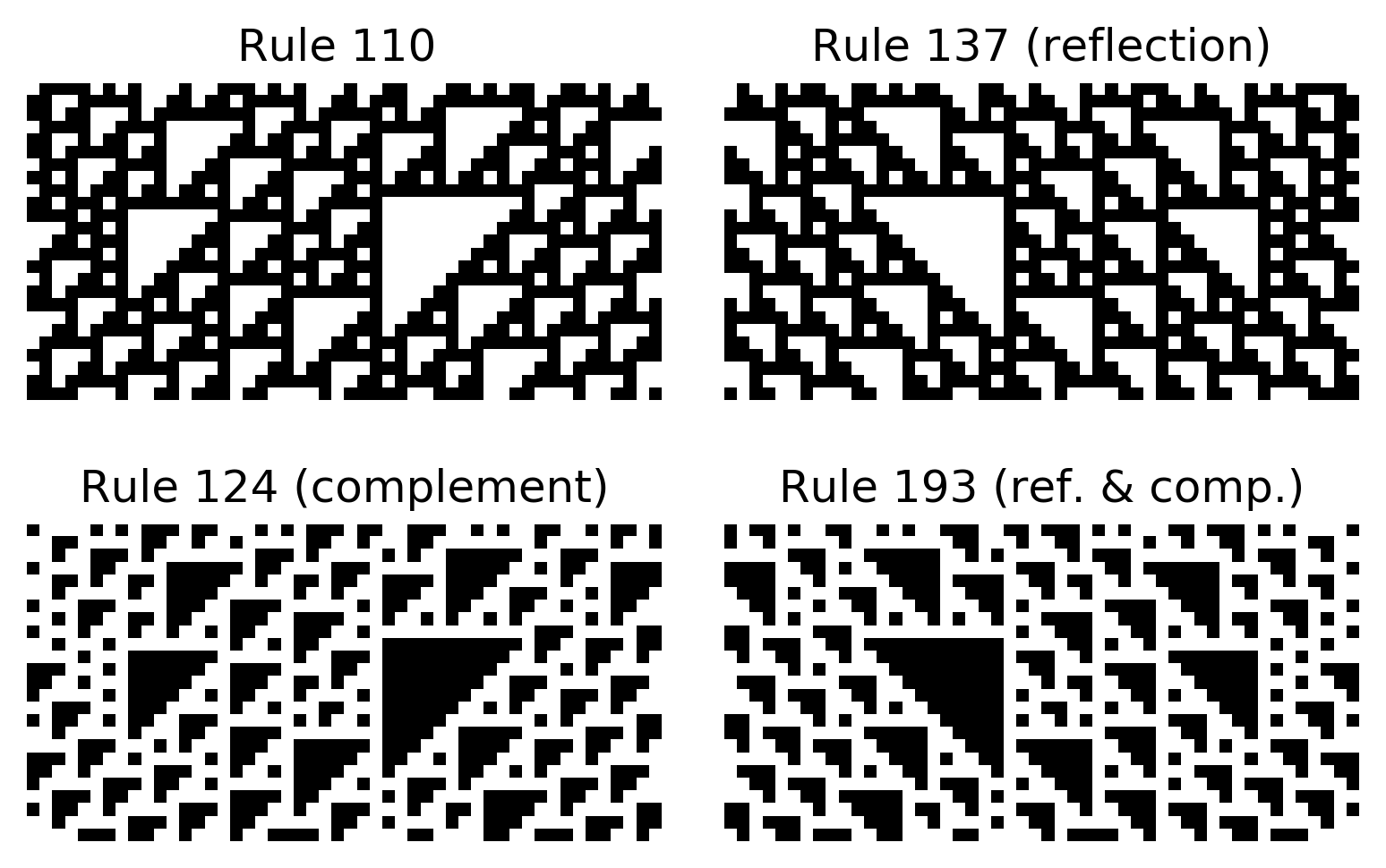}
        \caption{Reflection, complement and reflection complement transformation of rule 110 and equivalent with random initialisation. The reflection rule is initialised with a mirrored state and the complement rule with a flipped value state. }
        \label{fig:mnc_r}       
    \end{center}
\end{figure}

\begin{figure}[h]
    \begin{center}
        \includegraphics[width=1\columnwidth]{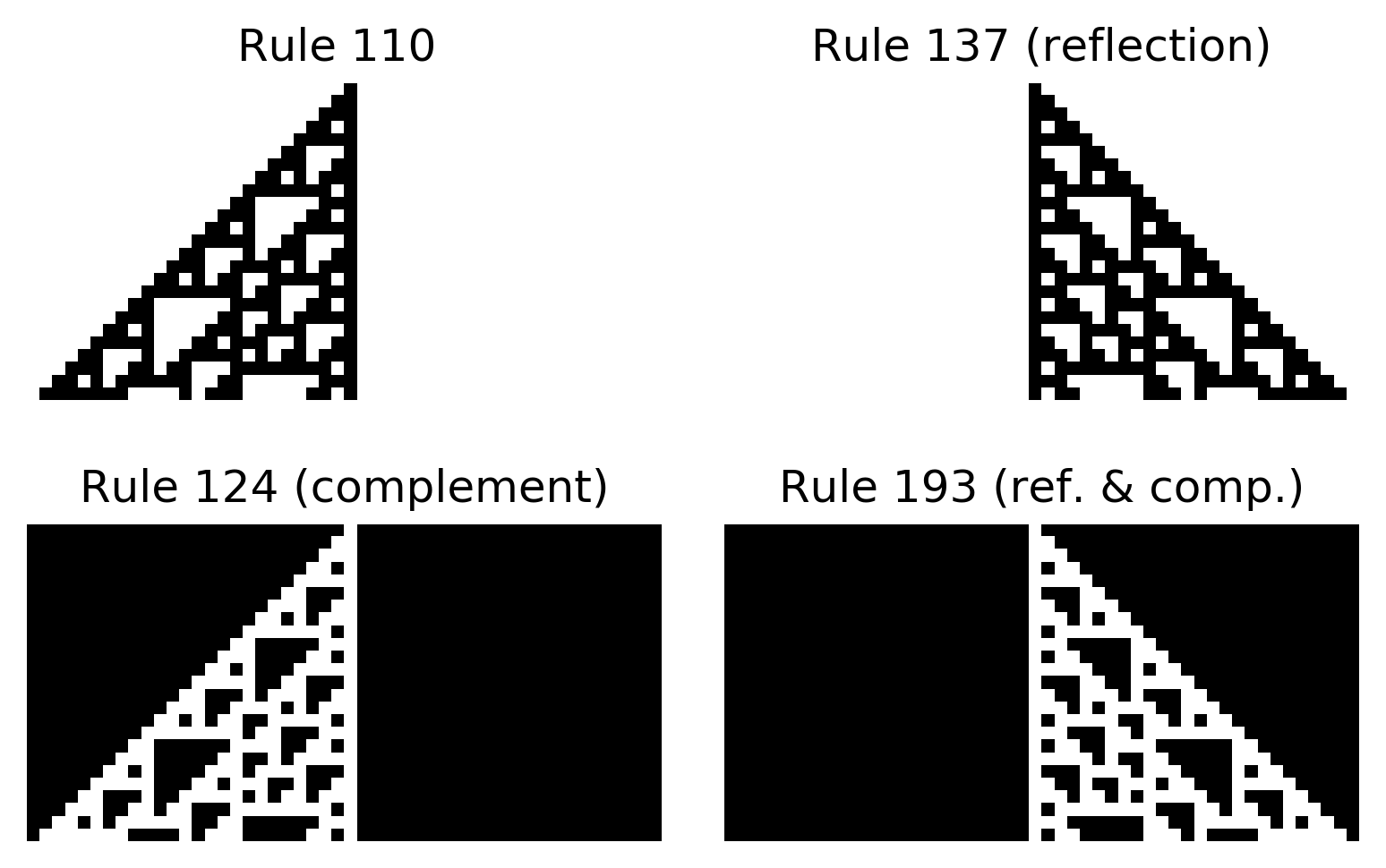}
        \caption{Reflection, complement and reflection complement transformation of rule 110 and equivalent with centroid initialisation. The reflection rule is initialised with a mirrored state and the complement rule with a flipped value state.}
        \label{fig:mnc_c}
    \end{center}
\end{figure}

An overview of the 88 rules can be found in Table \ref{tab:CAEQ}.
Figure \ref{fig:mnc_r} and \ref{fig:mnc_c} show examples of the transformed rules.

The concept of reflection and complement seems to originate already in \cite[p.~51, p.~176]{walker_study_1965}, and more densely explained by the same author in \citep{walker_behavior_1971}. In the previous source, the concept originated in an intermediate substrate between ECA and RBN, with two random neighbours and itself, aka (PL CA). The ME concept works the same in such a substrate, but the concept is perhaps best known when applied to ECA in \citep{wolfram_theory_1986,wolfram2018Tables,li_structure_1990,wuensche_global_1992}.
\subsubsection{Minimum Equivalence (ME) in Homogeneous Homogeneous Random Boolean Networks}

\begin{table}[h]
\scriptsize
\centerline{
    \begin{tabular}{|c|l||c|l|}
    \hline
    Rule & Equivalent & Rule & Equivalent \\
    \hline\hline
        0 & 255 & 44 & 56, 74, 88, 98, 100, 173, 185, 203, 217, 227, 229 \\ 
        1 & 127 & 45 & 57, 75, 89, 99, 101 \\ 
        2 & 4, 16, 191, 223, 247 & 46 & 58, 78, 92, 114, 116, 139, 141, 163, 177, 197, 209 \\ 
        3 & 5, 17, 63, 95, 119 & 60 & 90, 102, 153, 165, 195 \\ 
        6 & 18, 20, 159, 183, 215 & 62 & 94, 118, 131, 133, 145 \\ 
        7 & 19, 21, 31, 55, 87 & 104 & 233 \\ 
        8 & 32, 64, 239, 251, 253 & 105 &  \\ 
        9 & 33, 65, 111, 123, 125 & 106 & 108, 120, 169, 201, 225 \\ 
        10 & 12, 34, 48, 68, 80, 175, 187, 207, 221, 243, 245 & 110 & 122, 124, 137, 161, 193 \\ 
        11 & 13, 35, 47, 49, 59, 69, 79, 81, 93, 115, 117 & 126 & 129 \\ 
        14 & 50, 84, 143, 179, 213 & 128 & 254 \\ 
        15 & 51, 85 & 130 & 132, 144, 190, 222, 246 \\ 
        22 & 151 & 134 & 146, 148, 158, 182, 214 \\ 
        23 &  & 136 & 160, 192, 238, 250, 252 \\ 
        24 & 36, 66, 189, 219, 231 & 138 & 140, 162, 174, 176, 186, 196, 206, 208, 220, 242, 244 \\ 
        25 & 37, 61, 67, 91, 103 & 142 & 178, 212 \\ 
        26 & 28, 38, 52, 70, 82, 155, 157, 167, 181, 199, 211 & 150 &  \\ 
        27 & 29, 39, 53, 71, 83 & 152 & 164, 188, 194, 218, 230 \\ 
        30 & 54, 86, 135, 147, 149 & 154 & 156, 166, 180, 198, 210 \\ 
        40 & 72, 96, 235, 237, 249 & 168 & 200, 224, 234, 236, 248 \\ 
        41 & 73, 97, 107, 109, 121 & 170 & 204, 240 \\ 
        42 & 76, 112, 171, 205, 241 & 172 & 184, 202, 216, 226, 228 \\ 
        43 & 77, 113 & 232 & \\ 
        
    \hline
    \end{tabular}
 }
    \caption{The ME set for $K=3$, for the switching and complement transformations}
    \label{tab:hhrbnComp_ME}
    
\end{table}
In HHRBN, there is also an ME set, but the mirror complement changes in this substrate. Essentially, the mirror complement changes to a switching complement. In ECA, the transformation between left and right forms an equivalence class; in HHRBN, the transformation between any combination and set of combinations between left, centre and right forms an equivalence class. This new equivalence class further reduces the computational space to just 46 rules. The equivalence classes can be seen in Table \ref{tab:hhrbnComp_ME}.

A more in-depth explanation of this transformation can be found in \citep{glover2023minimum}.

\subsubsection{Rule Dependency}
Due to the properties of the ECA rule space, that space also includes rules that are invariant to one or more of their neighbours, therefore they strictly do not compute on 3 neighbours. Essentially, they do not differentiate on all the cells in the neighbourhood; e.g. rule 170 only differentiates on the left neighbour and is indifferent to the values of the right and central neighbour, or similarly, rule 90 only differentiates on the left and right neighbour, but not the central neighbour. This is called the rule dependency, and the 3 neighbour rules can be of either 0,1,2 or 3-dependency. An overview of the ME can be found in Table \ref{tab:dependancyECA} and a complete overview in \citep{wolframAtlasDependencies, wolfram2018Tables}. 

\begin{table}[h]
\centerline{
    \begin{tabular}{|c|l|}
    \hline
    Rule Dependency & Rules \\
    \hline\hline
        0 dep. & 0 \\
        1 dep. & 15,51,170,204 \\
        2 dep. & 3,5,10,12,34,60,90,136,160 \\
        3 dep. & the 74 Rules not included in the previous rows \\
    \hline
    \end{tabular}
 }
    \caption{The ECA ME rules by neighbour dependency}
    \label{tab:dependancyECA}
    
\end{table}

\subsection{Definition of chaotic behaviour in nonlinear systems}

To this day, the definition of chaotic behaviour is not mathematically univocal \citep{kolesov2009definition, touhey1997yet}. 
Still, we will work with a well-accepted definition of chaotic function.
The definition is as follows (see.~\cite[subsection 1.8]{devaney2022introduction}): 
a function or map $f:V \rightarrow V$, on a vector space $V$, is chaotic if it satisfies the following conditions:
\begin{itemize}
    \item $f$ has a sensitivity to initial conditions,
    \item $f$ must be topologically transitive,
    \item has dense periodic orbits (periodic points are dense in $V$).
\end{itemize}

In essence, a function or map is unpredictable as it is sensitive to initial condition, indecomposable (can not be decomposed into two or more subsystems) as it is topological transitive, and yet has an element of regularity as it has regular periodic orbits that are dense. 
Devaney also notes \citep{devaney2022introduction} that 
while there are stronger definitions of chaotic function, the one above is a good definition in that it is generally easy to verify and applies to a larger number of important examples. 

Indeed, if a system is topologically transitive and has dense periodic orbits, then it also has sensitivity to initial condition \citep{banks1992devaney}. 
This means one should view sensitivity to initial condition as a necessary condition for $f$ to be chaotic, whereas topological transitivity and the existence of dense periodic orbits together are sufficient conditions. 
This implication means it is possible to drop the first condition of Devaney's definition.
We will, however, keep it as part of the definition of chaotic function since it is easy to check in practice.

As mentioned above, the definition of chaotic behaviour usually applies to continuous variables. However, it can be adapted to systems whose time-evolution is intrinsically discrete, as is the case of CAs and RBNs. 
Chaotic behaviour can be considered in eight different situations, taking all possible combinations of continuous and discrete time, continuous and discrete space, and continuous and discrete state variables. 
We will consider the most common cases of such combinations, aiming to introduce the concept of chaotic behaviour in CA and RBN.


\subsubsection{Chaotic behaviour in continuous space, time and states}

The definition of chaotic function when space, time, and states are continuous variables is given above, as this is the common situation when defining chaotic behaviour.
The prototypical example of a chaotic system in such conditions is a turbulent fluid, described by the so-called Navier-Stokes equations \citep{pope2001turbulent}, which are partial differential equations, having therefore continuous time and space variables as well as continuous state functions (velocity of the fluid). 
\subsubsection{Chaotic behaviour with continuous time and states and discrete space}
The most well-known example of a chaotic system is the Lorenz system. 
This system comprehends a set of three ordinary differential equations to model forced dissipative hydrodynamic flow \citep{lorenz1963deterministic}. It is defined by $\dot{x} = -\sigma x + \sigma y$, $\dot{y} = -xz + rx - y$ and $ \dot{z} = xy - bz$. 
For the choice of parameter values $\sigma = 10, b = 8/3, r = 28$ \citep{elaydi2007discrete}, one gets an orbit resembling a butterfly when plotting the $x$ against $z$. Due to its odd features, this orbit was classified as a "strange attractor", a stable orbit showing chaotic behaviour. One such feature is that arbitrarily small difference in the initial condition leads to large deviations in the following orbits over time. In other words, the Lorenz system shows sensitivity to initial conditions. 

Note that this system does not have a space dimension - in our classification with time, space and state, the phase space, accounts for the state dimension. It is rather a low-dimensional example of chaos \citep[subsection 3.2.2]{kaneko2001complex}, not an example of spatiotemporal chaos. 

Yet, the same definition can be used when space is discrete. Indeed, a spatially extended (continuous) system can be discretized into a mesh or grid of points, each governed by a set of differential equations where both dependent and independent variables - state variable and time, respectively - are continuous. This would be the case of a (discrete) set of coupled Lorenz systems distributed in space, each described by continuous states and time.

\subsubsection{Chaotic behaviour with continuous states and discrete time and space}

Another famous example of chaos but with discrete time is the logistic map $x_{t+1}=ax_{t}(1-x_{t})$, used as a simple mathematical model to population dynamics \citep{elaydi2007discrete, may1976simple}. Yet, this simple equation is capable of producing chaotic dynamics. From $3<a<3.57$, the population starts to oscillate between more and more values until it reaches the parameter region of about $3.57<a<4$ where the oscillation explodes into the infinite (no longer oscillating). For this region, small initial values for the population yield large variations over time (sensitive to initial condition) \cite[Chapter 2]{mitchell2009complexity}. To note that this is not true for all regions of $3.57<a<4$, e.g. at $a=1 + \sqrt{8}$ there is a period-3 cycle. 

Similarly to the Lorenz system, the logistic map does not have an explicit space, but space can be introduced by considering again a discrete set of coupled maps for a so-called coupled map lattice \cite[subsection 3.2.2]{kaneko2001complex}.  

\subsubsection{"Chaotic" behaviour with discrete time, space and states}

This is the case of chaotic behaviour observed in CAs and RBNs, which have discrete state space (discrete number of accessible states, e.g., a binary set of "0" and "1"), evolve iteratively (discrete time), and are composed of a discrete set of (spatially localised) nodes or cells. We, therefore, call them fully discrete systems.

In a strict sense, taking the definition of chaotic function introduced above, this case can not show chaotic behaviour. 
To understand this, we can consider the concept of dense periodic orbits. The definition of dense can be stated as follows: 

Let \( A \) be a subset of a topological space \( X \). Then \( A \) is said to be \textbf{dense} in \( X \) if:
\[
\forall x \in X, \forall \epsilon > 0, \exists a \in A : d(x, a) < \epsilon \, ,
\]
where $d(x,a)$ is the Euclidean distance between $x$ and $a$.
This means that in order to be dense, it must be possible that elements of the two sets $X$ and $A$ are arbitrarily close.
This feature can not occur in intrinsically discrete systems, as distances are always multiples of an elementary (the smallest positive) distance between two distinct instants, states or spatially distributed nodes.
We shall continue this discussion in subsection \ref{sec:discdiscChaos}.
In the rest of this paper,
we will assume an "analogue" of chaotic behaviour based on the comparison of the number of accessible global states (set of individual states) and the periodicity of the overall trajectories.
More precisely, in the particular case of having a CA with $N$ cells, each one taking values $\{0,1\}$ (binary states), and then there are $2^N$ accessible configurations (global states). Being a finite number of accessible configurations, chaotic orbits can not occur strictly, as defined above: all orbits will be periodic with a period not larger than $2^N$. In this context, we define "chaotic" behaviour as the one typical of rules for which the periodicity scales geometrically with the size of the system.
In particular, a CA of size $N$ is considered to show "chaotic" behaviour if, when doubling its size to $2N$, the periodicity of its orbits increases quadratically, from $2^N$ to $2^{2N}$. 
Henceforth, behaviours observed in ECA are called "chaotic" with the quotation marks to more clearly separate it as an analogy of chaos in other systems. However, these quotation marks are often omitted. 

Perhaps the most famous example of "chaos" in systems with discrete time, space and states is the ECA Rule 30. Despite being a simple substrate (ECA) configuration, it can produce very complex and pseudo-random behaviour. In Mathematica, the central column of rule 30 is used for the pRNG \cite[p.~317]{wolfram2003new}. 

Work has also been done for a subset of CA called linear CA, in \citep{cattaneo2000ergodicity} it is claimed that for linear CA over $Z_m$ that ergodicity is equivalent to topological transitivity and that dense periodic orbits (regularity) are equivalent to surjectivity.

In \citep{manzini1999complete} Martinez worked on a method to determine if a linear CA over $Z_m$ has a behaviour that is equicontinutity, sensitive to initial condition, strong transitivity or is positive expansive. They take a hierarchical view of the definitions of chaos, from positively expansive to strong transitive to transitive to sensitive, being the weakest. 


\subsubsection{Chaos in terms of computational utility}

Beyond a scientific interest in the computational aspect of chaos, it has some very useful applications. Like rule 30, many other chaotic systems, such as the logistic map \cite[Chapter 2]{mitchell2009complexity}, can be used as a pseudorandom number generator (pRNG).  

To understand why, we begin with Shannon's concepts of confusion and diffusion \cite{shannon1945mathematical}. To mitigate simple statistical analysis, a cipher must have properties of confusion and diffusion. Confusion, as in the input and the output, should have a very complicated relation. Diffusion means that the input should affect the whole output. These concepts are also important for pRNG.
Let's view this in terms of chaotic behaviour; a system sensitive to initial conditions would mean that small changes to the initial condition would lead to large changes in the output. Therefore, one can not use a similar output to predict the input, enhancing confusion. Further, a topologically transitive system (cannot be decoupled) ensures that one cannot decouple the solution, leading over time to an effect on every part of the system, enhancing diffusion. Finally, a dense periodic orbit means the system can return to close but not the same configurations, ensuring a rich set of values that are close in output but distant in input, enhancing confusion and diffusion. 

Typically, when testing systems for pRNG, one does many statistical tests such as the ones in \citep{bhattacharjee2017pseudo}. Note that one of the tests in this article is non-linearity, which would imply that the exploration of linear CA as chaotic systems would unlikely result in a good pRNG. 
Furthermore, theoretical reasoning states that highest capacity computation lies on the edge of chaos (see subsection \ref{sec:EoC}). Assuming this is true, identifying chaotic behaviour is important for identifying useful computational substrates. 

\subsection{Identifying the edge of chaos and with a parameter}
\label{sec:EoC}

The parameter space of a complex system often has a phase transition between order and disorder; this phase transition region is often called "Edge of Chaos" It is theorised that this region commonly contains the highest capacity for computation defined as transformation, manipulation and storage of information.

Langton \citep{langton1990computation} explored this theory in 1-dimensional multi-state CA with enlarged neighbourhoods and found that the CA rule-space forms a phase transition between order and chaos when organised over a $\lambda$ (Lambda) parameter.
The $\lambda$ parameter starts by defining a state as the quiescent state. To generate a Transition Table (TT) with a given $\lambda$ value, one allocates to each TT entry a random number $\alpha$ uniformly distributed between $0$ and $1$ and attributes the quiescent state to all entries with $\alpha<\lambda$ and a non-quiescent state to the other. Using this method, Langton generated different candidate rules in several regions of the rule-space over the $\lambda$ parameter. He showed that the rules-space organises into a phase transition between order and chaos and that strong candidates for computation are more likely to be found there.  
Notably, this lambda method does not seem to work in the ECA rule space, as mentioned in \citep{langton1990computation} and previous work. 
\begin{table}[h]
\centerline{
 \newcommand\Cstrut{\rule[-1.4ex]{0pt}{4.2ex}}
    \begin{tabular}{|c|l|}
    \hline
    Rule $\lambda$ & Rules \\
    \hline\hline
        $ \lambda = \frac{0}{8}$ & 0 \Cstrut\\
        $ \lambda = \frac{1}{8}$ & 1, 2, 4, 8, 32, 128 \Cstrut\\
        $ \lambda = \frac{2}{8}$ & 3, 5, 6, 9, 10, 12, 18, 24, 33, 34, 36, 40, 72, 130, 132, 136, 160  \Cstrut\\
        $ \lambda = \frac{3}{8}$ & \makecell[l]{7, 11, 13, 14, 19, 22, 25, 26, 28, 35, 37, 38, 41, 42, 44, 50, 56, \\ 73, 74, 76, 104, 134, 138, 140, 146, 152,162, 164, 168, 200} \Cstrut\\
        $ \lambda = \frac{4}{8}$ & \makecell[l]{15, 23, 27, 29, 30, 43, 45, 46, 51, 54, 57, 58, 60, 77, 78, 90, 105, \\ 106, 108, 142, 150, 154, 156, 170, 172, 178, 184, 204, 232 } \Cstrut\\
        $ \lambda = \frac{5}{8}$ & 62, 94, 110, 122 \Cstrut\\
        $ \lambda = \frac{6}{8}$ & 126 \Cstrut\\
        $ \lambda = \frac{7}{8}$ &  \Cstrut\\       
        $ \lambda = \frac{8}{8}$ &  \Cstrut\\       
    \hline
    \end{tabular}
 }
    \caption{The ECA ME rules by $\lambda$}
    \label{tab:lambdaECA}
    
\end{table}

\subsection{Reservoir Computing (RC)}
Reservoir Computing (RC) is a substrate-independent framework for computing. RC is independent because it works on many different substrates, but to be clear, different substrates would, of course, have different capabilities. The RC framework consists of 3 parts: the input, the untrained reservoir and the output.

\begin{figure}[h]
\begin{center}
\includegraphics[width=0.5\textwidth]{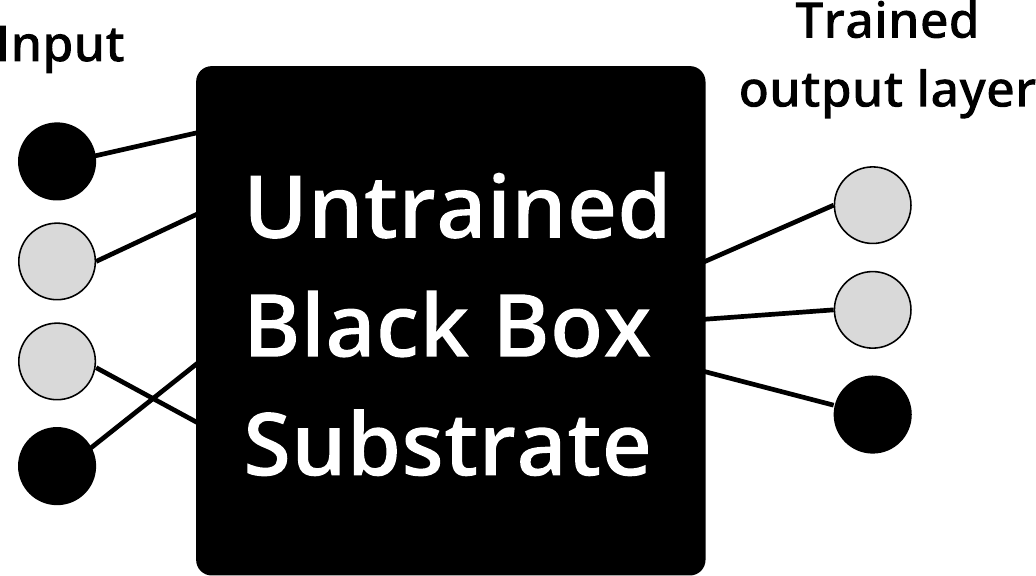}
\caption{RC as a substrate-independent framework }
\label{fig:RCBlackBoxModel}
\end{center}
\end{figure}

The input part encodes some information into the untrained reservoir and typically into higher dimensions. The untrained reservoir typically expands, modifies or changes the information, but could, in the context of the framework, be considered a black box as seen in Figure \ref{fig:RCBlackBoxModel}. The output part is typically linear, does dimensional reduction, and extracts useful features. 

\begin{figure}[h]
\begin{center}
\includegraphics[width=0.5\textwidth]{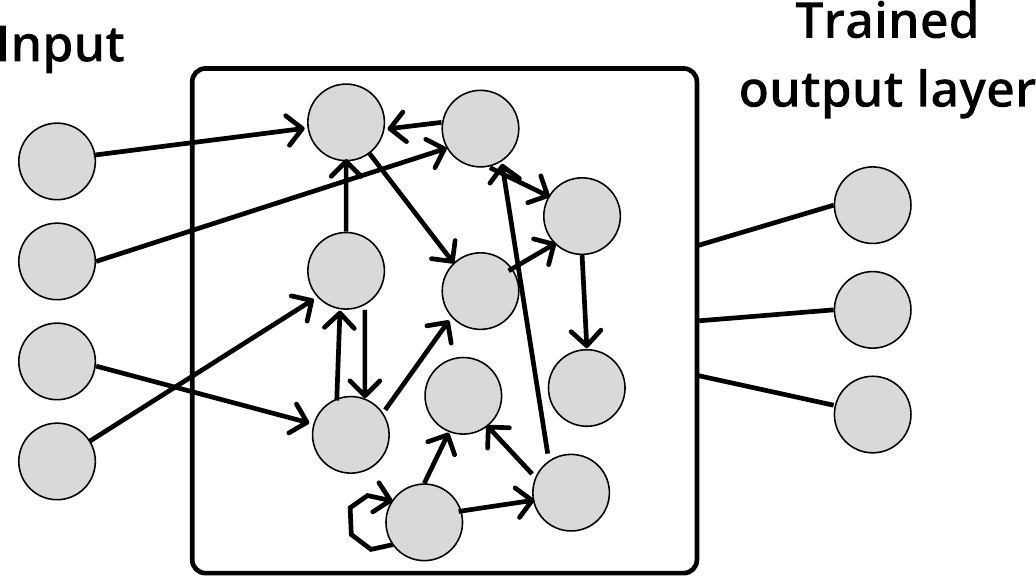}
\caption{Basic network Architecture of an ESN.}
\label{fig:ESNModel}
\end{center}
\end{figure}

The RC concept originated in echo state networks (ESN) using recurrent neural networks as a substrate \citep{jaeger2001echo} and in liquid state machines (LSM) using a spiking neural network for a substrate \citep{maass2002real}. Since then, both ESM and LSM and a host of other substrates have been put under the umbrella term of RC. Due to RC substrate-independent nature, many different substrates have been explored and/or compared  \citep{tanaka2019recent}. Some explore different topology configurations as in \citep{gallicchio2016deep}, where a deep layered sub-reservoirs were analysed instead of the typical one big reservoir. RC is also a very popular method with physical reservoirs \citep{tanaka2019recent}, as an extreme example in \citep{fernando2003pattern} it was demonstrated that RC can use the surface waves on a bucket of water as a reservoir and they successfully solved speech recognition and xor tasks using this substrate. One interesting substrate is real biological neural networks (BNN), specifically disassociated neurons that self-organise over a microelectronic array \citep{aaser2017towards}. 

There is also evidence that reservoir computing is a useful trick for computation (one of many) used in biology. \citep{nikolic2006temporal} shows that a linear classifier can extract information about the short-term past stimulus (images, xor) from the primary visual cortex of an anaesthetised cat. 
Additionally, there is some evidence of RC in other biological and computational processes. In \citep{dai2004genetic}, the ESN (RC) model was used to simulate an example of a known genetic regulation network (GRN) process and performed satisfactorily. Similarly, in \citep{jones2007there}, the LSM (RC) model was used. 

\subsection{Reservoir Properties}
An important property in ESN is the Echo State Property (ESP). Given some input signal, the reservoir must asymptotically remove the initial condition information to have this property. In \citep{jaeger2001echo}, it is shown that for a reservoir with specified conditions, it violates the ESP if the spectral radius of the weight matrix is larger than 1, and it was empirically observed that for Spectral radius below 1, the ESP is given. Note that in \citep{Jaeger2007ESN} Jaeger warns that this does not mean that ESP is granted for any system with a spectral radius of below 1 (asymptotically stable). It is not a necessary nor a sufficient condition. 

Similar to the ESP is the concept of the fading memory property. It states that an input/output system is said to have fading memory when the outputs associated with inputs that are close
in the recent past are close, even when those inputs may be very different in the distant past. \citep{grigoryeva2018echo, tanaka2019recent}.

\citep{maass2002real} determines two conditions for real-time computation on perturbations. The separation property (SP) is a necessary condition, and the approximation property (AP) is a sufficient condition. SP refers to the separation between trajectories based on differences in perturbations. AP refers to the capabilities of the readout mechanism. 


\subsection{Reservoir Computing with CA (ReCA)}
The first study that introduced CA as a substrate in reservoir computing is \citep{yilmaz2014reservoir}. This study investigated Game of Life and several ECA rules as reservoir substrates and tested on a 5-bit and 20-bit memory benchmark. In addition, it presents a theoretical comparison of CA vs ESN, using the metric of the number of operations needed to solve the benchmark, which documents a clear advantage of using CA.

As an ECA reservoir only relies on simple discrete binary interactions between cells (see \citep{wolframAtlasBoolean} for details), it affords a hardware-friendly substrate implementation. The problem (perhaps ironically) becomes how to implement the readout layer in hardware.
In \citep{moran2019energy}, ReCA using ECA with a max-pooling and softmax strategy was implemented on a Field Programmable Gate Array (FPGA).
In \citep{olin2019cellular}, a CA was implemented on Complementary metal-oxide-semiconductor (CMOS) combined with a custom hardware SVM implemented in resistive random-access memory (ReRAM).
In \citep{liang2021bloomca}, a synthesised hardware implementation of ReCA using ECA with a max-pooling and ensemble bloom filter classifier. Showing impressive results compared to "state-of-the-art" in terms of energy efficiency, memory usage and area(number of gates) usage, but with comparably poor accuracy \citep{moran2019energy}.

Other works have also studied ReCA using the 5-bit memory benchmark. \citep{nichele2017deep} changed the structure of the CA to a deep-layered architecture and compared it to a single layer, which resulted in noticeable performance improvements. Additionally, in \citep{nichele2017reservoir} the authors organised the CA substrate as consisting of two regions of different ECA rules. Different combinations of rules were explored, and some showed great promise. In \citep{margem2019reservoir}, an exploration was conducted of different cell history selection methods for the classification model on the 5-bit memory task, a temporal order task and arithmetic and logic operation tasks. 
In \citep{babson2019reservoir}, CA rules with multiple states and larger neighbourhoods were evolved and then tested on the 5-bit memory benchmark. In \citep{uragami2022universal} ECA and asynchronous ECA is tested and compared on the 5-bit memory benchmark, mainly in the context of the distractor period. 

In \citep{margem2020feed}, it was pointed out that the benchmark has no train test split. They modified the benchmark by training on just a few (2 or 3) of the 32 possible input streams, and some of the rules with more ordered behaviour could still solve this version of the benchmark.

In \citep{glover2021dynamical}, the full ECA set was tested using key parameters of number of bits ($N_b$), redundancies ($R$) and Grid size. \citep{glover2023investigating} extended this work to include more parameters such as Iterations ($I$) and Distractor Period ($D_p$). This paper also explained many of the unexpected results in the previous study, but perhaps as important, it similarly to \citep{margem2020feed} pointed out some weaknesses in the 5-bit memory benchmark. 

ReCA is also used on other benchmarks than the 5-bit memory benchmark. \citep{moran2019energy, liang2021bloomca, glover2024reservoir} implemented ReCA in hardware and tested using MNIST. An additional example is \citep{mcdonald2017reservoir}, where the authors solved tasks of sine and square wave classification, non-linear channel equalization, Santa Fe Laser Data and iris classification.

In \citep{kantic2024relicada} an method for Rule Selection for ReCA was presented. Limiting the search-space to only Linear rules that obey a list of specific mathematical properties (see paper for details), the paper demonstrates the method selects for rules in the high performance (95-80 percentile) bracket on several time-series prediction benchmarks compared to the full Linear CA space of same neighbourhood and number of states.

\subsection{Reservoir computing with Random Boolean Networks (ReRBN)}
In \citep{snyder2012finding, snyder2013computational}, ReRBN was explored on temporal parity and temporal density (temporal majority task). For the tasks and parameters explored, it was found that the heterogeneous  RBN (different in-degree RBN) reservoir worked best at a critical connectivity $K=2$ (in-degree of 2).
In contrast, \cite{burkow2015evolving} found that for homogeneous RBN, criticality was instead found at $K=3$. \cite{burkow2016exploring} extended this work, exploring different reservoir properties such as perturbation percentage, the relationship with attractor and performance and comparing a subset reading from a larger reservoir to a subset equal reservoir. 

In \citep{calvet2024connectivity}, the relationship between $N$ and $K$ was also studied with a balance $b$ between excitatory and inhibitory nodes. They find that $K$ is the most important of the control parameters, as it affords simpler fine tuning of the other parameters.

\subsection{Reservoir computing with Intermediate substrates}
RC explorations Between CA and RBN substrates are less common. This paper reports and extends on work done in a master thesis \citep{jahren2022comparison}, where Life-like CA, ECA, PLCA and HHRBN were explored using the 5-bit memory benchmark. 

In another master thesis \citep{johansson2024text}, Reservoir computing with cellular automata networks where explored on a simple text classification task. The study explores and compares different ways to construct the network and how that affects performance. The cellular automata networks described include fixed predecessors (in-degree); from the description, it seems they explored PLCA, confirmed by the lack of the same score for rules 204 and 170. Yet, we can not directly compare it with the work in this paper, as the study constructs the transitions rule differently.

\subsection{5-bit Memory Benchmark}
The 5-bit memory benchmark traces its root to the short long-term memory task introduced in \citep{hochreiter1997long}. Although often cited as the source \citep{yilmaz2014reservoir, nichele2017deep, nichele2017reservoir, babson2019reservoir}, none of the benchmarks in \citep{hochreiter1997long} are the 5-bit memory benchmark, but some of them are very similar in intention. The earliest source where the 5-bit memory benchmark is recognisable is in \citep{martens2011learning}, but named "noiseless memorisation",  corroborated with the clearer and more detailed explanation of the benchmark in \cite[p.~47]{sutskever2013training} and in \citep{jaeger2012long}.

\begin{table}[h]
    \centering
    \begin{tabular}{|c|c|c|c|c|c|c|c|c|}\hline
       Step & \multicolumn{4}{c|}{Input} & \multicolumn{3}{c|}{Output} & Stage\\ \hline\hline
        1 & 1 & 0 & 0 & 0 & 0 & 0 & 1 & \multirow{5}{4em}{Input bits to memorise}\\
        2 & 1 & 0 & 0 & 0 & 0 & 0 & 1 & \\
        3 & 0 & 1 & 0 & 0 & 0 & 0 & 1 & \\
        4 & 0 & 1 & 0 & 0 & 0 & 0 & 1 & \\
        5 & 1 & 0 & 0 & 0 & 0 & 0 & 1 & \\
        \hline
        6 & 0 & 0 & 1 & 0 & 0 & 0 & 1 & \multirow{3}{4em}{Distractor period}\\
        ... & 0 & 0 & 1 & 0 & 0 & 0 & 1  & \\
        
        204 & 0 & 0 & 1 & 0 & 0 & 0 & 1 &  \\
        \hline
        205 & 0 & 0 & 0 & 1 & 0 & 0 & 1 & Cue signal\\
        \hline
        206 & 0 & 0 & 1 & 0 & 1 & 0 & 0 & \multirow{5}{4em}{Output bits to memorise} \\
        207 & 0 & 0 & 1 & 0 & 1 & 0 & 0 & \\
        208 & 0 & 0 & 1 & 0 & 0 & 1 & 0 & \\
        209 & 0 & 0 & 1 & 0 & 0 & 1 & 0 & \\
        210 & 0 & 0 & 1 & 0 & 1 & 0 & 0 & \\

        \hline

    \end{tabular}
    \caption{Example of the 5-bit memory task with distractor period of 200 and Input of the number 25 in binary form. Artefact inspired by \citep{babson2019reservoir}}
    \label{tab:5bit}
\end{table}

The 5-bit memory benchmark's goal is to test whether a system is capable of memorising a 5-bit and reproducing it at a later stage. Table \ref{tab:5bit} shows an example of the memory task.
The benchmark has 4 input channels where only a single channel can be active at the same time. The first two input channels are dedicated to the 5-bits. The bits are fed into the system sequentially over 5 steps. One can view the first input channel as the "pure" 5-bits and the second as the reversed 5-bits. 
The 3rd input channel is dedicated to constantly feeding input into the system during the distractor period and the output stage. 
The 4th input channel is dedicated to the cue signal, signalling that the output is to be given. 
The benchmark has 3 output channels where one and only one should be active simultaneously. Note that some earlier examples have 4 output channels but one is dropped as it is never intended to give output. The first two are dedicated to the original 5-bits inserted into the system and should sequentially output them following the cue signal, the final output channel should give a signal in all other cases. Due to this output's nature, one can abstract and view the task as a temporal classification problem.

In this paper, we often call it the x-bit memory benchmark, as we have varied the number of bits to be memorised. Also, note that the 20-bit memory benchmark mentioned in some of the previous sources is not the same as the 5-bit memory benchmark but with 20 bits to memorise. The 20-bit memory benchmark uses 7 input channels, 5 for the input and a bit length of 10. 

\subsection{Small-world}
\label{sec:smallWorld}
In \citep{watts1998collective}, they explored graphs varying on p value where p = 1 meant random connectivity and p = 0 is regular connectivity. They demonstrated that small worldliness was achieved with a relatively low p-value. However, in relation to this work, they have a larger neighbour degree. Additionally, we work with fixed in-degree networks (all cells have 3 neighbours). For these reasons, we might not see the same level of small-worldness in our topologies, but naturally, in contrast to ECA, some is expected in PLCA and HHRBN. 

\subsection{Derrida Plots and the Derrida Coefficient}
\label{sec:derridaPlot}

Derrida Plots is named after the author of its origin in \citep{derrida1986evolution}. It is a tool primarily used to identify the behaviour of a particular RBN (critical, chaotic or ordered (frozen)). 
Derrida originally used it to compare a classical RBN (quenched) and an "annealed" RBN, in which every connection and activation function is randomly reassigned after each iteration. 
To construct a Derrida plot, one compares different initial conditions on the same system. 
Start with a random initial condition, flip 1-bit for one of the initial conditions, then evolve both the RBN one step, and calculate the $D_h$. Then do the same but for two flipped bits in the original initial condition, and so on until N bits. One plots the $D_h$ as a function of the number of flips in the initial condition. Typically, the hamming distance increases linearly with the number of flips until it saturates. If the linear increase has a slope larger than one, 
the behaviour is considered to be chaotic; 
if the slope is below one, the behaviour is ordered (frozen);
and if it is exactly one, the behaviour is critical ("complex")
\citep{fretter2009perturbation}\citep[p.~246]{wuensche2011exploring}.

This behaviour classification can be formalised with the Derrida coefficient, $D_c$, given the angle $\theta$ of the initial slope, namely 
$D_c = \log_2{( \tan{\theta})}$.
For $\theta > \pi/4$, 
$D_c$ is positive and negative for 
$\theta < \pi/4$
\citep[p.~250]{wuensche2011exploring}.

In \citep{adamatzky2013creativity} the Derrida coefficient of ECA was mapped together with the Generative morphological diversity $\mu$, to classify ECA on a spectrum of autistic, schizophrenic, and creative personality. 


\section{Methodology and Experimental Setup}
This section will detail the specific methods and experimental setup used in this paper. We begin with the experimental methods by documenting the x-bit memory benchmark details used for the 2D life-like (CA, HHRBN) and the 1D (CA, PLCA, HHRBN). Then, we will explain the details of the Temporal Derrida Plots (TDP) used to analyse the sensitivity. Then, we will give details on how we measured the rate of defect collapse (collapse rate). We continue with the network analysis method of the longest simple cycle and how it is estimated. Finally, we will document the source code and the dependencies with which the code was built. 

\subsection{x-bit memory benchmark}
The 2D life-like experiments use the same setup as \cite{martinuzzi2022lifeLikeBlog}. It uses a parameter of R, which in this specific experiment is the grid size, and in terms of full grid size, it is R x R. The Iterations I represent the number of iterations between encoding steps and the number of steps fed into the classifying model, chose to be a ridge regression model. The projection ratio $P_r$ is the ratio of cells that the input is encoded into, set to $P_r = 0.6$.

\begin{figure}[t]
\begin{center}
\includegraphics[width=\textwidth]{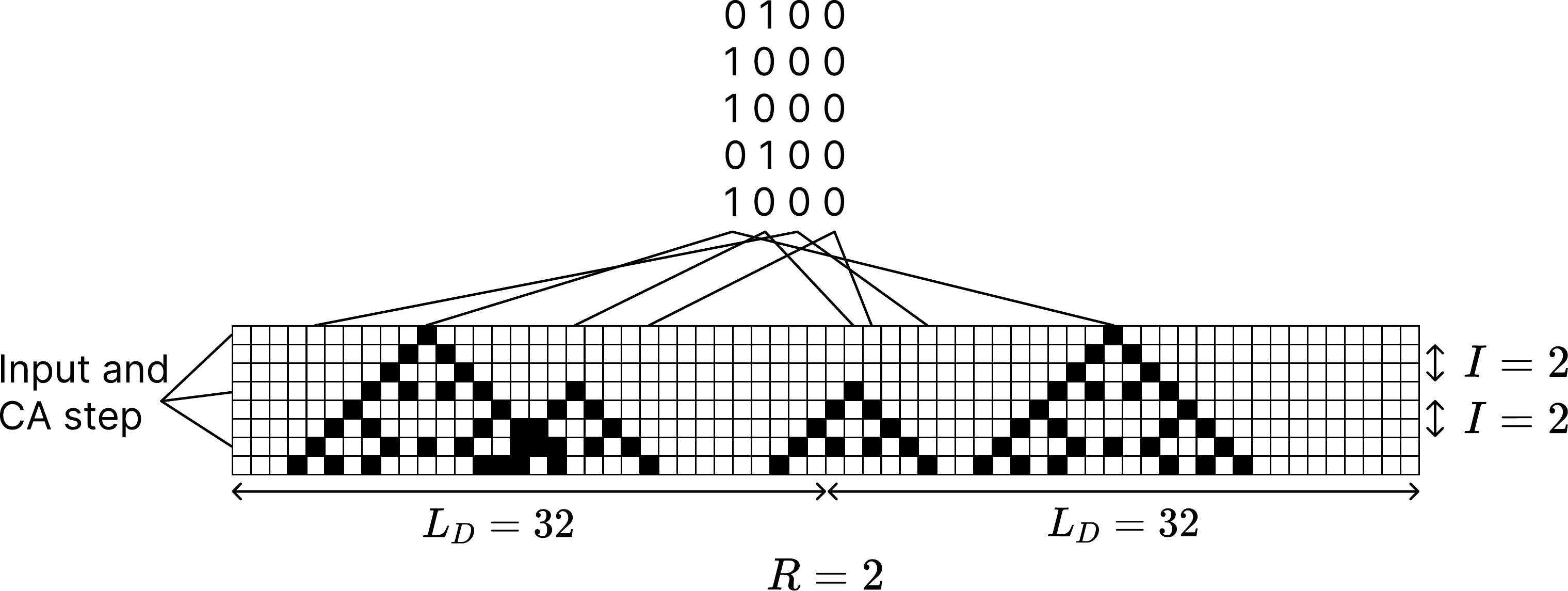}
\caption{ReCA example showing $R$, $I$ and $L_d$. Additionally, the top streams is an example of how input is encoded temporarily into the reservoir.}
\label{fig:ReCA params}
\end{center}
\end{figure}

For the 1D substrates 5-bit memory benchmarks, the experimental setup and the default parameters are the same in \cite{glover2023investigating}. Redundancy $R = 4$ is the number of connected "sub-reservoirs" with individual mapped input. Note how R in the life like experiment and in the 1D experiment signifies different things. The Iterations $I = 2$ represent the number of iterations between encoding steps and the number of iterations the classifying model had access to. The sub reservoir grid size $L_d = 40$, meaning the total number of cells(nodes) where $L_d * R = 160$. An example demonstrating $R, L_d$ and $I$ can be seen in Figure \ref{fig:ReCA params}. The classifying model, in this case, was an SVM with a linear kernel. 

\subsection{Temporal Derrida Plots (TDP)}

In this work, we introduce a variant of Derrida plots:
Instead of introducing a new defect at each step (see subsection \ref{sec:derridaPlot}), we follow the development of one or a few defects starting at $t=0$ and see how $D_h$ changes throughout time (i.e.~as a function of iterations).
In this way, one follows how $D_h$ diverges or converges to a specific value over multiple iterations.
Henceforth, we call these plots "temporal Derrida plots" (TDP). Derrida plots retrieve approximately the Lyapunov exponent in state space, whereas the temporal Derrida plot retrieves the Lyapunov exponent in time.

If we take the simple example of the rules 204 and 170, a simple inspection of these rules would tell you they are very ordered in their behaviour, simply propagating the initial condition. Yet, using the original method \citep{derrida1986evolution} and as described in \citep{wuensche2011exploring} (cf.~\citep[Appendix A]{adamatzky2013creativity}), Derrida plots for rule 204 and 170 yields $D_c=0$, meaning that they follow the 45 angle line. This interpretation is that rules 204 and 170 are complex/ critical in the Derrida Plot method. This is not the case with our TDP.  
Therefore, we argue that the Derrida plot method's weaknesses in ECA substrates are solved with our variant of TDP.
Furthermore, if one is identifying chaotic features for the purpose of harvesting the chaos for something directly useful, e.g. a Random Number Generator (as they are being used for \cite[p.~317]{wolfram2003new}). Then, it would be more beneficial to know the development through the substrate over time, as a single step would not be enough to diffuse the seed value. 
In contrast, a strength of using the Derrida plot rather than the TDP is that it allows one to sample a larger number of initial states of the state space.  

In addition to the $D_h$, we plot the Damerau–Levenshtein distance ($D_{dl}$). This can catch deceptive different-looking configurations like the aether in rule 110; this effect has a marked impact later in the results section with Figure \ref{fig:SensRule110}. 

Furthermore, we run these TDP beyond $N$ steps as we also want to see where the substrates settle. Which we argue tells us something of how "chaotic" the substrate truly is; a substrate that is "chaotic" to the idea of using it as an RNG should not have any preference for 1 or 0 (balanced), and where the substrate settles tells us this experimentally. If a system is "ergodic" to the sense that it covers the entire state space, then it should find every state equally likely and settle at a $D_h$ of half the grid size (half-max distance). 

We run this for five configurations, $1, 5$ and $9$-bit changed in the centre, and $5$ and $9$-bits changed randomly. The randomly placed defects are coded such that there are no collisions in placements, as introducing a defect in the same place twice would cancel each other out. Note that there is, in effect, little difference between centre and random in PLCA and HHRBN; the random defects were introduced to better compare between the substrates and kept throughout the experiments for PLCA and HHRBN for consistency. All the TDP experiments use grid size of 100 cells ($N = 100$).

\subsection{Defect collapse}
We explore whether the systems tend to collapse into the same attractor after a defect is introduced. This can happen in CA because the attractor basin is large enough to encompass the defect. However, in PLCA and HHRBN, this can also occur due to the neighbourhood itself, e.g. if you encode the information into a node that is not the in-node of any other cell, the information can not propagate anywhere. We inspect the substrate in two ways: via the Defect plots when all collapsed defects have been excluded from the data in Subsection \ref{sec:TDPresults}, and we look at the statistic of collapsing in Subsection \ref{sec:collapseRateResults}. 
\subsection{Longest Simple Cycle}
The difference between CA, PLCA, and HHRBN is essentially that of the topology. All the topologies can be reduced to graphs, and therefore, it is natural to apply some graph theory, yet the graph theory sub-field is broad, and the scope of this paper is already large.
 Therefore, we limit ourselves to finding the longest simple cycle in the topology. The longest simple cycle is the longest cycle without any repeating node (except the first and last). We picked this metric because it indicates how much information can be encoded into the network, as any oscillating pattern in the substrate would be limited by the longest simple cycle. 
Note that the longest simple cycle in CA would be equal to the number of cells ($N$) due to its regular neighbourhood configuration. It is Therefore, it is not necessary to run this analysis on CA. 

\subsection{Source code and Dependencies}
The source code for the project can be found at \citep{sourceCode}. The code relies primarily on Evodynamics \cite{pontes2019evodynamic} to run the ECA, PLCA and HHRBN and on scikit-learn \cite{pedregosa2011scikit} for the classification models. A more detailed list of dependencies can be found in \citep{sourceCode}.
\pagebreak
\section{Results}
In this section, we present the results of this paper in chronological order. Starting with the 5 bit memory benchmark experiments, then the TDP and collapse rate results, followed by the network analyse and finally a extended 3 and 4 bit memory benchmark results are presented. 
\subsection{Life-like 5-bit memory benchmark}
\begin{figure}[h]
\centering

\begin{subfigure}{0.45\textwidth}
\includegraphics[width=\textwidth]{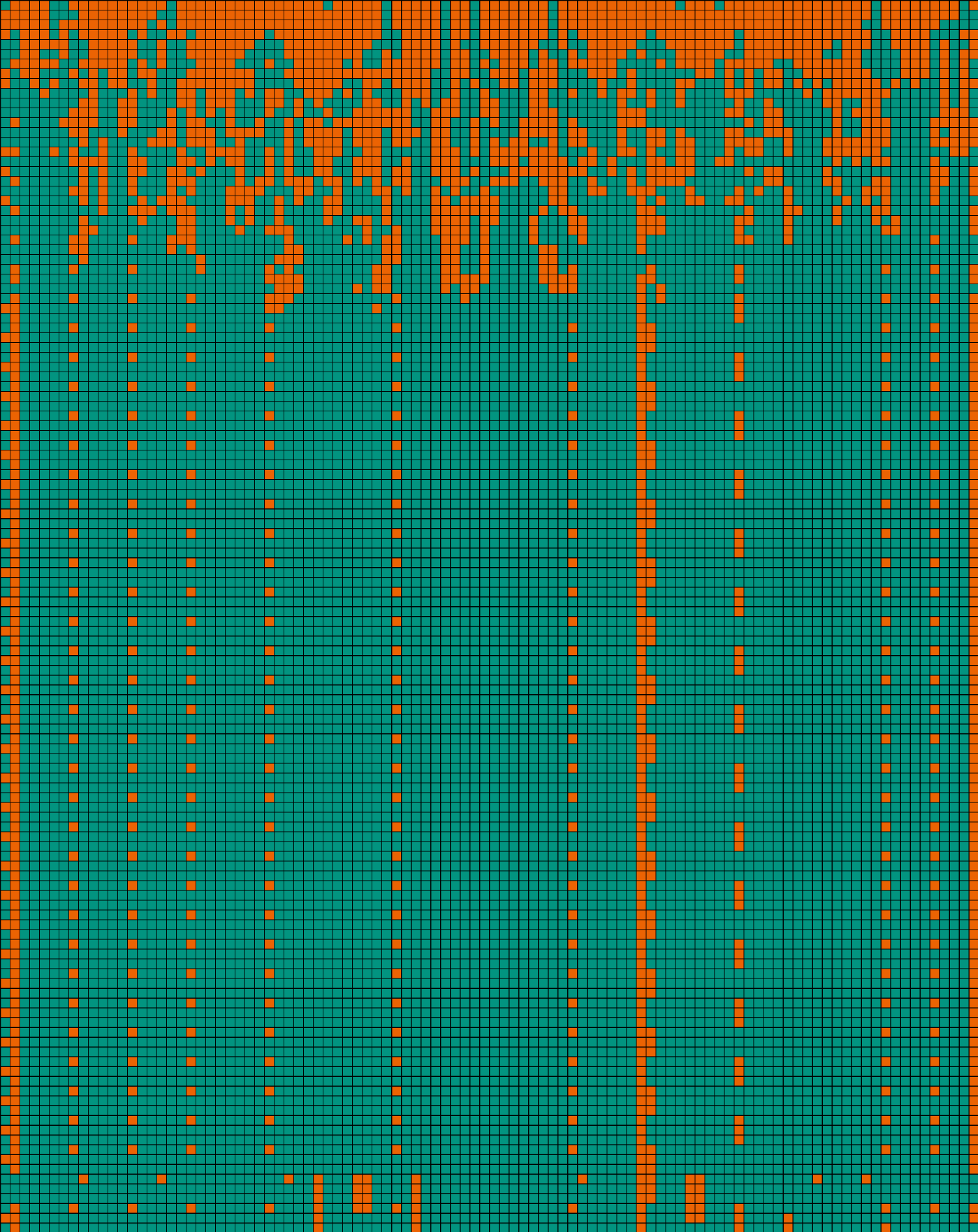}
\caption{CA}
\label{fig:history_B368S12578_10by10_ca}
\end{subfigure}
\begin{subfigure}{0.45\textwidth}
\includegraphics[width=\textwidth]{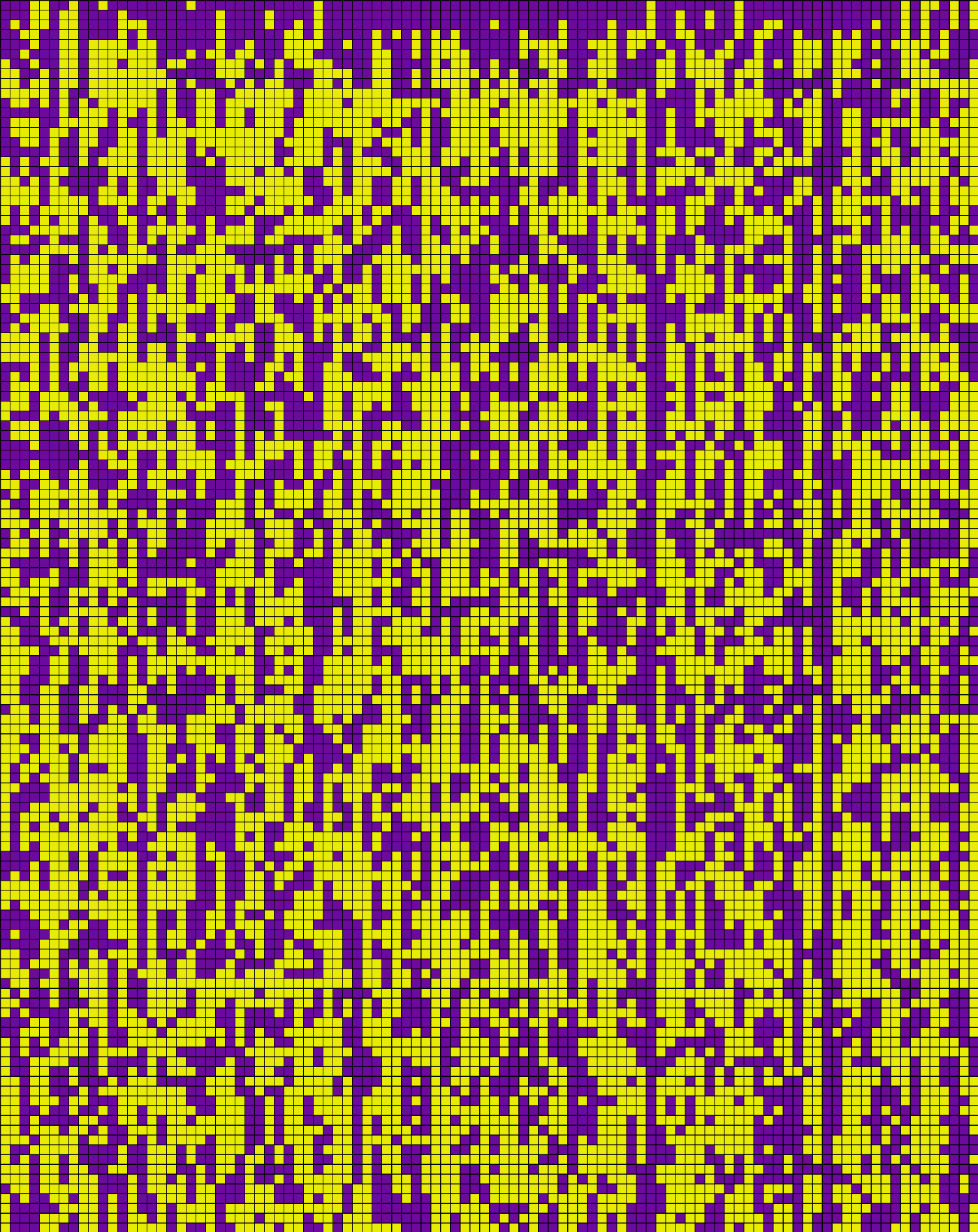}
\caption{HHRBN}
\label{fig:history_B368S12578_10by10_rbn}
\end{subfigure}
\label{fig:example_rbn_ca}
\caption{B368/S12578 reservoir with $N_b = 1$ and $D_p = 10$. Reservoir is flattened. We can still see some shadows and flashes of the same behaviour, but clearly, the CA and HHRBN behave differently.}
\end{figure}

We begin with a smaller experiment between 2D outer totalistic CA (life-like CA) and HHRBN (note that the concept of 2D breaks down in HHRBN). There are $2^{18} = 262 144$ different rules in the "life-like" rule space(see subsection \ref{sec:lifeLikeBackground}). Therefore, an exhaustive search was not practical. A subset of interesting behaving rules were selected from \citep{martinuzzi2022lifeLikeBlog, martinuzzi2022lifelike, pena2021life}. 
The results can be found in Table \ref{tab:lifelike_results}, and we see here that many of the rules perform well in the CA case but not in the HHRBN case, except for B368/S12578. We can quite clearly see from Figure \ref{fig:history_B368S12578_10by10_ca} and \ref{fig:history_B368S12578_10by10_rbn} that the behaviour of said rule changes. 
It is important to point out that there is a bias as these rules have been selected for their behaviour in a CA context. Therefore one can not conclude about the greater scope of ECA, and HHRBN reservoirs, we can at least say that the topology changes the behaviour. 
The CA results are better overall than in \cite{martinuzzi2022lifelike, martinuzzi2022lifeLikeBlog}, though not of a different scale, this might be explained by the difference in hyper-parameter or other implementation detail of the ridge regression model as one was implemented in Julia and the other in Python, we therefore still consider it a successful replication of the previous study.
\begin{table}[h]
\centering
\begin{tabular}{|l | r r| r r |r r|}
        \hline
     Model &  CA  &    HHRBN    &   CA  &    HHRBN    &   CA  &    HHRBN\\
     (R, I) & \multicolumn{2}{c|}{(24, 8)} & \multicolumn{2}{c|}{(28, 8)} & \multicolumn{2}{c|}{(30, 6)}   \\
        \hline
        \hline
     B3/S23         &  2\%    &   0\%    &   100\%   &   0\%     &   91\%    &   0\%   \\
     B35/S236       &  72\%   &   0\%    &   100\%   &   0\%     &   100\%   &   0\%   \\
     B368/S12578    &  60\%   &   48\%   &   100\%   &   100\%   &   100\%   &   100\%   \\
     B356/S23       &  67\%   &   0\%    &   100\%   &   0\%     &   100\%   &   0\%   \\
        \hline

\end{tabular}
\caption{Results from the replicated reservoir architecture of \citep{martinuzzi2022lifelike, martinuzzi2022lifeLikeBlog} with the addition of Dynamic Life (B356/S23) \citep{pena2021life}. Cross-referenced with the RBN reservoir architecture. $(R,I)$ where $R$ is the \emph{reservoir width}, and $I$ is both the length of the reservoir feature vector and the number of iterations between inputs.}
\label{tab:lifelike_results}
\end{table}

\subsection{ECA 5-bit memory benchmark}
\begin{figure}[t]
    \begin{center}
        \includegraphics[width=1\columnwidth]{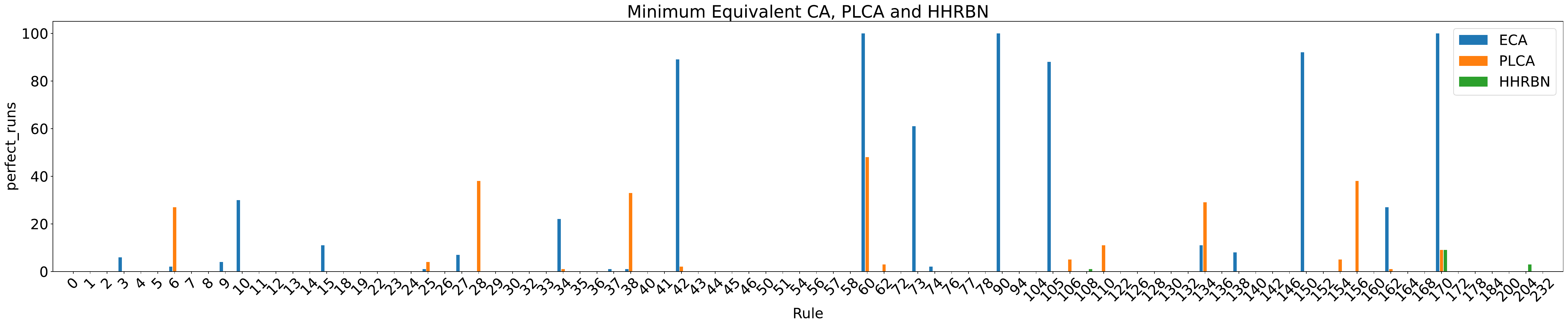}
        \includegraphics[width=1\columnwidth]{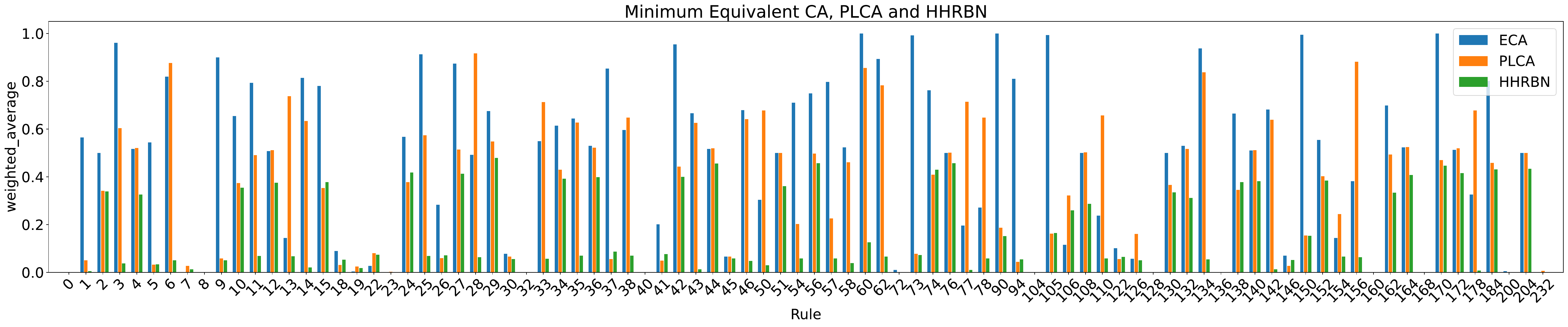}
        \caption{Breakdown of individual rule performance with 5-bit memory.}
        \label{fig:5bit_sens}       
    \end{center}
\end{figure}

A similar exploration of the full ME set of ECA rules for ECA, PLCA, and HHRBN was also conducted. The bias of selecting rules is removed by exploring the entire rule space. The rules have different ME sets \citep{glover2023minimum}, but the ECA ME set is a super-set of the HHRBN ME set. Therefore, we use the ECA ME set by default. 
In Figure: \ref{fig:5bit_sens}, we see these results. There is a clear trend that general performance goes down from CA to PLCA to HHRBN. In the perfect run metric, only three rules scored any perfect run for HHRBN (108, 170, 204). We see a similar trend in the weighted average metric. Note that these rules (108, 170, 204) are all ordered in behaviour and that rules 170 and 204 are equivalent in the HHRBN ME set, meaning in effect, only 2 behaviours of the 46 unique HHRBN managed to solve the task. 
More details and results can be found in \cite{jahren2022comparison}.

\subsection{Temporal Derrida Plots}

\label{sec:TDPresults}

In this subsection, we will present the TDP, all plots except Figure \ref{fig:ECAHammAcc} have the collapsed runs removed; the tally of the collapsed runs can be found later in Table \ref{tab:ruleECACollapse}, \ref{tab:rulePLCACollapse} and \ref{tab:ruleHHRBNCollapse}. They are separated because the collapse can greatly impact the temporal Derrida plots. In short, we still see this impact by removing them and displaying them alone, but we can compare sensitivity independent of collapse rate. 

\begin{figure}[H]
\begin{center}
\begin{subfigure}{0.45\textwidth}
        \includegraphics[width=1\columnwidth]{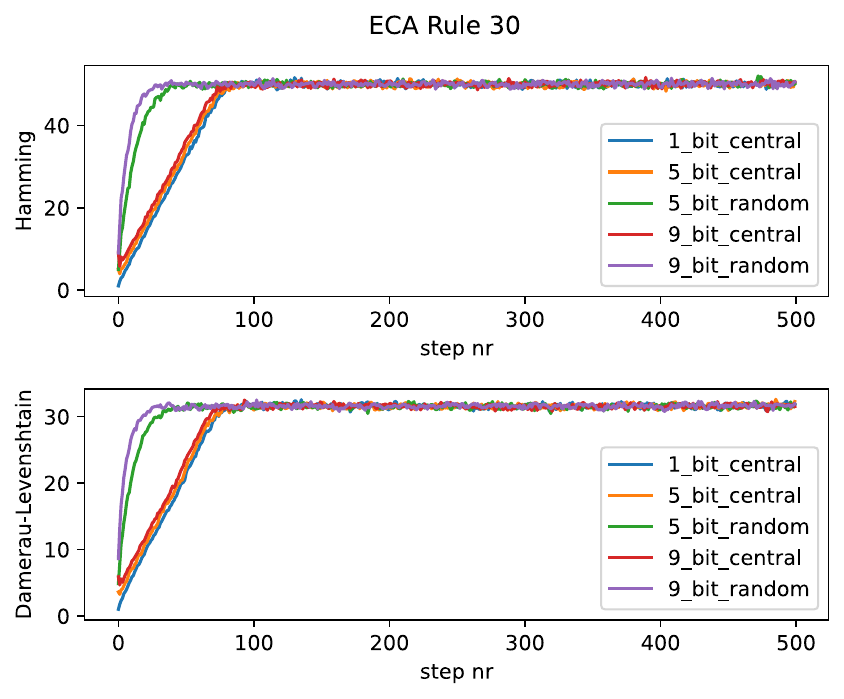}
        \includegraphics[width=1\columnwidth]{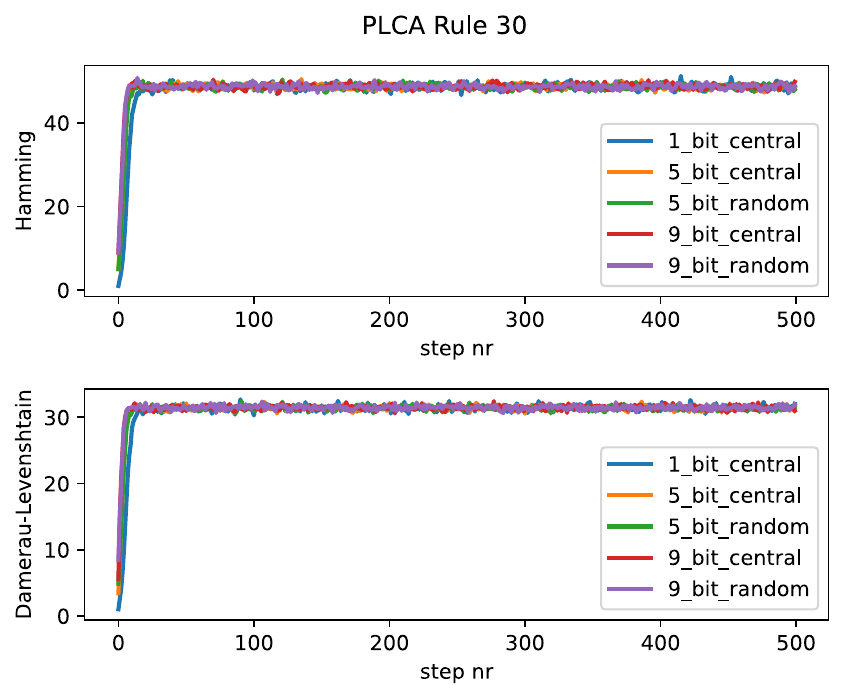}
        \includegraphics[width=1\columnwidth]{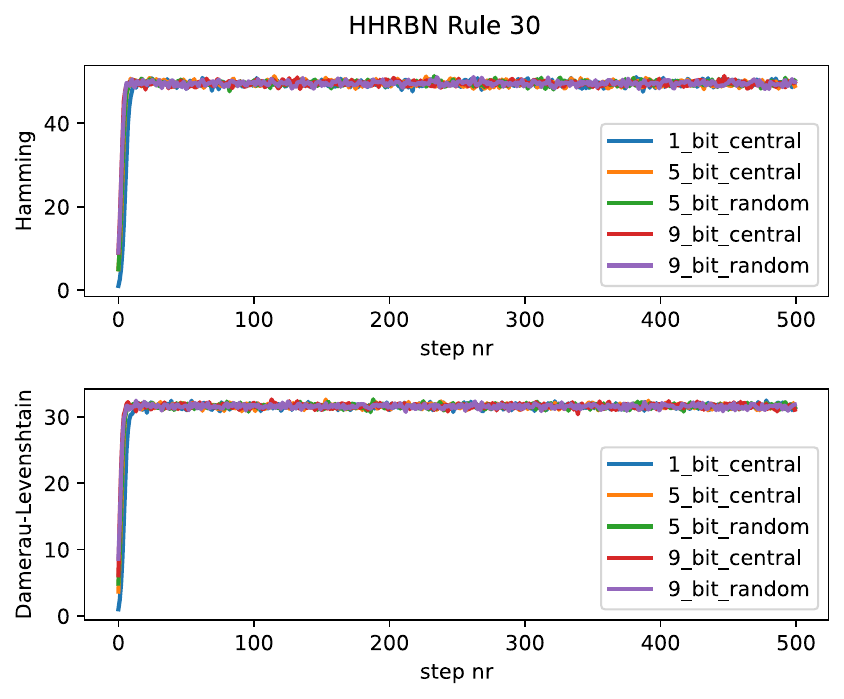}
\caption{ Rule 30}
\label{fig:SensRule30}
\end{subfigure}
\begin{subfigure}{0.45\textwidth}
        \includegraphics[width=1\columnwidth]{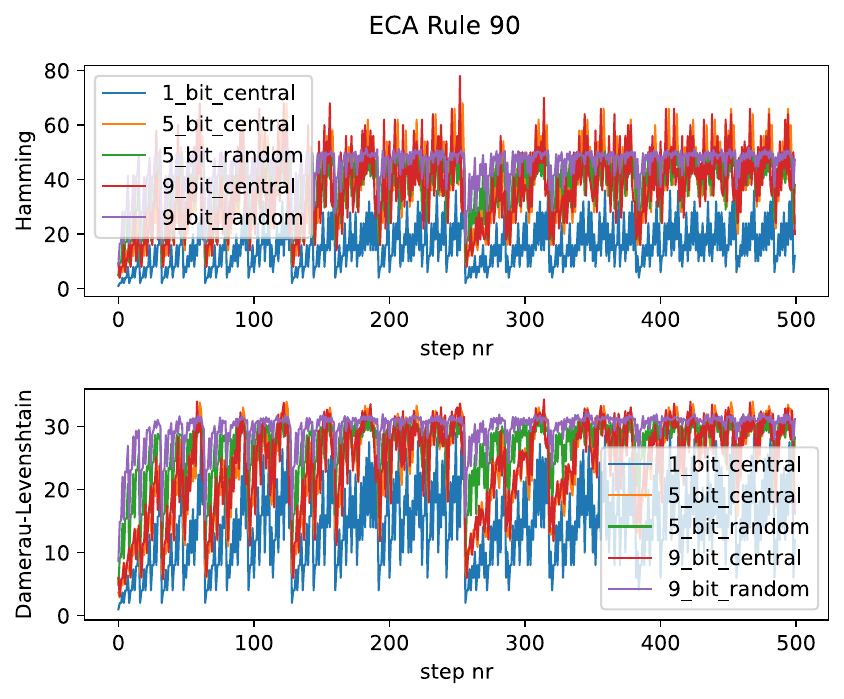}
        \includegraphics[width=1\columnwidth]{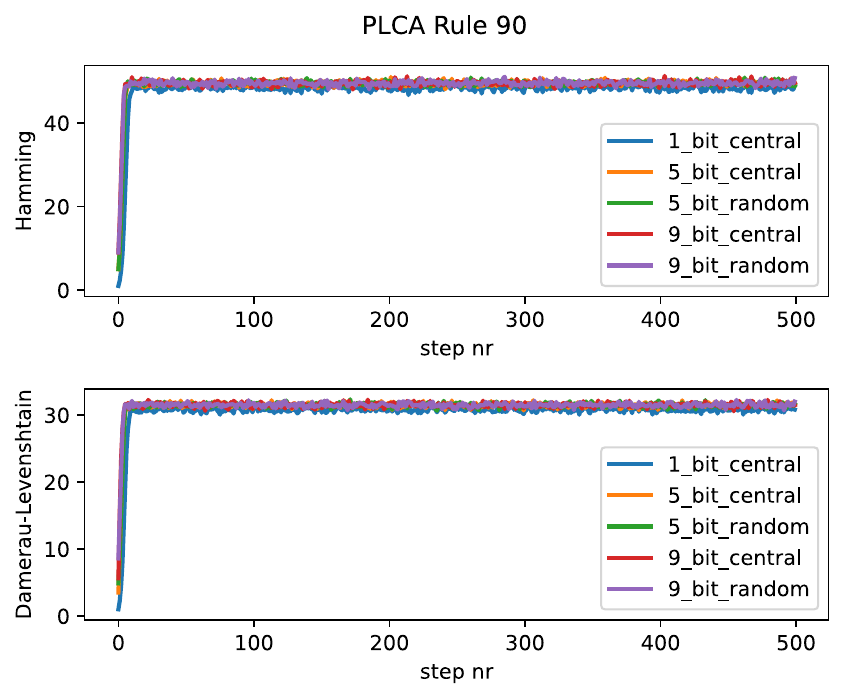}
        \includegraphics[width=1\columnwidth]{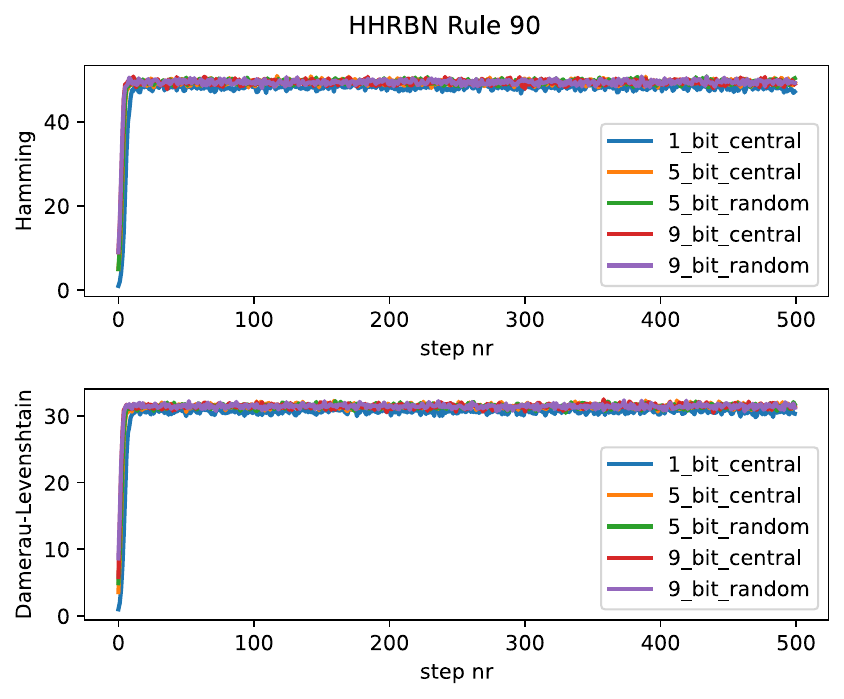}
\caption{ Rule 90}
\label{fig:SensRule90}
\end{subfigure}
\caption{Rule (left) 30 displaying ideal "chaotic" behaviour. Rule 90 (right) displays additive behaviour in CA and more ideal "chaotic" behaviour in PLCA and HHRBN.}
\end{center}
\end{figure}

We begin with an example of the ideal "chaotic" ECA rule 30 in Figure \ref{fig:SensRule30}. In ECA, we see as we expect, the randomly placed defects to be quicker to permute the substrate. The central defects naturally take longer to permute the substrate as they are limited by the CA's "speed of light" (in CA, information can only flow to direct neighbouring cells at every step, creating a speed limit for information often called the "speed of light"). For PLCA and HHRBN, much is the same, except the defects permute the substrates faster. The CA "speed of light" does not apply similarly to a random topology of PLCA and HHRBN; the random topology creates a certain level of "small worldness". We will discuss this further in subsection \ref{sec:smallWorld}
The defect configuration for many substrates settles towards the same $D_h$ of 50 (half-max distance), the normal average distance two random binary vectors would have between each other. That they all settle at the same distance is also a test of sensitivity, as the differences between configurations have expanded to the maximum probabilistic difference. 

For rule 90, seen in Figure \ref{fig:SensRule90}, it's clear that the ECA behaviour is very different. The distance might look erratic at first but is quite regular, and it is due to rule 90s additive behaviour. In rule 90 and all the other additive ECA, the defects permute in a way that is invariant to the initial condition of the different cells see \citep[subsubsections: 2.1.2-3 and 5.5.5-6 ]{glover2023investigating} for more detail. The PLCA and HHRBN behave much closer to rule 30 and "chaotic" behaviour. 

However, reflecting on this, we hypothesise that this is due to every run having a random topology rather than the regular Rule 90 behaviour somehow becoming "chaotic" in PLCA and HHRBN. We also see that the trajectories settle at roughly the same place but with a slightly higher amplitude in rule 90 than in rule 30, indicating a more significant variance.

\begin{figure}[H]
\begin{center}
\begin{subfigure}{0.49\textwidth}
        \includegraphics[width=1\columnwidth]{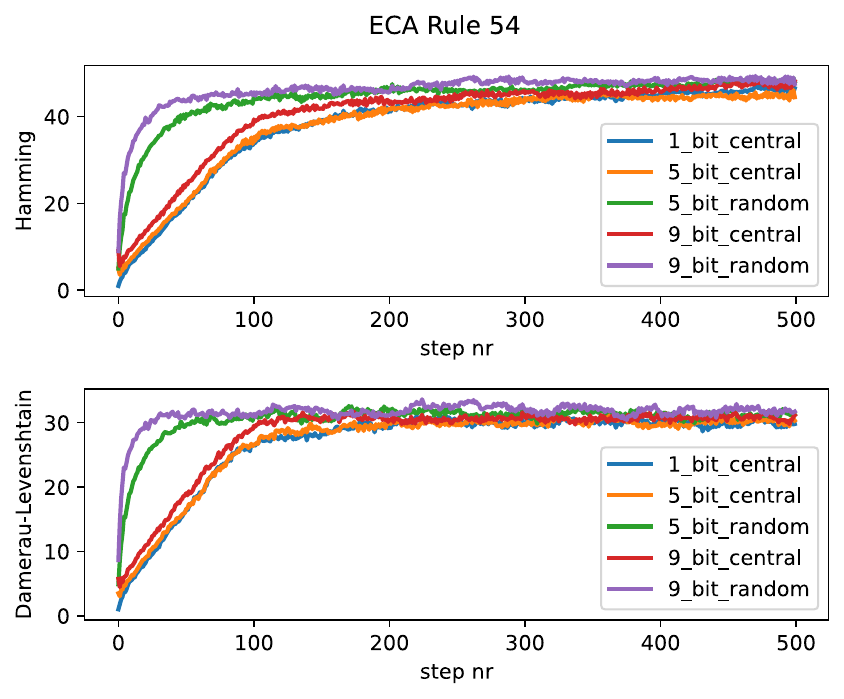}
        \includegraphics[width=1\columnwidth]{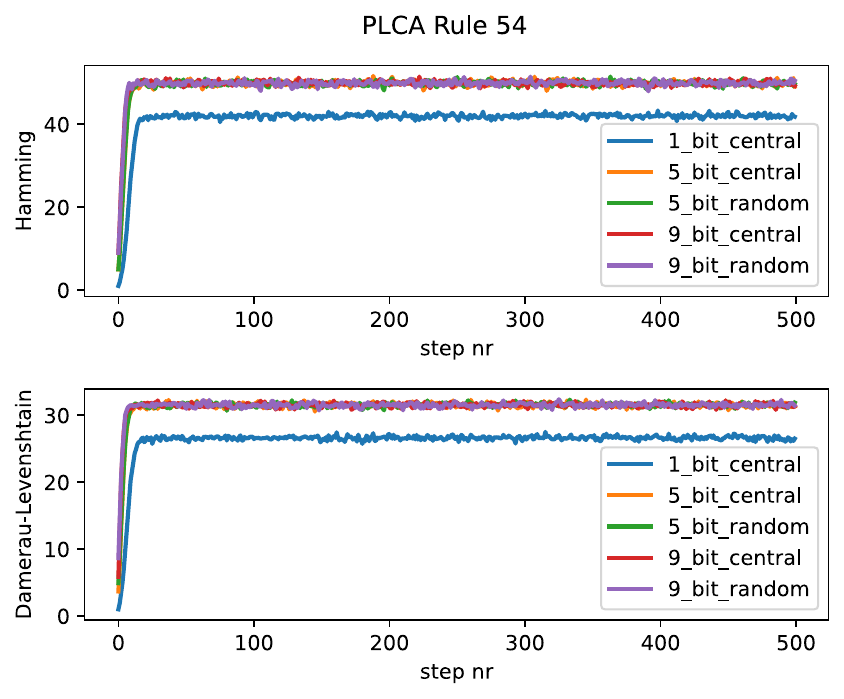}
        \includegraphics[width=1\columnwidth]{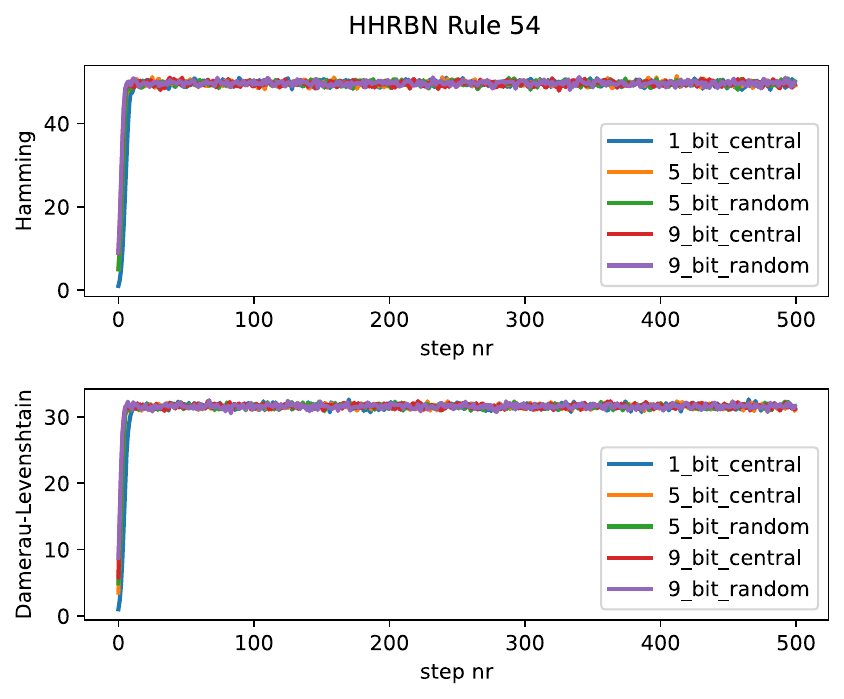}
\caption{ Rule 54}
\label{fig:SensRule54}
\end{subfigure}
\begin{subfigure}{0.49\textwidth}
        \includegraphics[width=1\columnwidth]{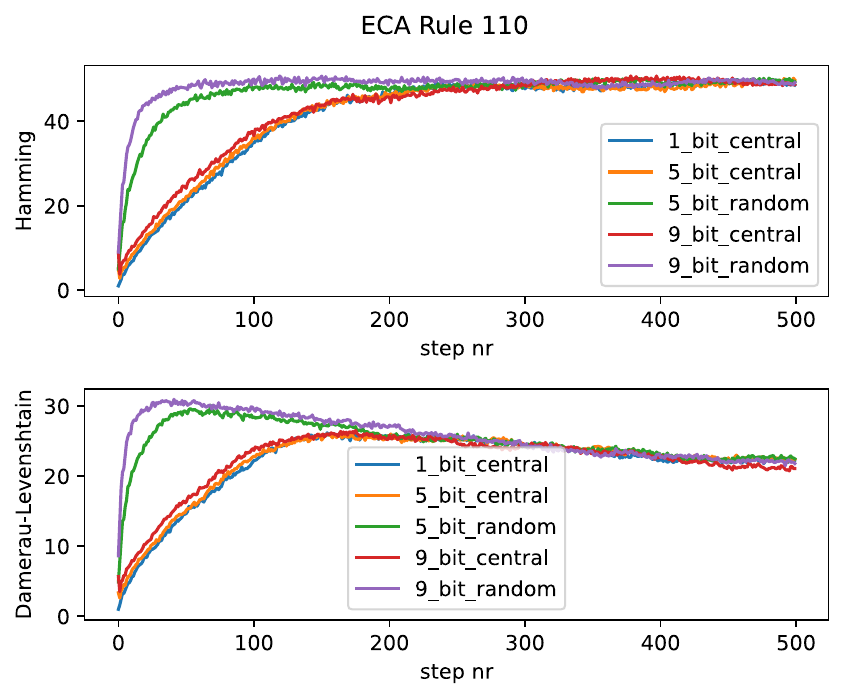}
        \includegraphics[width=1\columnwidth]{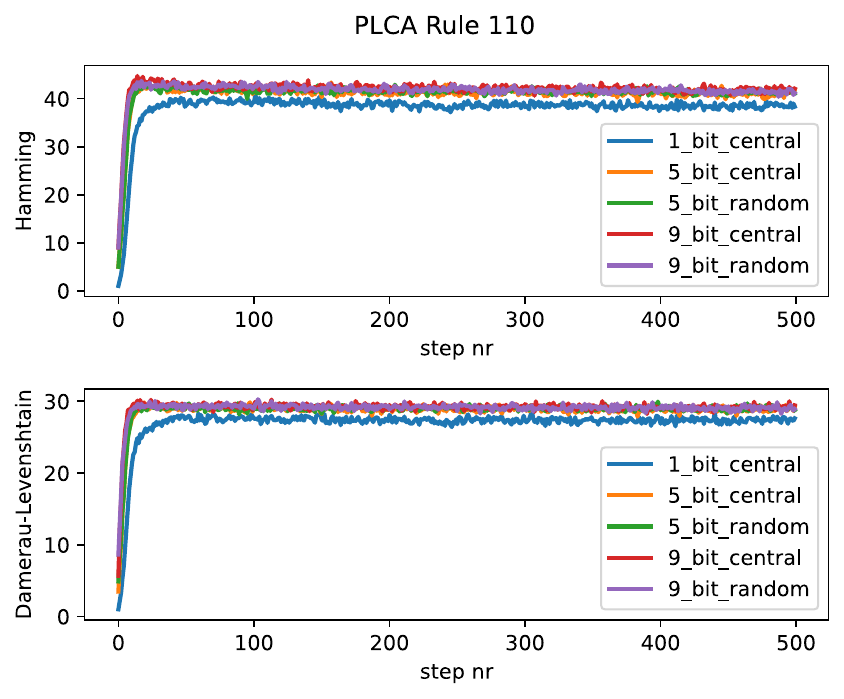}
        \includegraphics[width=1\columnwidth]{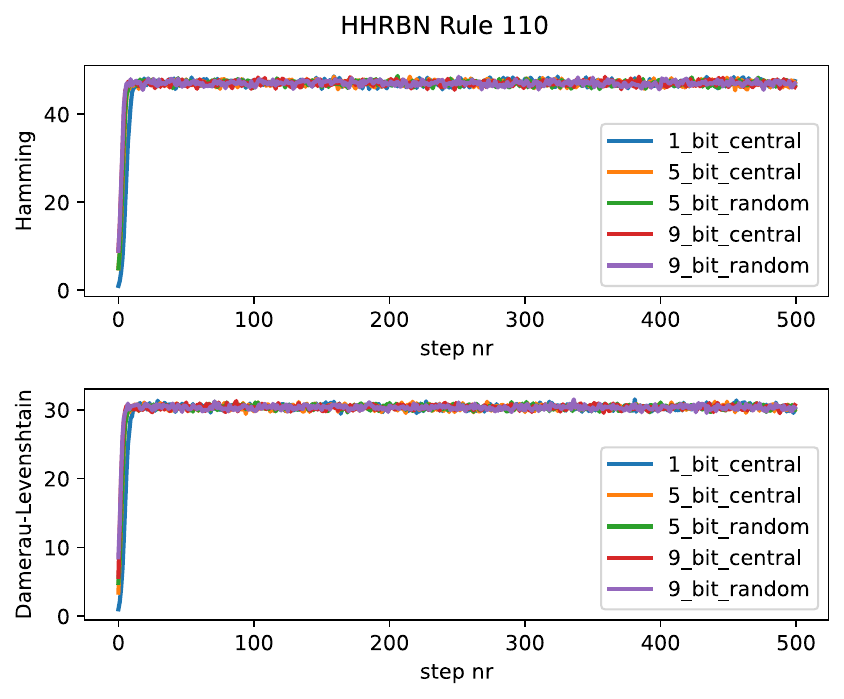}
\caption{ Rule 110}
\label{fig:SensRule110}
\end{subfigure}
\caption{ Rule 110 and 54 displaying more complex behaviour in CA, and more sensitivity in PLCA and HHRBN }
\end{center}
\end{figure}

\begin{figure}
    \centering
    \includegraphics[width=0.6\columnwidth]{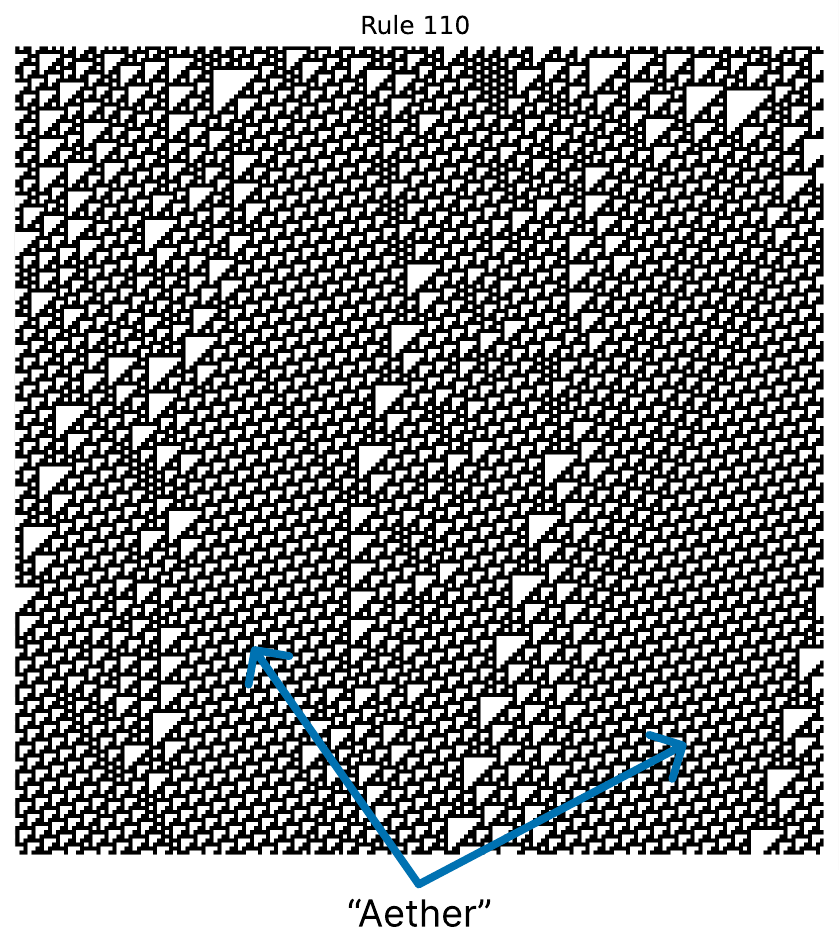}
    \caption{Example of 200x200 rule 110 displaying an example of aether}
    \label{fig:aether}
\end{figure}
Figure \ref{fig:SensRule54} shows one of the "complex" rule 54. Its behaviour would seem to fit such a classification as the defects permute the substrate but in a prolonged manner; even after 500 steps, it seems it has not fully settled at a distance in some runs. Fitting with the edge of chaos theorem, this behaviour matches expectations that its behaviour should not be trivial (ordered) and not quite chaotic. In PLCA, the effect is closer to the chaotic behaviour, yet we can see that 1-bit defects settle at different distance then the other defects. Though typically 1, bit defects are more likely to collapse, as this result excludes runs that have collapsed, that is not the reason. As this does not happen in HHRBN, we hypothesise this must have something to do with the computation of the central cell. It might be able to erect, at the very least, a weak local barrier. 

In Figure \ref{fig:SensRule110}, we see another "complex" rule 110 that, similarly to rule 54, takes long to saturate the difference distance. In contrast to rule 54, rule 110 in the $D_{dl}$ distance first grows but then shrinks again. We hypothesise this is due to the feature of Rule 110 settling in large regions "aether" as seen in Figure \ref{fig:aether}.
The aether in rule 110 is regions of regular small triangles, as the CA develops the more likely they are to show up and the larger they will be. If two configurations of CA have large regions of this aether, the configurations can be shifted to line up with each other. A shift operation is not possible in simple $D_h$, but with $D_{dl}$, a shift can be constructed using a delete and insert operation. This way, the $D_{dl}$ can create a significantly shorter edit distance than the $D_h$. Rule 54 similarly has an aether, but in contrast to rule 110, the aether areas are smaller and local in space. See \cite{wolframAtlasRule54} for an example. Therefore, we don't seem to see the same effect there. 
Similarly to Rule 54, we see a separation in PLCA where 1-bit defects do not settle on the same distance. 

\begin{figure}[H]
\begin{center}
\begin{subfigure}{0.49\textwidth}
        \includegraphics[width=1\columnwidth]{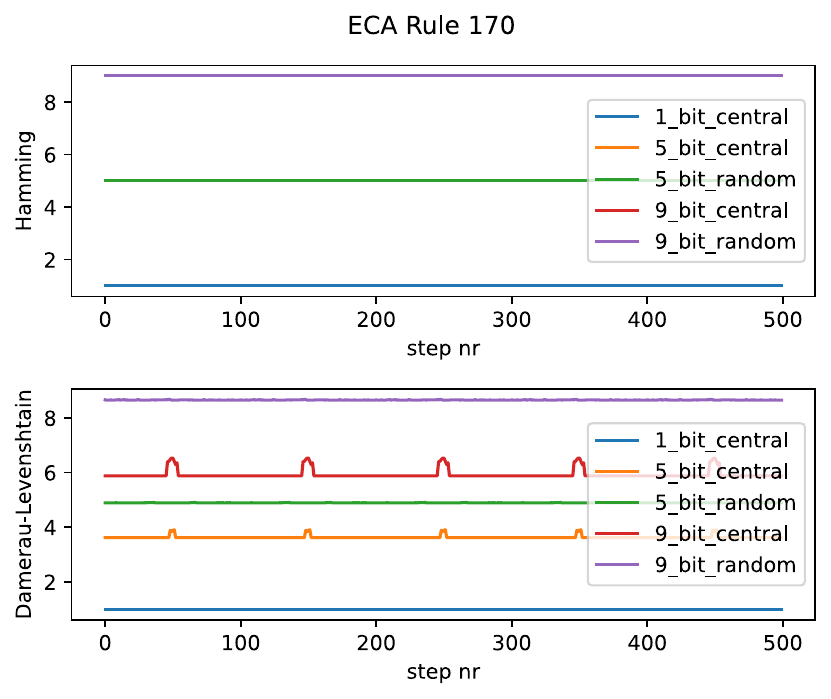}
        \includegraphics[width=1\columnwidth]{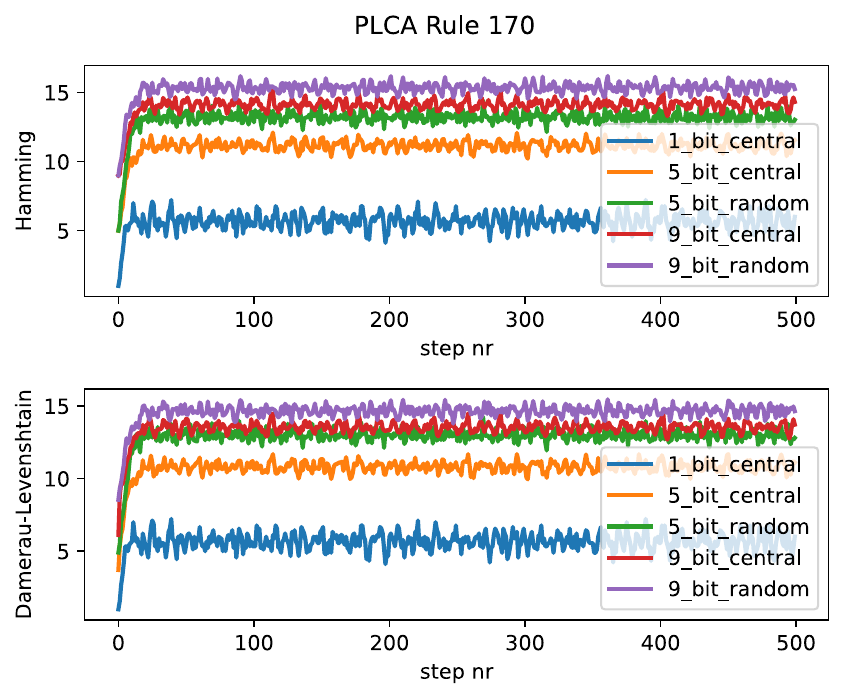}
        \includegraphics[width=1\columnwidth]{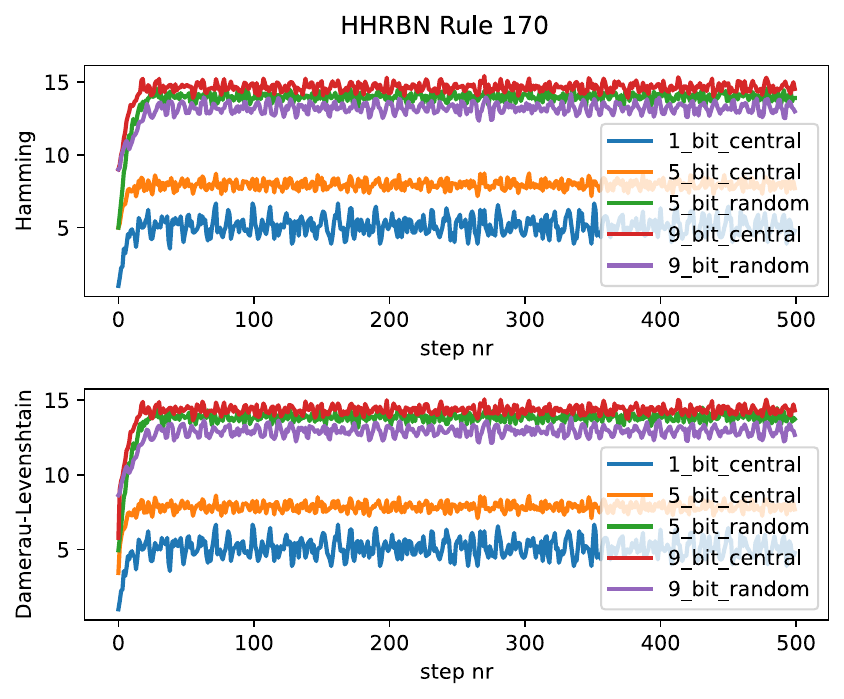}
\caption{ Rule 170}
\label{fig:SensRule170}
\end{subfigure}
\begin{subfigure}{0.49\textwidth}
        \includegraphics[width=1\columnwidth]{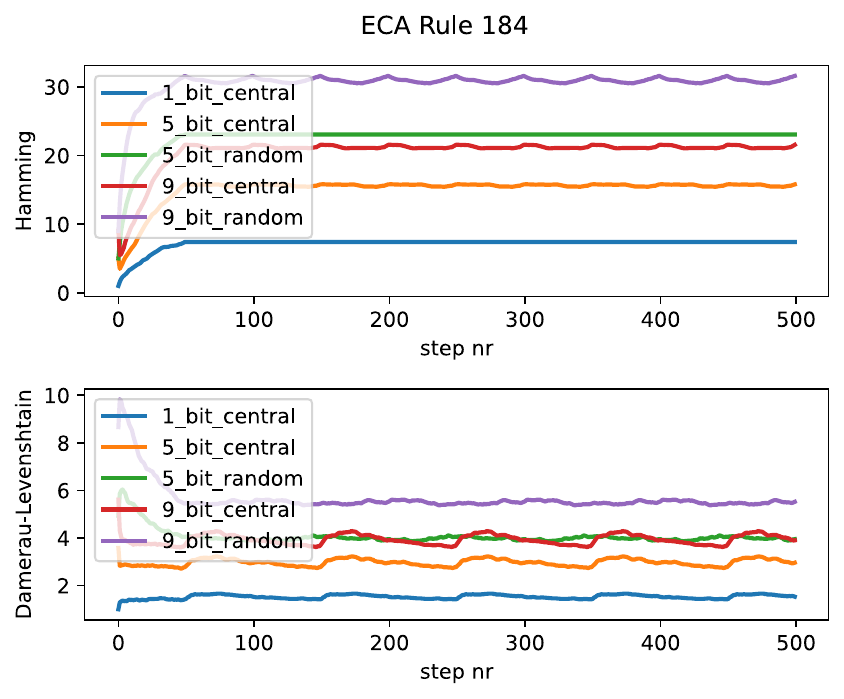}
        \includegraphics[width=1\columnwidth]{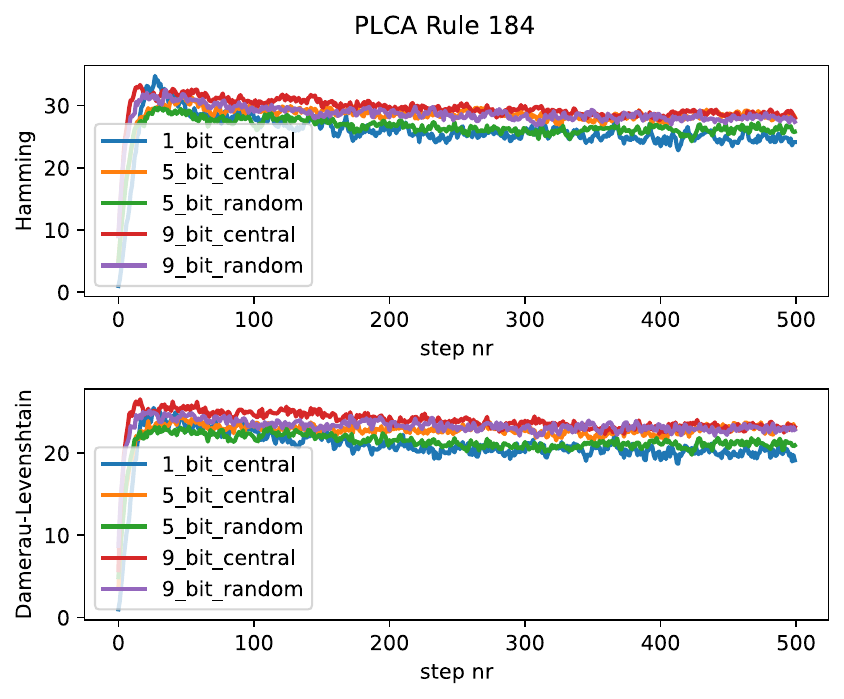}
        \includegraphics[width=1\columnwidth]{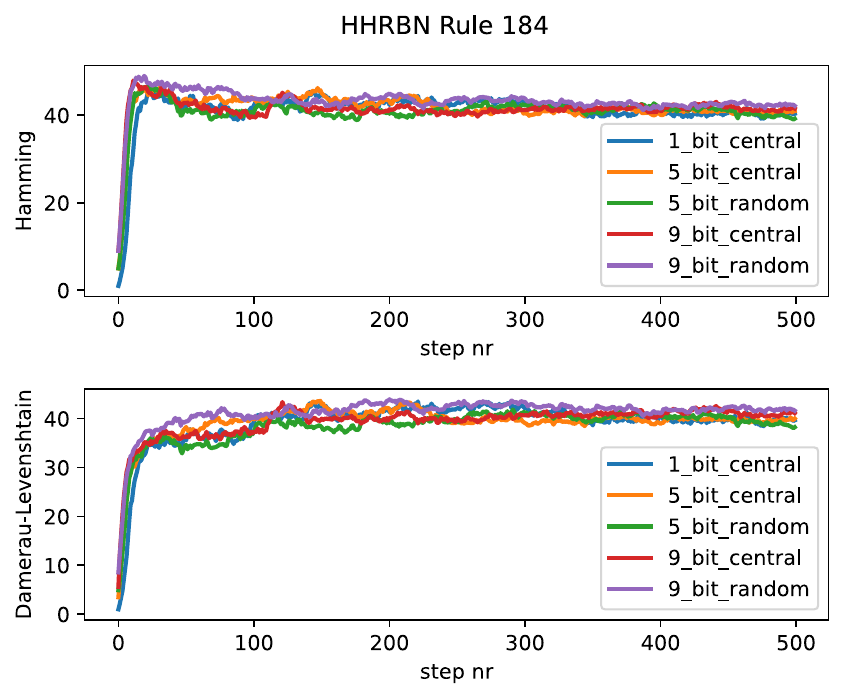}
\caption{ Rule 184}
\label{fig:SensRule184}
\end{subfigure}
\caption{ Rule 170 and 184 displaying more ordered behaviour}
\end{center}
\end{figure}

In Figure \ref{fig:SensRule170}, we see rule 170. For ECA, as we expect, the distance does not change in relation to time. In Rule 170, every cell updates based on a copy of the left neighbour, which simply shifts the previous configuration one step to the left every iteration. Therefore, a fixed distance is expected. The small blips in the $D_{dl}$ are caused by the defect hitting the CA boundary, making it less effective to use the insert-delete shift method explained earlier in rule 110 to shorten the edit distance. In rule 170s PLCA and HHRBN results, we see something unexpected. Firstly, there are different results between PLCA and HHRBN, as rule 170 only computationally depends on its right neighbour (1 cell dependant); there should be no difference. Secondly, there seems to be a separation between the random and the central defects; thirdly, in HHRBN 9, 9-bit random is lower in distance than 5-bit random. This can at least partly be explained with Table \ref{tab:rulePLCACollapse} and \ref{tab:ruleHHRBNCollapse}. As there are many collapsed runs, the remaining runs do not have as much statistical power, causing larger variance. This is further enforced by how 1-dependant rules are more vulnerable to imperfect topology. 

In Figure \ref{fig:SensRule184}, we see rule 184. We see very ordered behaviour in ECA, though not quite as trivial as Rule 170. Rule 184, similarly to rule 110, often forms large aether regions, so we see a shrinking $D_{dl}$ here. In PLCA and HHRBN, we see more dynamic behaviour, yet the distances do not settle at a half-max distance. Interestingly rule 184 has a $\lambda = \frac{4}{8}$ yet settle below half-max and rule 110 has a $\lambda = \frac{5}{8}$ yet settle at roughly half-max. This decorrelation indicates that if it is useful for a rule to be balanced in output, it is not sufficient for a rule to be balanced in the TT. Still, it also needs to be balanced at the computational scale independently of its $\lambda$. 

\subsection{Collapse rate}
\label{sec:collapseRateResults}
We will begin with the collapse rate for the selection of rules in the previous section; these are found in Table \ref{tab:ruleECACollapse}, \ref{tab:rulePLCACollapse} and \ref{tab:ruleHHRBNCollapse}. 

\begin{table}[]
    \centering
    \begin{tabular}{|c|c|c|c|c|c|c|}
      \hline
    defect & Rule 30 & Rule 90  & Rule 54 & rule 110 & rule 170 & rule 184\\
    \hline\hline
    
        1-bit c &0\% &0\% &2\% &3\% &0\% &0\% \\
        5-bit c &0\% &0\% &2\% &1\% &0\% &0\% \\
        9-bit c &0\% &0\% &2\% &0\% &0\% &0\% \\
        5-bit r &0\% &0\% &0\% &0\% &0\% &0\% \\
        9-bit r &0\% &0\% &0\% &0\% &0\% &0\% \\
        \hline
    \end{tabular}
    \caption{ECA Rule collapse rate of individual rules}
    \label{tab:ruleECACollapse}
\end{table}

\begin{table}[]
    \centering
    \begin{tabular}{|c|c|c|c|c|c|c|}
      \hline
    defect & Rule 30 & Rule 90  & Rule 54 & rule 110 & rule 170 & rule 184\\
    \hline\hline
    
        1-bit c &33\% &14\% &1\% &29\% &91\% &72\% \\
        5-bit c &0\% &0\% &0\% &0\% &62\% &29\% \\
        9-bit c &0\% &0\% &0\% &0\% &41\% &19\% \\
        5-bit r &0\% &0\% &0\% &0\% &56\% &28\% \\
        9-bit r &0\% &0\% &0\% &0\% &41\% &21\% \\
        \hline
    \end{tabular}
    \caption{PLCA Rule collapse rate of individual rules}
    \label{tab:rulePLCACollapse}
\end{table}

\begin{figure}[H]
\centering
\begin{subfigure}{0.49\textwidth}
        \includegraphics[width=1\columnwidth]{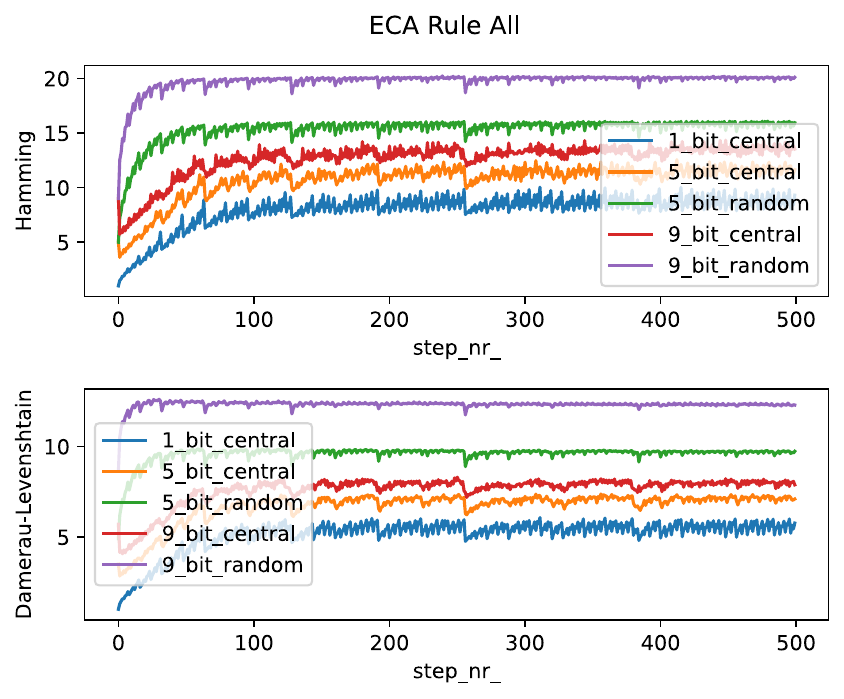}
        \includegraphics[width=1\columnwidth]{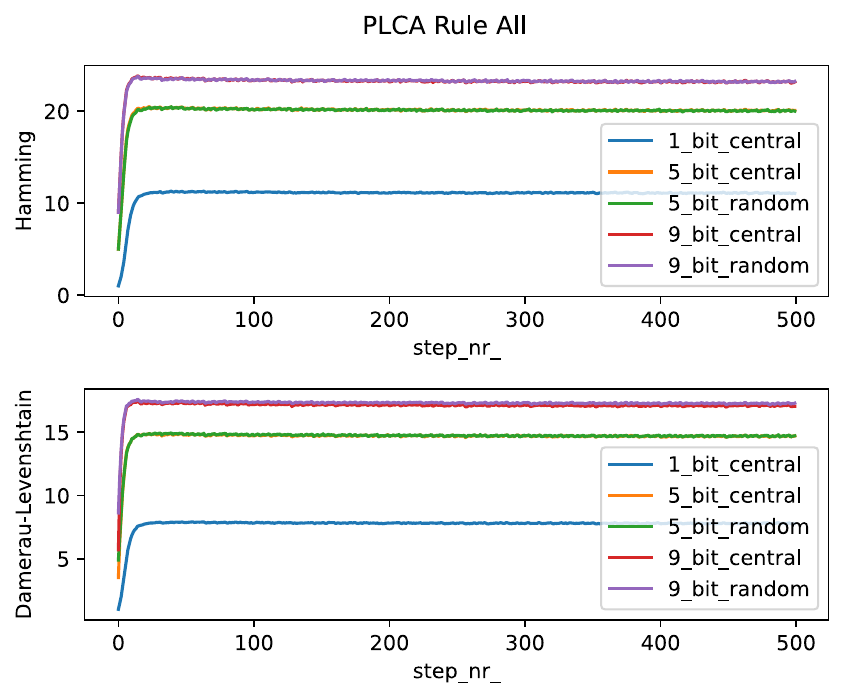}
        \includegraphics[width=1\columnwidth]{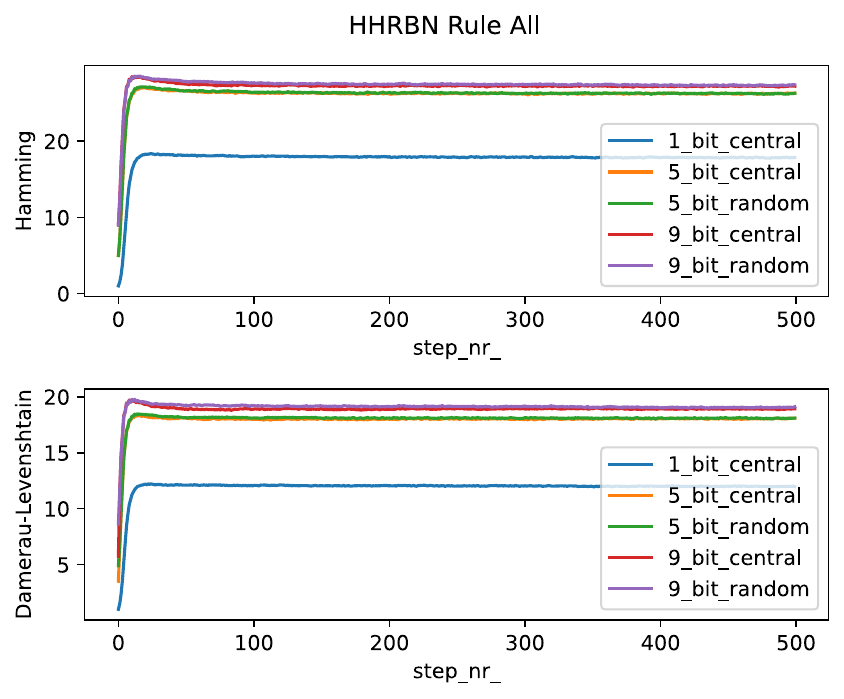}
\caption{ TDT including collapsed runs}
\label{fig:ECAHammAcc}
\end{subfigure}
\begin{subfigure}{0.49\textwidth}
        \includegraphics[width=1\columnwidth]{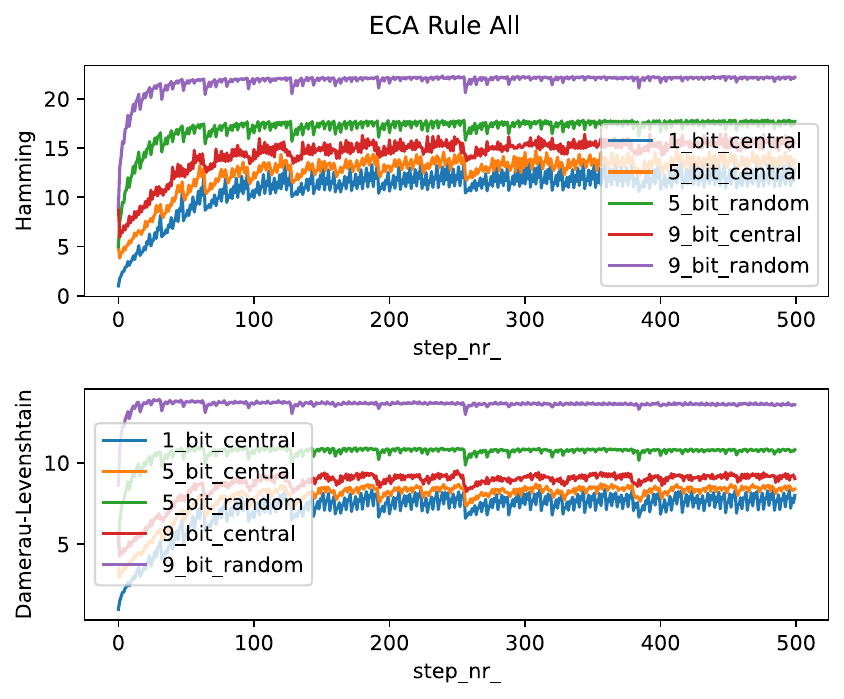}
        \includegraphics[width=1\columnwidth]{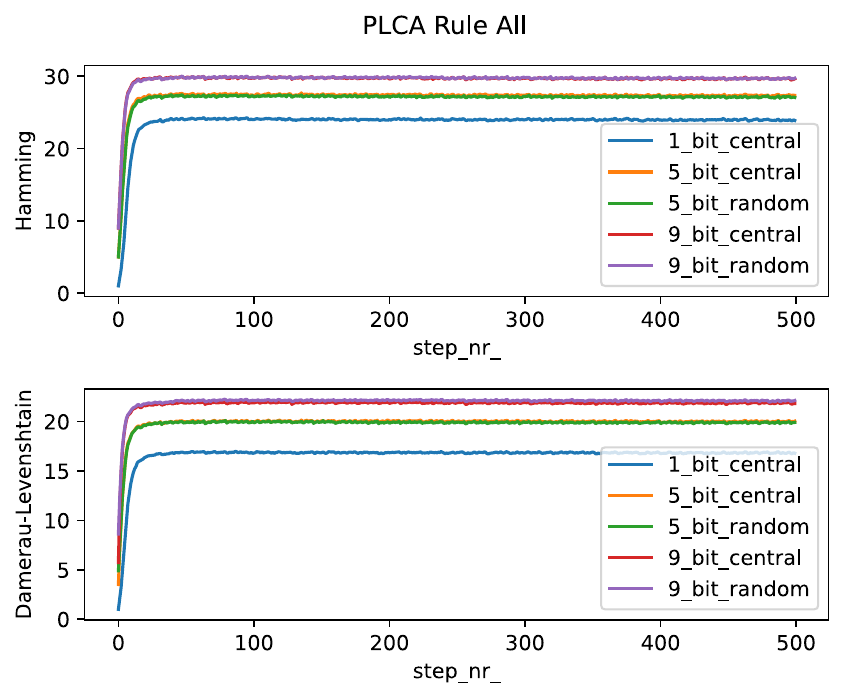}
        \includegraphics[width=1\columnwidth]{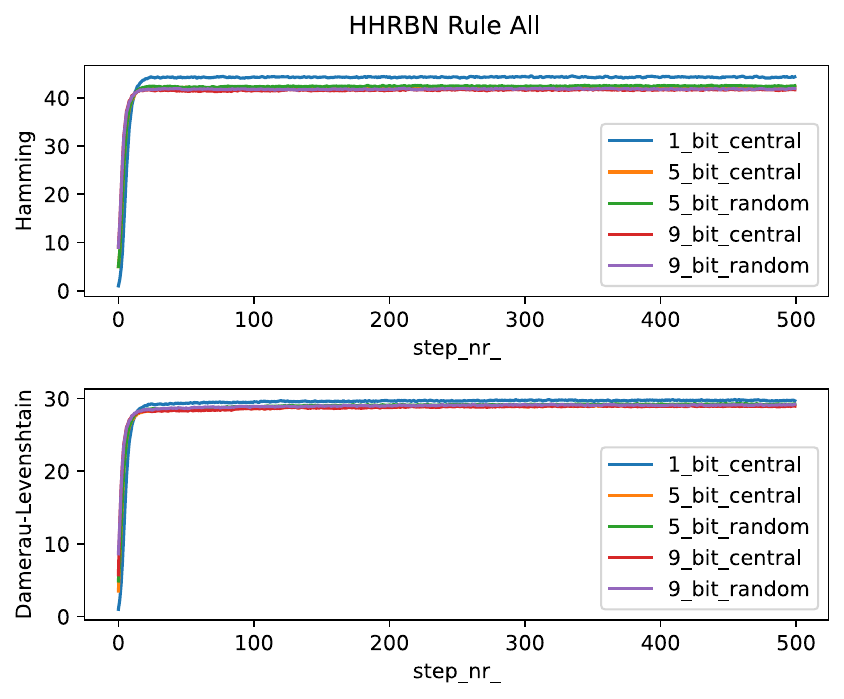}
\caption{ TDT excluding collapsed runs }
\label{fig:ECAHammAccDrop}
\end{subfigure}
\caption{TDT comparing with or without collapse of the run. There is a noticeable increase in $D_h$ and $D_{dl}$ when excluding collapsed runs, especially with the 1-bit defect.}
\end{figure}

\begin{table}[]
    \centering
    \begin{tabular}{|c|c|c|c|c|c|c|}
      \hline
    defect & Rule 30 & Rule 90  & Rule 54 & rule 110 & rule 170 & rule 184\\
    \hline\hline
    
        1-bit c &21\% &21\% &17\% &21\% &91\% &59\% \\
        5-bit c &0\% &0\% &0\% &0\% &56\% &29\% \\
        9-bit c &0\% &0\% &0\% &0\% &35\% &26\% \\
        5-bit r &0\% &0\% &0\% &0\% &61\% &31\% \\
        9-bit r &0\% &0\% &0\% &0\% &46\% &21\% \\
        \hline
    \end{tabular}
    \caption{HHRBN Rule collapse rate of individual rules}
    \label{tab:ruleHHRBNCollapse}
\end{table}
We see that for these rules, there is almost no collapse in ECA, but for PLCA, 1-bit defects regularly collapse except for Rule 54. In contrast, rules 170 and 184 have a large number of collapses. In the case of rule 170, we will explain, in subsection \ref{sec:network topology}, how the rule dependency is one reason for the high collapse rate. For the HHRBN, much is the same compared to PLCA, except for rule 54, which has a higher collapse rate for 1-bit defects.
In Figure \ref{fig:ECAHammAccDrop}, we see the TDP for the full ECA ME results with and without collapsed runs included. Firstly, we can see that the additive rules, such as rule 90 and its frequencies, impact the data such that it is still visible in the total data consisting of all the 88 ME rules. We can also separate the different defect initialisations in inclusive and exclusive collapsed runs. Similarly, in PLCA, we see a separation between the defect sizes but not the defect types (random, central); this is as expected as they are functionally the same in PLCA and HHRBN substrates. In HHRBN, excluding collapsed runs, we see something that looks mainly like sensitive behaviour, though it does not settle at a roughly half-max distance. This might be due to how the ME set is derived because of the complement transformation many of the later rules in the rule-set are excluded, and the later rules have on average a higher $\lambda$, in fact, the average $\lambda = 0.386$. See Table \ref{tab:lambdaECA} for why that is. Additionally, we observed that a 1-bit defect has a slightly higher settling distance. Looking at many individual rules, a few rules, such as rule 140, have this effect of a significantly higher average for the 1-bit difference.  

\begin{table}[h]
\centerline{
    \begin{tabular}{|r|r|r|r|}
    \hline
    init & ECA & PLCA  & HHRBN\\
    \hline\hline
        1 central & 2450 (27.8\%) & 4722 (53.7\%) & 5253 (59.7\%) \\
        5 central & 1352 (15.4\%) & 2338 (26.6\%) & 3328 (37.8\%) \\
        9 central & 1123 (12.8\%) & 1918 (21.8\%) & 3049 (34.6\%) \\
        5 random  &  872 (9.9\%)  & 2295 (26.1\%) & 3355 (38.1\%) \\
        9 random  &  828 (9.4\%)  & 1919 (21.8\%) & 3043 (34.6\%) \\
        
    \hline

    \end{tabular}
 }
    \caption{The number of times in the whole 8800 runs (100 runs for 88 rules) per substrate rule population that collapsed}
    \label{tab:distance_drop}
\end{table}

Table \ref{tab:distance_drop} provides a comprehensive overview of the collapsed runs by substrate. The data reveals a clear trend: the likelihood of collapse increases from ECA to PLCA to HHRBN and from more defects to less. 
In Table \ref{tab:ECA_distance_drop_dep}, we see the same ECA data but grouped by dependencies. See Table \ref{tab:dependancyECA} for details on which rule belongs where. For the ECA, we can, with the support of theory, understand the results quite reliably. 0-dependency always collapses as rule 0 (the only 0-dependency rule in the ME) always has the same attractor of an entirely quiescent configuration (all cells become 0 state). For 1-dependency (rule 15,51,170,204), they will never collapse in ECA because they are all of a simple behaviour that propagates the information without much (if any) transformation. Rules 170 and 204 are additive and rules 15 and 51 are negative versions of additive rules. Negative transformation is simply changing the output bit of the TT to the opposite (not to be confused with complement transformation). This transformation leads to similar behaviour but is not typically considered part of the ME transformations. The 2-dependency rules are 3,5,10,12,34,60,90,136,160. For these rules, rules 136 and 160 dominate the collapse rate as they always collapse. These rules are such that they always turn quiescent after a fixed number of iterations. As is for rule 0, it is more accurate to describe it as all initial conditions collapse rather than the defects collapse. In contrast, rules 60 and 90 should almost never collapse due to their additive nature. However, some configurations will collapse for these rules, such as for specific grid sizes \cite[subsubsecton 5.5.5]{glover2023investigating}. Similarly, we can go through the 3-dependency rules, but for the sake of brevity, we will not. 

\begin{table}[h]
\centerline{
    \begin{tabular}{|r|r|r|r|r|}
    \hline
    init. & 0 dep. & 1 dep.  & 2 dep. & 3 dep.\\
     \hline \hline
    1 central & 100 (100.0\%) & 0 (0.0\%) & 315 (35.0\%) & 2035 (27.5\%)\\
    5 central & 100 (100.0\%) & 0 (0.0\%) & 204 (22.7\%) & 1048 (14.2\%)\\ 
    9 central & 100 (100.0\%) & 0 (0.0\%) & 200 (22.2\%) & 823 (11.1\%)\\
    5 random  & 100 (100.0\%) & 0 (0.0\%) & 200 (22.2\%) & 573 (7.7\%)\\
    9 random  & 100 (100.0\%) & 0 (0.0\%) & 200 (22.2\%) & 528 (7.1\%)\\
    \hline
    \end{tabular}
 }
    \caption{Number of times the ECA rules difference died out grouped by neighbour dependencies}
    \label{tab:ECA_distance_drop_dep}
\end{table}

\begin{table}[h]
\centerline{
    \begin{tabular}{|r|r|r|r|r|}
    \hline
    init. & 0 dep. & 1 dep.  & 2 dep. & 3 dep.\\
     \hline \hline
    1 central & 100 (100.0\%) & 182 (45.5\%) & 626 (59.6\%) & 3814 (51.5\%)\\
    5 central & 100 (100.0\%) & 112 (28.0\%) & 487 (54.1\%) & 1639 (22.1\%)\\ 
    9 central & 100 (100.0\%) & 68 (17.0\%) & 424 (47.1\%) & 1326 (17.9\%)\\
    5 random  & 100 (100.0\%) & 104 (26.0\%) & 476 (52.9\%) & 1615 (21.8\%)\\
    9 random  & 100 (100.0\%) & 72 (18.0\%) & 414 (46.0\%) & 1333 (18.0\%)\\
    \hline

    \end{tabular}
 }
    \caption{Number of times the PLCA rules difference died out grouped by neighbour dependencies}
    \label{tab:PLCA_distance_drop_dep}
\end{table}

\begin{table}[h]
\centerline{
    \begin{tabular}{|r|r|r|r|r|}
    \hline
    init. & 0 dep. & 1 dep.  & 2 dep. & 3 dep.\\
     \hline \hline
    1 central & 100 (100.0\%) & 354 (88.5\%) & 715 (79.4\%) & 4084 (55.2\%)\\
    5 central & 100 (100.0\%) & 217 (54.3\%) & 618 (68.7\%) & 2393 (32.3\%)\\ 
    9 central & 100 (100.0\%) & 152 (38.0\%) & 583 (64.8\%) & 2214 (29.9\%)\\
    5 random  & 100 (100.0\%) & 210 (52.5\%) & 621 (69.0\%) & 2424 (32.8\%)\\
    9 random  & 100 (100.0\%) & 151 (37.8\%) & 572 (63.5\%) & 2220 (30.0\%)\\
    \hline
    \end{tabular}
 }
    \caption{Number of times the HHRBN rules difference died out grouped by neighbour dependencies}
    \label{tab:HHRBN_distance_drop_dep}
\end{table}

In PLCA and HHRBN, we expected a strong trend of lower dependency, meaning more collapse, mainly due to network properties, which we discuss in subsection \ref{sec:network topology}. 
In Table \ref{tab:PLCA_distance_drop_dep} and \ref{tab:HHRBN_distance_drop_dep}, we see some trend towards more collapse the lower the dependency, but only if we ignore the 2-dependency results. Except for 1-bit defects, 2-dependency has a higher collapse rate than 1-dependency in both cases. One hypothesis is that because the rules in 1-dependency are all $\lambda = \frac{4}{8}$ (balanced), this dominates the dependency effect in 1-dependency rules, and all of the 2-dependency rules except rules 60 and 90 are $\lambda = \frac{2}{8}$. This could be tested by grouping by $\lambda$.

In Table \ref{tab:ECA_distance_drop_lambda}, \ref{tab:PLCA_distance_drop_lambda} and \ref{tab:HHRBN_distance_drop_lambda} we do just this. Group our results by $\lambda$ value of the TT. In this case, we see a clear trend of the balance of the rule affecting the collapse rate. This $\lambda$ value organise the collapse rate clearly, for all parameters the collapse rate grows when the $\lambda$ is closer to the edge values 0 and 1. This also gives some reason as to why the dependency value was not so neatly described.

\begin{table}[h]
\centerline{
    \newcommand\Hstrut{\rule[-1.4ex]{0pt}{4.2ex}}
    \begin{tabular}{|r|r|r|r|r|r|}
    \hline
    
    init. & $\lambda = \frac{0}{8}\lor\frac{8}{8}$ & $\lambda = \frac{1}{8}\lor\frac{7}{8}$  & $\lambda = \frac{2}{8}\lor\frac{6}{8}$ & $\lambda = \frac{3}{8}\lor\frac{5}{8}$ & $\lambda =  \frac{4}{8}$ \Hstrut\\
     \hline \hline
    1 central & 100 (100.0\%) & 422 (70.3\%) & 655 (36.4\%) & 787 (23.1\%) & 486 (16.8\%)\\
    5 central & 100 (100.0\%) & 342 (57.0\%) & 420 (23.3\%) & 337 (9.9\%) & 153 (5.3\%)\\ 
    9 central & 100 (100.0\%) & 318 (53.0\%) & 378 (21.0\%) & 224 (6.6\%) & 103 (3.6\%)\\
    5 random  & 100 (100.0\%) & 307 (51.2\%) & 312 (17.3\%) & 131 (3.9\%) & 23 (0.8\%)\\
    9 random  & 100 (100.0\%) & 300 (50.0\%) & 302 (16.8\%) & 108 (3.2\%) & 18 (0.6\%)\\
    \hline

    \end{tabular}
 }
    \caption{ Number of times the ECA rules difference died out group by Lambda}
    \label{tab:ECA_distance_drop_lambda}
\end{table}

\begin{table}[h]
\centerline{
    \newcommand\Hstrut{\rule[-1.4ex]{0pt}{4.2ex}}
    \begin{tabular}{|r|r|r|r|r|r|}
    \hline
    
    init. & $\lambda = \frac{0}{8}\lor\frac{8}{8}$ & $\lambda = \frac{1}{8}\lor\frac{7}{8}$  & $\lambda = \frac{2}{8}\lor\frac{6}{8}$ & $\lambda = \frac{3}{8}\lor\frac{5}{8}$ & $\lambda =  \frac{4}{8}$ \Hstrut\\
     \hline \hline
    1 central & 100 (100.0\%) & 542 (90.3\%) & 1236 (68.7\%) & 1739 (51.1\%) & 1105 (38.1\%)\\
    5 central & 100 (100.0\%) & 481 (80.2\%) & 783 (43.5\%) & 662 (19.5\%) & 312 (10.8\%)\\ 
    9 central & 100 (100.0\%) & 463 (77.2\%) & 679 (37.7\%) & 489 (14.4\%) & 187 (6.4\%)\\
    5 random  & 100 (100.0\%) & 470 (78.3\%) & 777 (43.2\%) & 652 (19.2\%) & 296 (10.2\%)\\
    9 random  & 100 (100.0\%) & 459 (76.5\%) & 679 (37.7\%) & 494 (14.5\%) & 187 (6.4\%)\\
    \hline

    \end{tabular}
 }
    \caption{ Number of times the PLCA rules difference died out group by Lambda}
    \label{tab:PLCA_distance_drop_lambda}
\end{table}

\begin{table}[h]
\centerline{
    \newcommand\Hstrut{\rule[-1.4ex]{0pt}{4.2ex}}
    \begin{tabular}{|r|r|r|r|r|r|}
    \hline
    
    init. & $\lambda = \frac{0}{8}\lor\frac{8}{8}$ & $\lambda = \frac{1}{8}\lor\frac{7}{8}$  & $\lambda = \frac{2}{8}\lor\frac{6}{8}$ & $\lambda = \frac{3}{8}\lor\frac{5}{8}$ & $\lambda =  \frac{4}{8}$ \Hstrut\\
     \hline \hline
    1 central & 100 (100.0\%) & 593 (98.8\%) & 1404 (78.0\%) & 1873 (55.1\%) & 1283 (44.2\%)\\
    5 central & 100 (100.0\%) & 566 (94.3\%) & 1091 (60.6\%) & 1070 (31.5\%) & 501 (17.3\%)\\ 
    9 central & 100 (100.0\%) & 562 (93.7\%) & 1026 (57.0\%) & 940 (27.6\%) & 421 (14.5\%)\\
    5 random  & 100 (100.0\%) & 576 (96.5\%) & 1096 (60.9\%) & 1064 (31.3\%) & 519 (17.9\%)\\
    9 random  & 100 (100.0\%) & 544 (90.7\%) & 1023 (56.8\%) & 950 (27.9\%) & 426 (14.7\%)\\
    \hline
    \end{tabular}
 }
    \caption{ Number of times the HHRBN rules difference died out group by Lambda}
    \label{tab:HHRBN_distance_drop_lambda}
\end{table}

\subsection{Network topology, longest simple cycle}
\label{sec:network topology}
We can see from examples such as Figure \ref{fig:SensRule30} that the original ECA "speed of light" is still there in a sense also for PLCA and HHRBN, though vastly different. Defects can still only affect neighbouring cells, but the pathways through the network that the defects can propagate are more small-world than ECA. This is unsurprising as they are randomly generated in contrast to ECA. We want to create a sense of how this affects the computations. We do this by finding a network's longest simple cycle. The longest simple cycle indicates the theoretical memory size that can be encoded into the network and the substrate's ability to retain a memory in the cycle. 

We begin by considering a 1-dependency rule, such as rule 170. A ECA rule that depends only on 1 neighbour means by definition that 2 of the neighbours do not affect the computation. This effectively reduces the in-degree from 3 to just 1. Though our collapse results do not show a clear trend based on this dependency, we still argue that it must affect the computation because the network properties of a 1-in-degree network are majorly different from a 3-in-degree network. The following evidence may persuade the reader of this conclusion. 

\begin{figure}[h]
    \begin{center}
        \includegraphics[width=1\columnwidth]{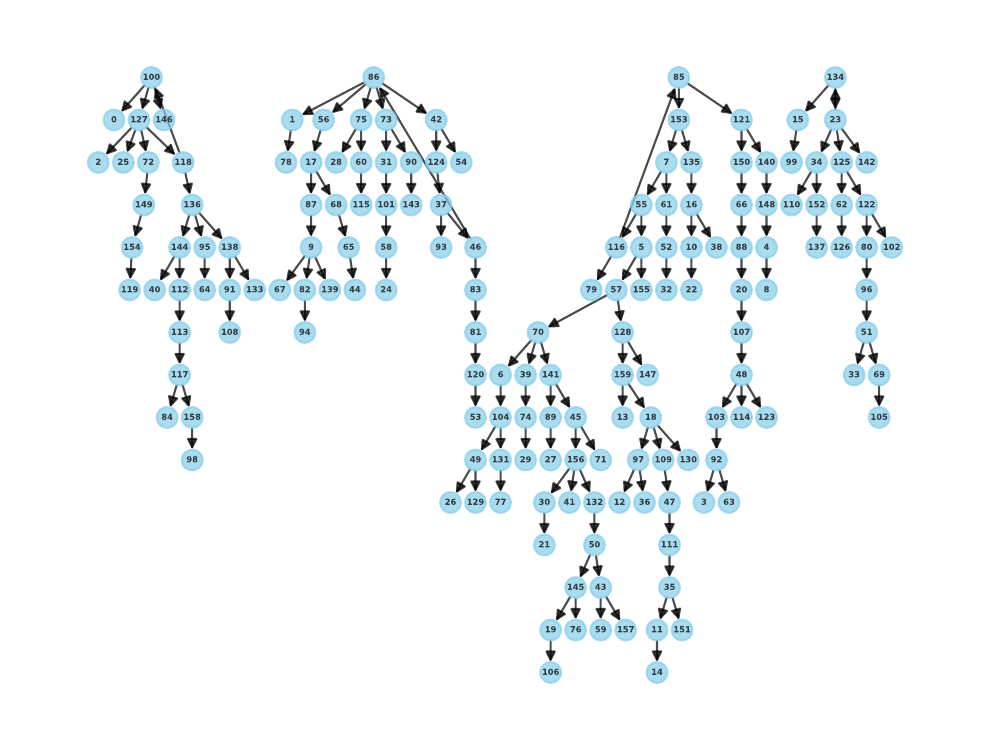}
        \caption{Tree graph visualisation of an example of a generated connection graph for Rule 170 PL CA. Showing four isolated subgraphs and very few cycles. }
        \label{fig:R170GraphT}       
    \end{center}
\end{figure}

\begin{figure}[h]
    \begin{center}
        \includegraphics[width=1\columnwidth]{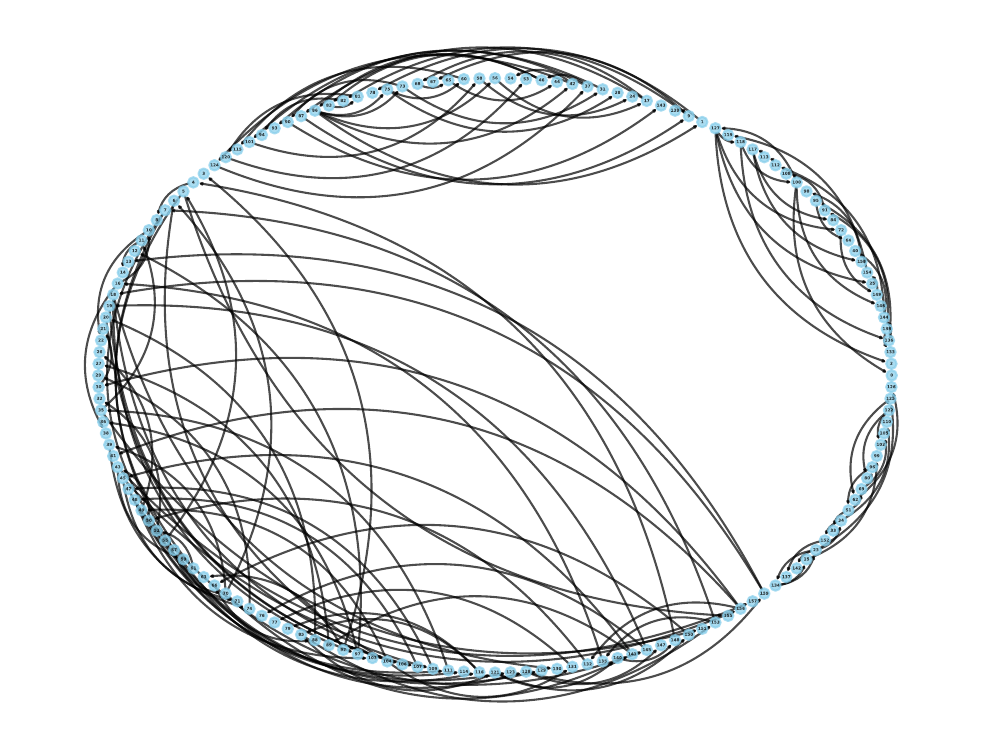}
        \caption{Circle graph visualisation of an example of a generated connection graph for Rule 170 PL CA. Showing four isolated subgraphs.}
        \label{fig:R170GraphC}       
    \end{center}
\end{figure}

We take the first topology generated for rule 170 from the data in Figure \ref{fig:5bit_sens} and only keep the nodes that compute (the third in node neighbour). The network can be seen in Figure \ref{fig:R170GraphT}, and \ref{fig:R170GraphC}. We see that there are 4 isolated subgraphs in this topology and very few and very short cycles. 63\% of the networks generated had at least one isolated subgraph. On average, the networks had 2.26 isolated subgraphs. Note, that this means the example is skewed towards a more affected network, it still demonstrates well the issues with the networks. 

\begin{figure}[h]
    \begin{center}
        \includegraphics[width=1\columnwidth]{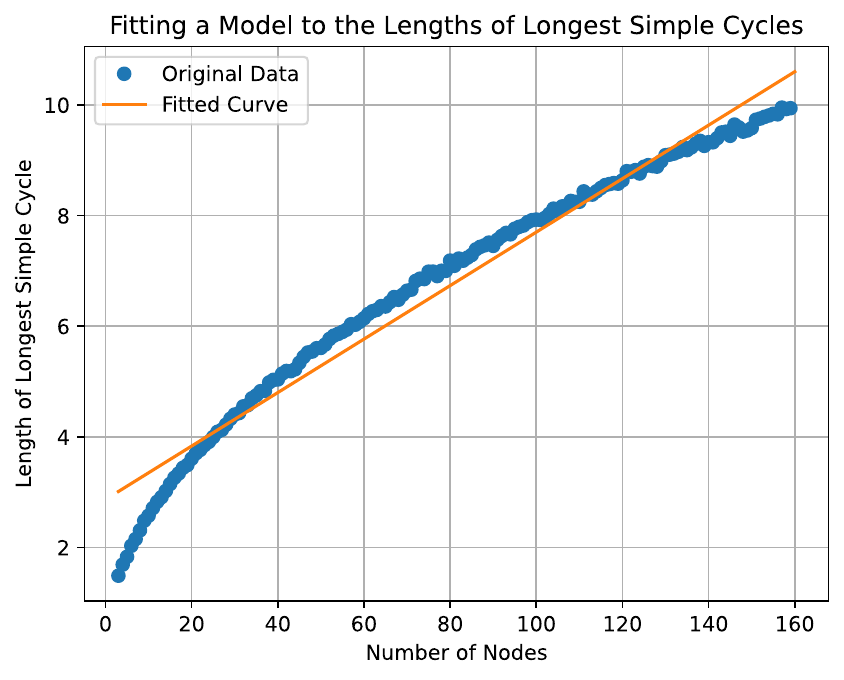}
        \caption{ From samples of random graphs between 2 and 160 nodes with in-degree of 1. Every x value is the average of 10000 random graphs. We fit this data to a linear function using linear regression for the sake of comparison to Figure \ref{fig:LSC_id2} and \ref{fig:LSC_id3}}
        \label{fig:LSC_id1}       
    \end{center}
\end{figure}

\begin{figure}[h]
    \begin{center}
        \includegraphics[width=1\columnwidth]{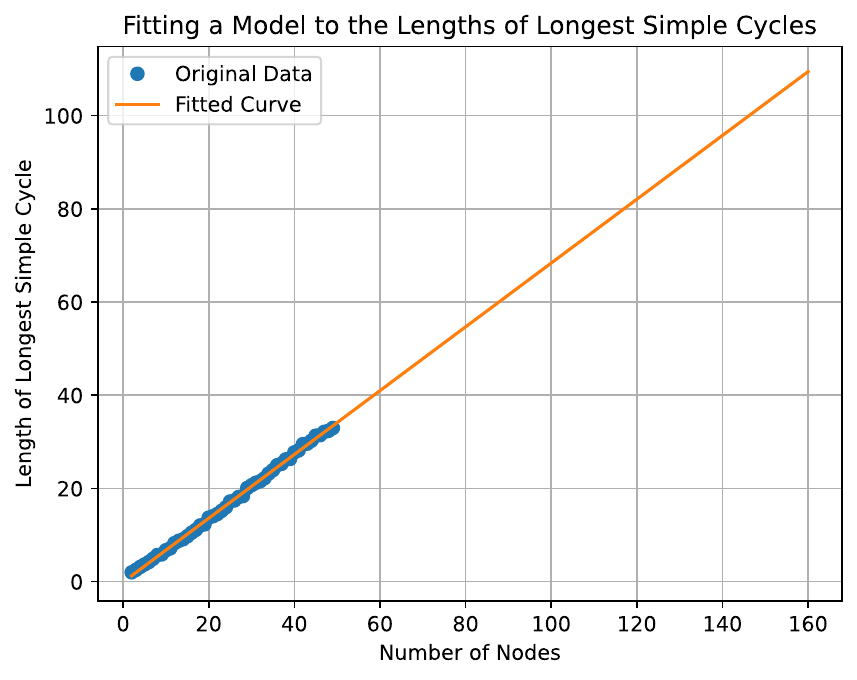}
        \caption{Predicted average length of the longest simple cycle using Simple linear regression fitting. From samples of random graphs between 2 and 50 nodes with in-degree of 2. Every x value is the average of 100 random graphs. Assuming the true function is asymptotic as in Figure \ref{fig:LSC_id1}, the true value for the portion after the data should be somewhere below the fitted line}
        \label{fig:LSC_id2}       
    \end{center}
\end{figure}

\begin{figure}[h]
    \begin{center}
        \includegraphics[width=1\columnwidth]{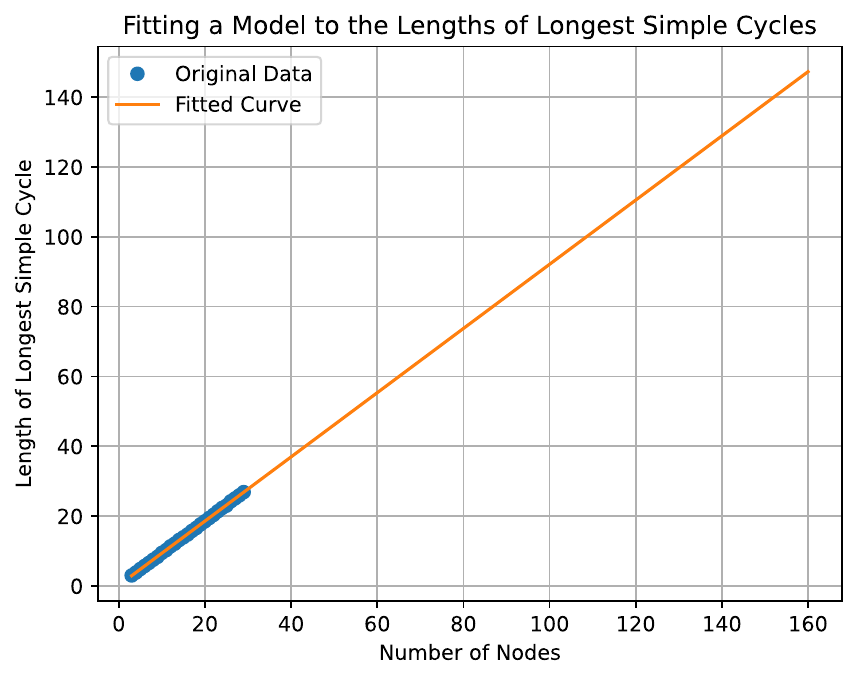}
        \caption{Predicted average length of the longest simple cycle using Simple linear regression fitting. From samples of random graphs between 3 and 30 nodes with in-degree of 3. Every x value is the average of 100 random graphs. Assuming the true function is asymptotic as in Figure \ref{fig:LSC_id1}, the true value for the portion after the data should be somewhere below the fitted line  }
        \label{fig:LSC_id3}       
    \end{center}
\end{figure}

We made many networks with 1-in-degree for $N$ (3 ... 160) nodes and found the longest simple cycle in these networks; see Figure \ref{fig:LSC_id1}. The computing cost of running this for 1-in-degree is trivial. Therefore, we run this 10000 per $N$. This data shows that the trend is asymptotic and that for $N = 100$, the average longest simple cycle is about 8 and 10 for $N = 160$. 
For 2-in-degree networks, finding the values for 100 and 160 nodes is vastly different; the number of cycles in this graph grows quickly, and the times to compute them were exponential. Therefore, we only computed for 3 to 50 nodes. Assuming this is asymptotic in the same manner as for 1-in-degree, we can fit a line to the data and know that the true values for the longest simple cycle are below that line. This can be seen in Figure \ref{fig:LSC_id2}. From this estimate, the longest simple cycle should be below 70 for $N = 100$ and below 110 for $N = 160$. This is substantially larger than 1-in-degree, but still significantly below ECA, which the longest simple cycle will always equal the number of cells(nodes). 
For 3-in-degree, see Figure \ref{fig:LSC_id3}. It is still below the max, as for $N = 100$, it must be below 95 and for $N = 160$, it must be below 150. Note that this is a theoretical max, and assuming the data is asymptotic, the true value must be below this value. To be clear, even if they are not asymptotic and simply linear, we still expect this to have some effect on the computation. Also, note that we could have fitted these values to an asymptotic curve but opted not to. This is because the asymptotic degree would greatly affect where the curve ends up even with minor changes, and we do not know what it is for 2 and 3 degrees. Therefore, we opted for a theoretical max with linear fitting of the form $y = mx + b$.

\subsection{Sensitivity in the x-bit memory benchmark}
\begin{figure}[h]
    \begin{center}
        \includegraphics[width=1\columnwidth]{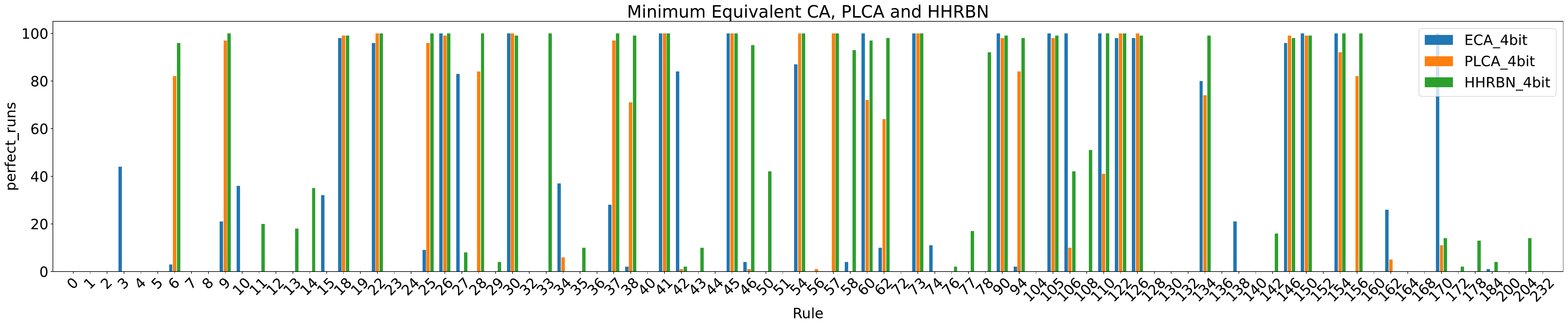}
        \includegraphics[width=1\columnwidth]{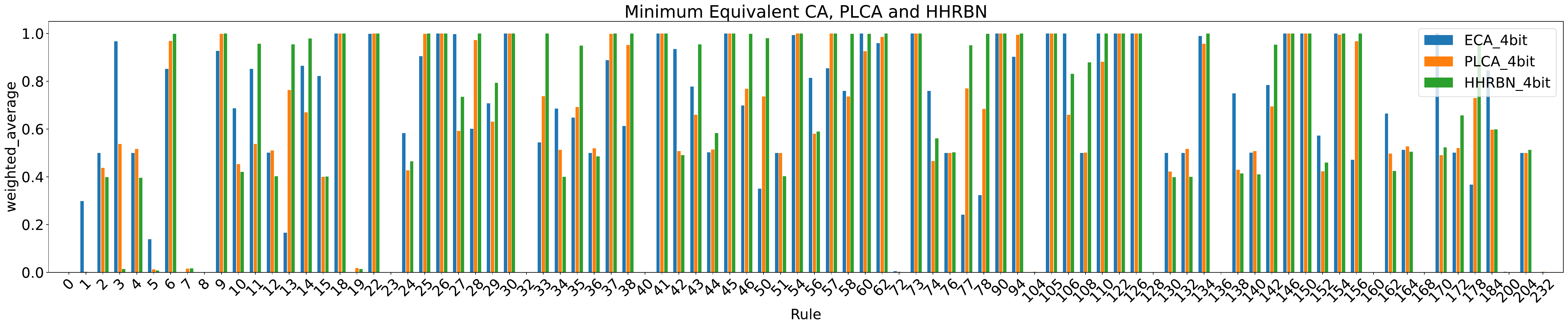}
        \caption{Breakdown of individual rule performance with 4-bit memory}
        \label{fig:4bit_sens}       
    \end{center}
\end{figure}

\begin{figure}[h]
    \begin{center}
        \includegraphics[width=1\columnwidth]{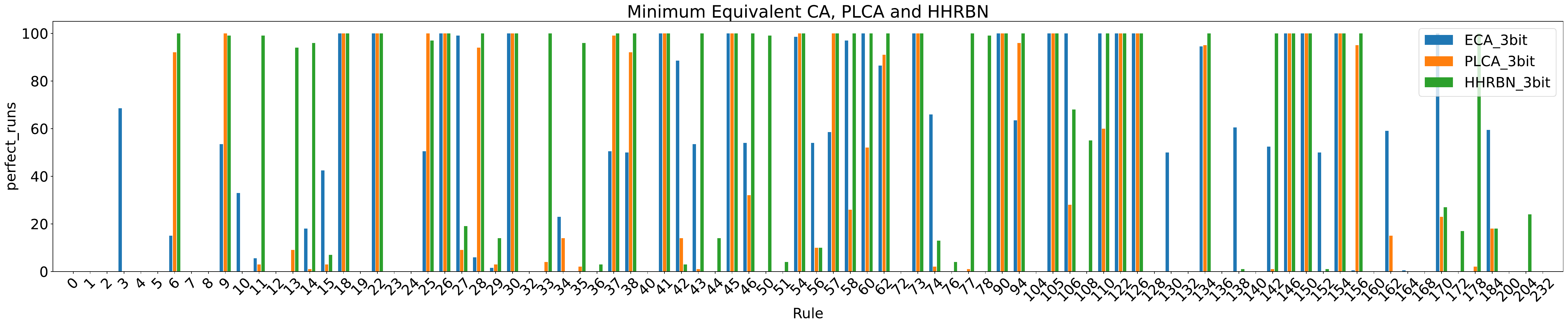}
        \includegraphics[width=1\columnwidth]{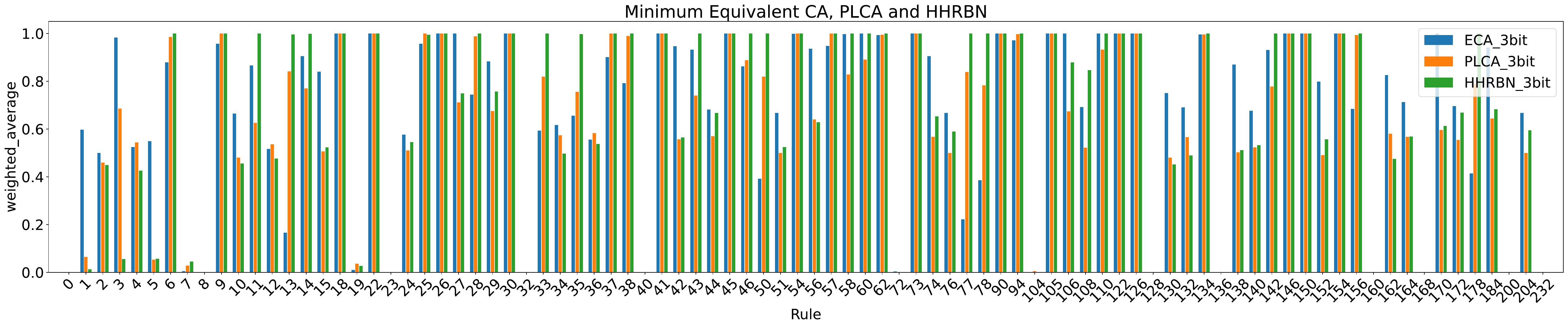}
        \caption{Breakdown of individual rule performance with 3-bit memory}
        \label{fig:3bit_sens}       
    \end{center}
\end{figure}

As was established in \citep{glover2023investigating}, the 5-bit memory benchmark can be solved with a simple random vector given sufficient dimensions. Therefore, it would be interesting to see if, as we compare CA, PLCA and HHRBN, as the tasks become easier, more and more rules can solve this issue and whether that would be true for HHRBN than for PLCA and PLCA compared to ECA. We expect HHRBN and PLCA to be more sensitive, and we expect them to perform better on the 3 and 4-bit memory benchmarks. We see from Figure \ref{fig:5bit_sens}, \ref{fig:4bit_sens} and \ref{fig:3bit_sens} that yes, more rules are capable of solving the easier tasks, but it is not entirely clear that a more significant portion of the rule-space of HHRBN or PLCA is finding the task trivial. If we view this through the average performance across the rule space, as seen in Table \ref{tab:Sen_avg}, there does seem to be a degree of this behaviour, but there are clear exceptions such as 3-bit $W.avg.$. The results corroborate the previous presented topology effects for the longest simple cycle and collapse rate in PLCA and HHRBN.  

\begin{table}[h]
\centerline{
    \begin{tabular}{|r|c|c|c|}
    \hline
    $N_b$ & Substrate & W.avg  & Perf. runs\\
    \hline\hline
        5 & ECA & 47.42 $\pm$ 33.29 & 8.67 $\pm$ 24.72  \\
        5 & PLCA & 35.64 $\pm$ 27.60 & 2.89 $\pm$ 9.17\\
        5 & HHRBN & 15.83 $\pm$ 16.65 & 0.15 $\pm$ 1.01\\
        5 & (ME) HHRBN & 14.00 $\pm$ 15.90 & 0.20 $\pm$ 1.31\\
    \hline
        4 & ECA & 60.89 $\pm$ 35.42& 27.40 $\pm$ 40.93\\
        4 & PLCA & 59.11 $\pm$ 35.02&  29.13 $\pm$ 42.58\\
        4 & HHRBN & 63.30 $\pm$ 38.93& 39.58 $\pm$ 45.70\\
        4 & (ME) HHRBN & 60.01 $\pm$ 41.28 & 39.78 $\pm$ 46.03\\
        \hline
        3 & ECA & 67.16 $\pm$ 35.77  & 39.36 $\pm$ 41.96\\
        3 & PLCA & 62.53 $\pm$ 35.12&  32.81 $\pm$ 43.56\\
        3 & HHRBN & 66.01 $\pm$ 38.24& 49.78 $\pm$ 47.49\\
        3 & (ME) HHRBN & 62.47 $\pm$ 40.93 & 48.87 $\pm$ 47.69\\
        \hline
    \end{tabular}
 }
    \caption{The Average and standard deviation performance over the rule-space in percentage. (ME) HHRBN is the average over only the HHRBN ME set (not the ECA ME set)}
    \label{tab:Sen_avg}
\end{table}

\section{Discussion}
This section focuses on discussing the results and, how they relate to each other, and how they relate to the field in a larger context. We will also give a short account of how we would approach the problem of a more natural definition of fully discrete chaos. 
\subsection{A complicated relationship between disordered topology and its computation}
We see through the collapse rate that in random topologies, the collapse rate significantly increases in PLCA and HHRBN, meaning random topology leads instead to more orderly computation. In contrast, if we look at results that don't collapse, we see a significant increase in sensitivity; we, in other words, have conflicting computational "forces". If we look at our results that would be affected by both "forces", we still see a slight trend towards disorder. However, if we consider individual rules such as rule 30, we would argue that as the collapse rate goes from 0\% to 33\% (PLCA) and 21\% (HHRBN), this rule, on average, becomes more orderly. 
Therefore, this relationship is complicated. 
\subsection{Implications for RBN and RBN reservoirs}
Our HHRBN has the same topology as Classical RBN, so implications beyond HHRBN can be derived. We demonstrated how different fixed in-degree networks should have a lower than max longest simple cycle. As RBN has random rules per node, the average dependency of RBN is slightly less than three. Therefore, the degree of RBN is also, in practice, less than three. This means that when applied to RBN, the results of the 3-in-degree random networks are, in practice, overestimates. It is likely already overestimated as we fit a line to a (hypothesised) asymptotic function. Therefore, the RBN's longest simple cycle in practice should be even smaller. We also see a significant collapse rate even in the most "chaotic" rule (rule 30). The natural assumption would be to see similar effects when applied to RBN. Therefore, we see many effects that reduce the sensitivity or disorderly nature of the computation that we would also hypothesise to apply RBN. 
\subsection{Weaknesses of this study}
Though this work is quite extensive, the work has some limitations. Due to computational constraints, we limit ourselves to CA grid sizes of a specific size, and we know that CA can exhibit significant behaviour changes on different grid sizes \cite{glover2023investigating}. 
\subsection{Size of the Edge of Chaos}
In \cite{sanchez2022heterogeneity, lopez2023temporal} several examples where introducing more heterogeneity extended the critical area. As we introduce topological heterogeneity to our networks, it would be interesting to consider if we are observing the same phenomenon of an increasing critical range. We begin by considering Wolfram's well-known ECA classification \cite{wolfram1984universality, martinez_note_2013}; there are only 11/88 rules in the "chaotic" class. We might also argue that the additive rules 60, 90 and 150 should not be considered "chaotic", making the space even smaller. Also, consider that there are only 4 complex rules, indicating that the ECA rule space is skewed towards ordered behaviour. As we see a small indication that introducing heterogeneity moves the space towards more sensitive "chaotic" behaviour, we should also expect to see more rules exhibit critical behaviour.
Nevertheless, the same evidence indicates the opposite of an extended criticality: a shrinking criticality. If we go by the collapse rate, we can say that the region of ordered behaviour has extended. If we control for the collapse rate, the remaining runs behave closer to sensitive "chaotic" behaviour; this indicates a shrunk critical range. If we go by the 5-bit memory benchmark, we again see signs of a shrunk criticality as very few runs managed to solve the benchmark. 
If we go by individual rules, one can argue that rules 170 and 184 behaved closer to critical, though the collapse rate is highly significant for these rules.
As our experiments were not specifically designed to identify the size of the critical area, we can not definitively conclude that our findings apply universally. A thorough examination of all individual rules might yield a different conclusion. It would not be surprising if heterogeneity could extend criticality. Still, the study of complexity often reveals that the relationship between critical behaviour, substrates, and hyper-parameters is intricate and complex. This rather indicates a rich field worthy of much study.  

\subsection{Implications for ReCA, ReRBN and intermediates}
We can consider some implications for reservoir computing with CA, PLCA, and HHRBN reservoirs from our results. 
If one considers the tool in practical terms, the ECA substrate seems superior as its regular topology lends reliability to the implementation, but also the localised neighbourhood affords easier implementation into FPGA, as the other substrates would naturally create more issues with the transfer of information due to the placement of neighbours. 

It is common practice in RC to have redundant mappings for encoding the input. As we observe a higher collapse rate in PLCA and HHRBN, reservoir usage of these substrates will benefit more from higher input redundancy than the ECA. 
Alternatively, one could more carefully select encoding placement into the network as there are likely to be nodes that afford better distribution of the perturbation than others.

\subsection{A "discrete" version of chaos}
\label{sec:discdiscChaos}

In this paper, we pointed out that the definition of chaos breaks down when applied to fully discrete systems, i.e., systems intrinsically discrete in space and time and with a discrete number of accessible states. 
The sensitivity on initial conditions required in the definition of chaotic function has 
a natural analogy for discrete systems, but what about the concept of dense periodic orbits and topological transitivity?

Here, we propose a definition of the meaning of a dense set of trajectories of a discrete system, with some additional mathematical formalism.
The set of accessible trajectories of a discrete system is dense in the full phase space of possible configurations if, for each configuration in phase space, the minimal Hamming distance, $\min{(D_h)}$ to an accessible trajectory converges to zero not slower than the size $N$ of the system: 
\begin{equation}
    \lim_{N \to \infty} N\min{(D_{h})} = 1 \, .
\end{equation}
In practice, this definition can be read as
\begin{equation}
    \min{(D_{h})} \sim \frac{1}{N} \, ,
\end{equation}
so graphically, plotting minimum Hamming distances as a function of the inverse of the system's size should hold a line with a unitary slope.
If we consider the orbit, the natural analogy is the attractor in binary systems, A cyclic trajectory that the deterministic discrete system must eventually converge to. Therefore, we conclude that a natural analogy for a dense periodic orbit is a long attractor that periodically expands and contracts the $D_h$ between previous states without finding the exact previous state. Expanding and contracting are important because a long attractor does not necessarily mean chaos. Consider the example of a binary vector that does iterative counting upwards by 1. This would have full coverage over the state space and the longest possible attractor, but it should not be considered chaotic behaviour.

As for the topological transitivity, it is similarly defined in continuous space \cite[subsection 1.8]{devaney2022introduction}:
an iterative map $f: J \to J$ is said to be topological transitivity if, for any pair of non-empty sets $U,V \subset J$, there exists $k\ge 0$ such that $f^k(U) \cap V \neq \emptyset $. 
In other words, it still implies that a system can not be decomposed into two subsystems. It also means that it must be possible from a given state to transition to any other state in the system. This condition is again reflected in the attractor, where there should be only one possible attractor in the system. The presence of two separate attractors would imply that the system can be decomposed, contradicting the notion of topological transitivity. 

\section{Conclusion}

This work investigated computational differences between ECA, PLCA, and HHRBN. It explores what happens with the simplest computational universe when introducing topological heterogeneity. We investigated using a simple 5-bit memory benchmark, sensitivity metric and collapse rate of the different substrates. We see how, in PLCA and HHRBN, performance on the 5-bit memory benchmark is substantially worse. That collapse rate increases substantially, which counterintuitively means that a more disordered topology can sometimes mean more ordered computation. In general, we see a weak sign of increased sensitivity, and if the collapse rate is controlled for, we see a strong sign of increased sensitivity. This indicates that we are observing a shrinking critical range. We see evidence consistent with the previous observations when we make the 5-bit memory benchmark easier by solving a 4 and 3-bit memory benchmark. 
Our results conclude that ECA is, at least with current hardware, the better reservoir for edge AI. We also try to address the issue of "chaos" in a fully discrete system and attempt to define a condition for the natural analogy. 
\section{Future Work}
Many future work projects can naturally extend this work. As we identified in the intro (Figure \ref{fig:SubstrateOverview}), there are many steps between ECA and BBN, which can be explored. Additionally, there are different paths between ECA and BNN, so other orders of steps could be explored. For example, would we get the same results if we introduce other forms of heterogeneity, such as mixed-rule CA? Furthermore, our networks have a random topology beyond what is typically true in most biological systems. It would be interesting to see how the regular topology (ECA) and the irregular topology (HHRBN) perform when compared to evolved networks such as the connectome of a C. elegans. Alternatively, if we explore networks with increasing locality of connections, this might even be a suitable control parameter for reservoir quality. 

\section{Contributions}

\begin{figure}[H]
    \begin{center}
        \includegraphics[width=1\columnwidth]{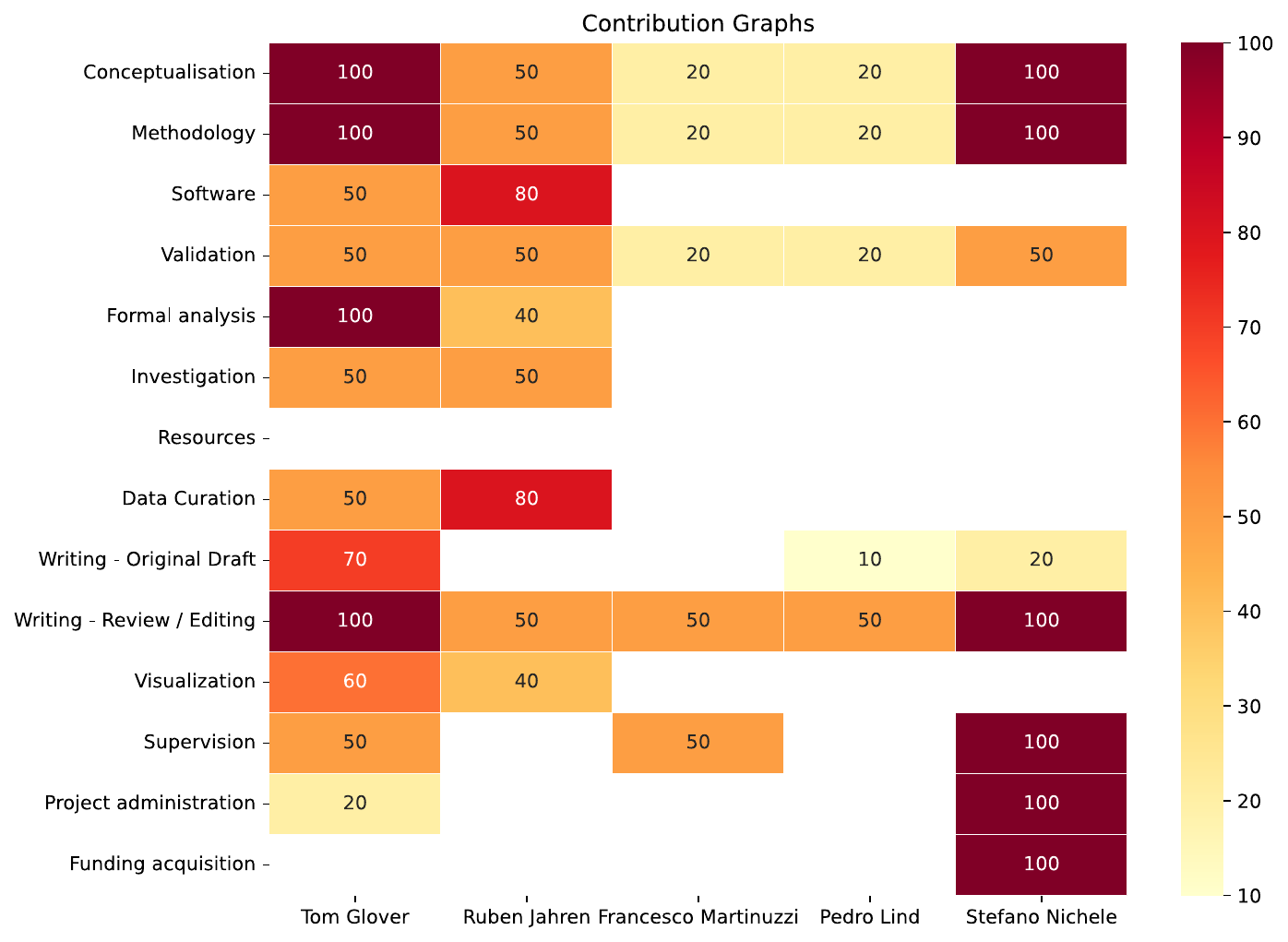}
        \caption{The contribution graph based on recommended categories from \cite{credit2024}. Contributions are based on degree of involvement. }
        \label{fig:contribGraph}       
    \end{center}
\end{figure}


\section{Acknowledgement}
This work was partially financed by the Research Council of Norway’s DeepCA project, grant agreement 286558.

\bibliography{bib}

\end{document}